\documentclass[manuscript]{acmart}

\AtBeginDocument{%
  }

\usepackage{cases}
\usepackage{amsfonts}
\usepackage{amsbsy}
\usepackage{bbm}
\usepackage{amsthm}
\usepackage{mathtools}
\usepackage{times}
\usepackage{array}
\usepackage{enumerate}
\usepackage{enumitem}
\usepackage{physics}
\usepackage{amsmath}
\usepackage{paralist}
\usepackage{graphicx}
\usepackage{subcaption}

\newtheorem*{theorem*}{Theorem}
\allowdisplaybreaks

\newcommand{\maxmargin}{\operatorname{Max-Margin}}
\newcommand{\sparse}{\operatorname{Sparse}}

\newcommand{\argmax}{\operatorname{argmax}}
\newcommand{\argmin}{\operatorname{argmin}}

\newcommand{\actpol}{\eta}

\newcommand{\stateset}{\ensuremath{\mathcal{X}}} 
\newcommand{\dataset}{\ensuremath{\mathcal{D}}}
\newcommand{\ndataset}{\ensuremath{\widehat{\mathcal{D}}}}
\newcommand{\actionset}{\ensuremath{\mathcal{A}}} 
\newcommand{\prior}{\ensuremath{\pi_0}} 

\newcommand{\reals}{\ensuremath{\mathbb{R}}} 
\newcommand{\grosspayoff}{\ensuremath{G}}

\newcommand{\prob}{\mathbb{P}}

\newcommand{\action}{a} 
\renewcommand{\state}{x}

\newcommand{\obs}{y}

\newcommand{\obsset}{\mathcal{Y}}

\newcommand{\RIcost}{C}
\newcommand{\RIcostest}{\widehat{\RIcost}}

\newcommand{\runcostinst}{c}

\newcommand{\utility}{\utilitysymbol(\state,\action)}
\newcommand{\utilitysymbol}{u}
\newcommand{\utilitysymbolbig}{\boldsymbol{U}}

\newcommand{\actselectagent}[1]{p_{#1}(\action\vert\state)}
\newcommand{\BDmat}{\boldsymbol{Q}}

\newcommand{\actselectagentnoise}[1]{\tilde{p}_{#1}(\action\vert\state)}
\newcommand{\actselectagentnoisesymb}[1]{\tilde{p}_{#1}}

\newcommand{\belief}{\pi}

\newcommand{\agent}{m}
\newcommand{\numagents}{M}

\newcommand{\utilitysymbolagent}[1]{\utilitysymbol_{#1}}

\newcommand{\exputilsymb}{J}
\newcommand{\exputil}{\exputilsymb_{\prior}}

\newcommand{\eps}{\varepsilon}

\newcommand{\actiontwo}{b}

\renewcommand{\dataset}{\mathcal{D}}

\newcommand{\dpiter}{k}
\newcommand{\dpitertwo}{j}
\newcommand{\dpset}{\mathcal{K}}

\newcommand{\attfunsymb}{\boldsymbol{\alpha}}
\newcommand{\attfunagent}[1]{\attfunsymb_{#1}}

\newcommand{\numdp}{K}

\newcommand{\blue}[1]{\textcolor{blue}{#1}}

\newcommand{\crp}{\operatorname{RP}}

\newcommand{\noise}{w}

\newcommand{\nonlinb}{g}
\newcommand{\nonlinbest}{\hat{\nonlinb}}
\newcommand{\response}{\boldsymbol{\beta}}
\newcommand{\nresponse}{\hat{\boldsymbol{\beta}}}

\renewcommand{\dim}{m}
\newcommand{\price}{\boldsymbol\alpha}

\newcommand{\utilityog}{u}

\newcommand{\BDsymb}{\mathcal{B}}
\newcommand{\BD}{\geq_{\mathcal{B}}}

\newcommand{\RPmod}{\operatorname{RP}_{\operatorname{mod}}}
\newcommand{\RImod}{\operatorname{RI}_{\operatorname{mod}}}

\newtheorem{theorem}{Theorem}
\newtheorem{assumption}{Assumption}
\newtheorem{definition}{Definition}%
\newtheorem{lemma}{Lemma}
\newtheorem{corollary}{Corollary}

\begin{document}

\title[Unifying Revealed Preference and Revealed Rational Inattention]{Unifying Revealed Preference and Revealed Rational Inattention}

\author{Kunal Pattanayak}
\email{kp487@cornell.edu}
\affiliation{%
  \institution{Cornell University}
  \streetaddress{Dept. of Electrical and Computer Engineering, Rhodes Hall, 136 Hoy Road}
  \city{Ithaca}
  \state{New York}
  \country{USA}
  \postcode{14850-14853}
}
\orcid{0000-0002-6684-4148}
\author{Vikram Krishnamurthy}
\email{vikramk@cornell.edu}
\affiliation{%
  \institution{Cornell University}
  \streetaddress{Dept. of Electrical and Computer Engineering, Rhodes Hall, 136 Hoy Road}
  \city{Ithaca}
  \state{New York}
  \country{USA}
  \postcode{14850-14853}
}
\orcid{0000-0002-4170-6056}

\renewcommand{\shortauthors}{Pattanayak and Krishnamurthy}
\begin{abstract}
  This paper unifies two key results from economic theory, namely,  revealed rational inattention~\cite{CD15} and classical revealed preference~\cite{AF67,FO09}. Revealed rational inattention tests for rationality of information acquisition for Bayesian decision makers. On the other hand, classical revealed preference tests for utility maximization under known budget constraints.  Our first result is an equivalence result - we unify revealed rational inattention~\cite{CD15} and revealed preference~\cite{AF67,FO09,BC15} through an equivalence map over decision parameters and partial order for payoff monotonicity over the decision space in both setups.
  Second, we exploit the unification result computationally to extend robustness measures for goodness-of-fit of revealed preference tests in the literature to revealed rational inattention. This extension facilitates quantifying how well a Bayesian decision maker's actions satisfy rational inattention. Finally, we illustrate the significance of the unification result on a real-world YouTube dataset comprising thumbnail, title and user engagement metadata from approximately 140,000 videos. We compute the Bayesian analog of robustness measures from revealed preference literature on YouTube metadata features extracted from a deep auto-encoder, i.e., a deep neural network that learns low-dimensional features of the metadata. The computed robustness values show that YouTube user engagement fits the rational inattention model remarkably well. All our numerical experiments are completely reproducible.
\end{abstract}




\maketitle

\section{Introduction} 
Afriat's theorem~\cite{SM38,HO50,AF67,VR82} in revealed preference theory gives necessary and sufficient conditions for a finite sequence of linear budget constraints and consumption bundles to  be rationalized by a monotone concave utility function. \cite{FO09} generalized Afriat's theorem to general (non-linear) budget sets and provided feasibility conditions for utility maximization under monotone budget constraints. More recently, in a Bayesian context in information economics, authors in \cite{CD15} address costly information acquisition and give necessary and sufficient conditions for a finite sequence of 
utility functions and action selection policies to be consistent with expected utility maximization with an information acquisition cost. 
In this paper, we refer to testing for costly information acquisition in \cite{CD15} as ``{\em revealed rational inattention}''.

Revealed preference and revealed rational inattention, respectively,  test for economics based rationality (optimal decision making under resource constraints) in a non-Bayesian and Bayesian sense, respectively.
So it is intuitively plausible that there exists a one-to-one correspondence between the two results. 
{\em Our key finding is that the NIAC (No Improving Attention Cycles) condition of \cite[Theorem~1]{CD15} in revealed rational inattention is a special case of the General Axiom of Revealed Preference (GARP)~\cite{VR82} used widely in revealed preference.}\footnote{Specifically, the NIAC condition~\cite{CD15} is equivalent (under an appropriate variable map) to the feasibility of Afriat inequalities~\cite{AF67} for GARP with the additional constraint that the feasible Lagrange multipliers are constant across all problem instances. On a related note, in Sec.\,\ref{sec:main_result}, we also discuss the equivalence between the NIAC condition~\eqref{eqn:niac-vanilla} and the cyclical monotonicity condition for testing quasi-linear utility maximization~\cite{BR07}.} To the best of our knowledge, this result is new\footnote{\cite{CD15} allude to the cyclical monotonicity condition of \cite{RO15} in the discussion of the NIAC condition. This provided us with additional motivation to investigate the connection between revealed preference and revealed rational inattention. Also, \cite{CH17-nonseparable} generalize the setup in \cite{CD15} and consequently propose a GARP-type condition for testing rational inattention. In Sec.\,\ref{sec:existing-works}, we argue how our unification result differs from that of \cite{CH17-nonseparable}.}, and stated formally in Theorem~\ref{thrm:eq}. To prove this result, we first develop a revealed preference to test for cost minimization subject to utility constraints, and then
extend the test to probability vectors in the unit simplex equipped with the Blackwell partial order~\cite{BW53}. 

Theorem~\ref{thrm:eq} states that GARP for non-linear budgets is a generalization of the NIAC condition~\cite{CD15}. Indeed, GARP is an acyclic condition due to unconstrained Lagrange multipliers (marginal utility values) and is a less restrictive condition compared to the the cyclical monotonicity structure of NIAC~\eqref{eqn:niac-vanilla}. To complete the connection between NIAC and GARP, we generalize the revealed rational inattention result of \cite{CD15} to test for expected utility maximization subject to a bound on the information acquisition cost (the result Lagrange multipliers need not be a constant across decision problems unlike \cite{CD15}). The NIAC generalizes to a condition we term `GARRI' (Generalized Axiom of Revealed Rational Inattention), and show GARRI is equivalent to GARP under the variable map of Theorem~\ref{thrm:eq}.

\begin{figure}[t]
    \centering
    \includegraphics[width=0.825\columnwidth]{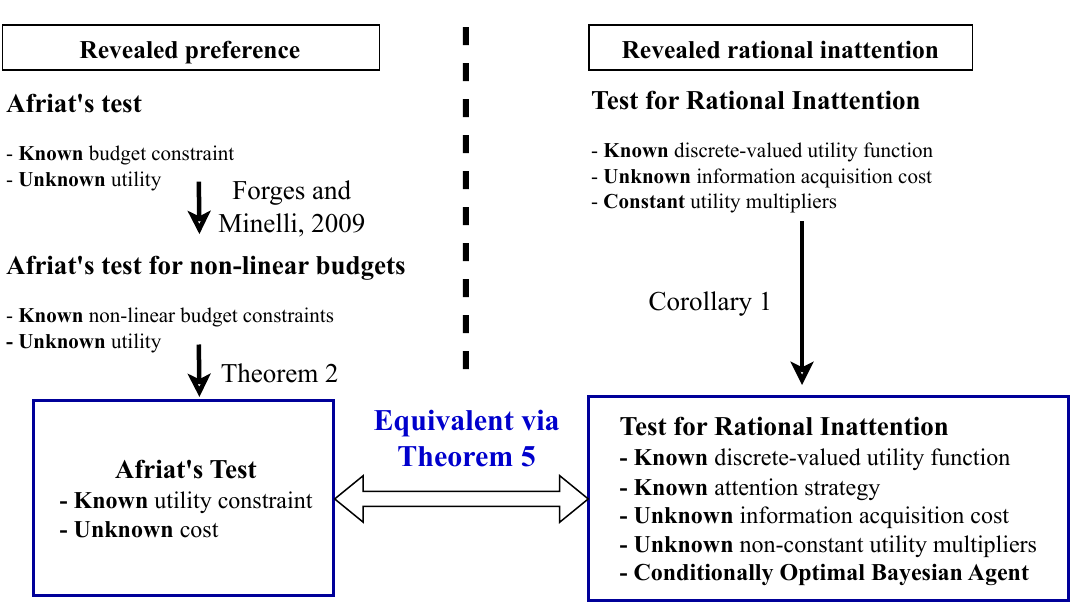}
    \caption{Schematic illustration of the main result in this paper. In Theorem~\ref{thrm:argmin}, we devise a revealed preference test for known utility constraints but unknown cost. In Corollary~\ref{corl:niac-to-garp}, we propose necessary and sufficient conditions for a costly information acquisition of a decision maker, where the decision maker follows a generalized decision model compared to that \cite{CD15}. Finally, in our main result, Theorem~\ref{thrm:eq}, we construct a one-to-one equivalence map between the revealed preference test of Theorems~\ref{thrm:argmin} and the revealed rational inattention test of Corollary~\ref{corl:niac-to-garp}, and show that the NIAC condition for revealed rational inattention is a special case of the GARP condition in revealed preference.}
    \label{fig:schematic}
\end{figure}

Since we will unify revealed preference and revealed rational inattention, the reader might wonder: how to abstract Bayes rule into the  revealed preference formulation? It is here that the usage of Blackwell partial order is critical. In the Bayesian framework, the decision maker computes the posterior belief of the state of nature via Bayes rule using a private measurement unknown to the analyst, and then takes an action observed by the analyst. From the analyst's perspective, the decision maker chooses analyst-observable probabilistic information structures that map the state to a distribution over actions. The observed information structures are termed as action selection policies in this paper, that link the decision maker's prior belief to its posterior belief given a chosen action. As a result, the action selection policy also determines the decision maker's expected utility. We will show that the consumption bundle in the revealed preference test translates to the action selection strategy (which is a probability distribution) in the revealed rational inattention test. In revealed preference, an element-wise higher consumption good yields a larger utility for the decision maker. Thus, the utility function is a monotonically increasing function of the consumption bundle with respect to the natural  (element-wise) partial order on the Euclidean space (space of consumption bundles).
{\em In complete analogy, for the Bayesian case, a more accurate action selection policy (in the Blackwell sense) results in a higher expected utility of the Bayesian decision maker. Equivalently, the expected utility in the Bayesian setup is monotone in the action selection policy with respect to the} \textbf{\textit{Blackwell order}}.\footnote{For revealed rational inattention, it suffices to ensure weak monotonicity of the expected utility with respect to attention strategies. One well-known partial order that satisfies this condition is the Blackwell~\cite{BW53} order.} 
This analogy is crucial for the main unification result of this paper, Theorem~\ref{thrm:eq} and is schematically shown in Fig.\,\ref{fig:schematic}. We formally discuss the Blackwell order and the monotonicity of expected utility wrt the Blackwell order in Lemma~\ref{lem:partialorder_relax}. To convey the key ideas early on in the paper, we present below an information version of our unification result, Theorem~\ref{thrm:eq} below:
\begin{theorem*}[Unification Result (Informal)]
(S1.) The NIAC condition~\cite{CD15} for revealed rational inattention is a special case of GARP~\cite{FO09} under the Blackwell partial order~\cite{BW53} and an appropriate variable map.\\
(S2.) The minimum modification needed in the decision model of \cite{CD15} so that a GARP-type condition is necessary and sufficient for revealed rational inattention is the addition of a multiplier that scales the decision maker's expected utility. We term the rational inattention analog of the GARP condition as `GARRI'~\eqref{eqn:GARRI}, defined formally in Corollary~\ref{corl:niac-to-garp}.
\end{theorem*}

Several reasons  motivate our paper. Revealed preference and revealed rational inattention are developed  largely independently in the literature; an exception being works like \cite{Var83} that use  revealed preference ideas to identify maximization of the mean (not Bayesian) utility. However, unlike \cite{CD15}, the expected utility maximization problem assumes the probability distribution over states of nature as an exogenous variable. With our  unification result, the results in these two areas can enrich each other. In Sec.\,\ref{sec:extension-robustness}, we extend the concept of robustness measures for goodness-of-fit in revealed preference literature to revealed rational inattention. We also illustrate the rational inattention analog of robustness measures to show rationally inattentive user engagement behavior in a massive YouTube dataset.\footnote{The term `user engagement' is used widely in the literature~\cite{KH17,user-eng-twitter,user-eng-orkut} to describe user interaction on online multimedia platforms.} Apart from applications in economics, revealed preference methods have also been applied in areas like machine learning, specifically for inverse reinforcement learning~\cite{NG00,LOP09,DIM11,HKP20,PK23}, adversarial signal processing~\cite{PKB22} and interpretable machine learning~\cite{PK21}.

\subsection*{Related Work}
Extending the revealed preference test of \cite{AF67} to more general partially ordered sets of consumption bundles dates back to \cite{RI66}, and more recently, to \cite{NIS17} where the consumption bundles are partially ordered via first-order stochastic dominance. \cite{FR16,FR21} generalize the  revealed preference test to the partial order over probability distributions (mixed strategies). Unlike the problem setting in this paper, the decision maker in \cite{FR21} does not update its belief via Bayes rule. The subtle distinction between \cite{FR21} and our work is that the decision maker's choice in this paper lies in the Cartesian product of probability simplices and thus requires a different partial order.  \cite{CR20} consider a generalized decision model (compared to \cite{CD15}) for the Bayesian agent, and give necessary and sufficient conditions for Bayesian rationality, namely, NIAS and GACI (Generalized Axiom of Costly Information) that generalize Theorem 1 in \cite{CD15}. In this paper, we focus primarily on the result of \cite{CD15} and its connection to  revealed preference. In spite of a unification flavor in the result of \cite{CR20} where GACI and GARP are discussed in a similar vein, the variable map in the unification result of this paper is distinct from that used by \cite{CR20} to formulate GACI. Finally, \cite{FRE22} unify multiple approaches in  revealed preference theory under an algebraic axiom of revealed preference. Our result builds on \cite{FRE22} in that we connect  revealed preference to revealed rational inattention~\cite{CM15,CD15} where the consumer's response takes the form of attention strategies and action selection policies, and the aim is to test for costly information acquisition. Also, we discuss relevant works that bridge revealed preference and revealed rational inattention in Sec.\,\ref{sec:existing-works} in more detail.

\subsection*{Terminology}
We use the terms `costly information acquisition', `rationally inattentive utility maximization', `rational inattention' and `Bayesian rationality' interchangeably in the paper. We also use the terms `decision maker' and `agent' interchangeably.

\begin{figure}
    \centering
    \includegraphics[scale = 0.5]{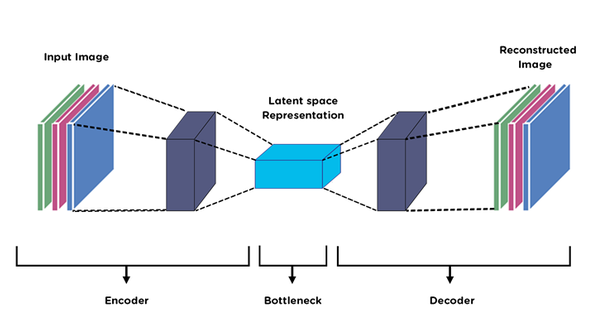}
    \caption{Schematic of the deep auto-encoder architecture (Image taken from \href{https://medium.com/@birla.deepak26/autoencoders-76bb49ae6a8f}{Medium}). The auto-encoder takes in as input and outputs an RGB image (high-dimensional data). The key feature of the auto-encoder is the `latent space representation' or the `bottleneck' in the above architecture, and is interpreted as a low-dimensional embedding of the input image. Auto-encoders are widely used in image processing and machine learning to generate compact representations of high-dimensional data~\cite{autoencoder-1,autoencoder-2,autoencoder-3,autoencoder-4}. In this paper, we train a auto-encoder to convert YouTube videos' thumbnails into 16 dimensional features, that are further mapped to one of 6 feature bins. The video thumbnail's feature bin, appended with the video title's sentiment (positive, negative,neutral) generated by an NLP-based sentiment analyzer is the `state' of the YouTube user in the rationally inattentive utility maximization model~\eqref{eqn:optimal-terminal-action},~\eqref{eqn:optimal-attention-strategy}.}
    \label{fig:autoencoder}
\end{figure}
\subsection*{Outline of Results}
{\em $\bullet$ Revealed preference background:}\\
In Sec.\,\ref{sec:background}, we introduce the key results of revealed preference and revealed rational inattention. We also propose: (i) the test $\RPmod$, an extension of the classical revealed preference test to identify the existence of a rationalizing cost subject to utility constraints on the decision maker, and (ii) the test, $\RImod$, a modification of the revealed rational inattention that introduces one extra degree of freedom in the objective function of the Bayesian decision maker.\\ {\em $\bullet$ Unification result - Relating GARP and NIAC:}\\
Theorem~\ref{thrm:eq} in Sec.\,\ref{sec:main_result} is our key unification result that says that $\RPmod$ is equivalent to $\RImod$ when the Bayesian decision maker is assumed to be conditionally optimal - it chooses the action that maximizes its expected utility conditioned on its posterior belief. Our second result, Corollary~\ref{corl:niac-to-garp}, introduced the minimal set of assumptions that facilitates testing for rational inattention via a GARP-type inequality, we term this inequality as {\em GARRI} (Generalized Axiom of Revealed Rational Inattention).\\
{\em $\bullet$ Discussion of related works:}\\
Sec.\,\ref{sec:existing-works} discusses existing works in the literature that bridge the methodologies of revealed preference and revealed rational inattention, and also highlights the differences between existing unification results and the key results of Sec.\,\ref{sec:main_result}.\\
{\em $\bullet$ Extending robustness measures in revealed preference to revealed rational inattention:}\\
Building on the unification results in Sec.\,\ref{sec:main_result}, Sec.\,\ref{sec:extension-robustness} introduces robustness measures for revealed rational inattention that measure how far a dataset is from being consistent with rationally inattentive behavior. We introduce Bayesian analogs of well-studied robustness measures in the literature, for example, the Afriat's efficiency index~\cite{AFT72-efficiency}, the Houtman-Maks index~\cite{HTM85-consistency} and the minimal perturbation test~\cite{VAR85-measurement}. To the best of our knowledge, robustness measures for revealed rational inattention have not been explored in the literature.\\
{\em $\bullet$ Robustness analysis of revealed rational inattention on a real-world YouTube dataset:}\\
Finally, in Sec.\,\ref{sec:real-world}, we perform a revealed rational inattention test on a real-world YouTube dataset comprising meta-data from approximately $140,000$ videos. The metadata in the YouTube dataset comprises the video thumbnail, title, viewcount, number of comments, and number of likes and dislikes on each video, recorded after 20 days of posting the video. The numerical experiments on the YouTube dataset extend our recent works~\cite{HKP20,PK23}  in the following aspects: (i) we present a novel deep auto-encoder and natural language processing (NLP)-based feature extraction procedure for  YouTube metadata (video thumbnail and title), (ii) we use the equivalence results developed in the paper to conduct  a systematic robustness analysis of YouTube metadata by computing robustness measures adapted from revealed preference theory, and (iii) we test for a generalized rational inattention model compared to that proposed by \cite{CD15}. Testing YouTube metadata for a generalized model of rational inattention is useful because YouTube user engagement literature~\cite{YT-diff-1,YT-diff-2,YT-diff-3} shows that different groups of YouTube users have different attention spans. In the rational inattention context, different attention spans translates to different marginal costs of information acquisition. The forward optimization model of \cite{CD15} is restrictive in that the decision maker has the same marginal cost in all decision problems (video categories in the YouTube context). The generalized model proposed in Corollary~\ref{corl:niac-to-garp} allows for non-constant attention spans in different decision problems. In the context of the YouTube dataset for our numerical experiments, we use the terms `user engagement' and `commenting behavior' interchangeably.
Our numerical results show that 
YouTube metadata features output pass the revealed rational inattention test by a large margin, where the goodness-of-fit to the rational inattention model is measured by computing the robustness metrics defined in Sec.\,\ref{sec:extension-robustness}. All our numerical results are completely reproducible and can be accessed from the GitHub repository \url{https://github.com/KunalP117/YouTube-Commenting-Analysis}.

\section{Background} \label{sec:background}
To set the stage for our unification result that relates revealed preference and revealed rational inattention, we start with a review of the key results of~\cite{FO09} (revealed preference for non-linear budgets) and~\cite{CD15} (revealed rational inattention).
 
\subsection{Revealed preference (non-linear budget)}
\begin{theorem}[\cite{FO09}] Consider a decision maker that, at time $\dpiter = 1,2,\ldots,\numdp$, chooses a consumption bundle $\response_\dpiter\in\reals_+^m$ subject to a non-linear budget constraint $\nonlinb_\dpiter(\response)\leq 0$. Assume that $\nonlinb_\dpiter(\cdot)$ is continuous, monotone and known to the external analyst, the set $\{\response\vert\nonlinb_{\dpiter}(\response)\leq 0\}$ of feasible bundles is compact, and the budget constraint is active at $\response_\dpiter$, that is, $\nonlinb_\dpiter(\response_{\dpiter})=0$. Then, the following statements are equivalent:
\begin{compactenum}[1.]
\item There exists a monotone, continuous utility function $\utilityog:\reals_+^m\rightarrow\reals$ that
rationalizes the data set $\{\response_\dpiter, \nonlinb_\dpiter(\cdot)\leq 0\}_{\dpiter=1}^\numdp$:
\begin{equation}
    \response_{\dpiter} \in \argmax_{\response\in\reals_+^\dim} \utilityog(\response),~\text{s.t.  } \nonlinb_\dpiter(\response)\leq 0. \label{eqn:utilitymaximization_nonlinear}
\end{equation}
\item The data set $\{\response_\dpiter, \nonlinb_\dpiter(\cdot)\leq 0\}_{\dpiter=1}^\numdp$ satisfies GARP:
\begin{align}\label{eqn:GARP_nonlinear}
    \response_\dpiter\geq_H\response_\dpitertwo \Rightarrow \nonlinb_\dpitertwo(\response_{\dpiter})\geq 0,~\forall\dpiter,\dpitertwo\in\{1,2,\ldots,\numdp\},~\dpiter\neq\dpitertwo,
\end{align}
where the relation $\response_\dpiter\geq_H\response_\dpitertwo$ (`revealed preferred to') means there exists indices $i_1,i_2,\ldots,i_L$ such that $\nonlinb_\dpiter(\response_{i_1})\leq 0,~\nonlinb_{i_1}(\response_{i_2})\leq 0,~\ldots,\nonlinb_{i_L}(\response_{\dpitertwo})\leq 0$.
\item There exist positive scalars $\utilityog_\dpiter,\lambda_\dpiter>0,~\dpiter=1,2,\ldots,\numdp$ such that the following inequalities hold:
\begin{equation}\label{eqn:Lagrange_nonlinearafriat}
    \utilityog_s-\utilityog_t-\lambda_t \nonlinb_t(\response_s)\leq 0\quad \forall t,s\in\{1,2,\ldots \numdp\}
\end{equation}
The reconstructed utility function $\hat{\utilitysymbol}$ defined in \eqref{eqn:construct_nonlinearafriat} below in terms of the feasible variables in \eqref{eqn:Lagrange_nonlinearafriat} is monotone, continuous and rationalizes $\{\response_\dpiter, \nonlinb_\dpiter(\cdot)\leq 0\}_{\dpiter=1}^\numdp$~\eqref{eqn:utilitymaximization_nonlinear}:
\begin{equation}\label{eqn:construct_nonlinearafriat}
    \utilityog(\response) = \min_{\dpiter\in\{1,\ldots,\numdp\}} \{\utilityog_\dpiter + \lambda_\dpiter \nonlinb_{\dpiter}(\response)\}
\end{equation}
\end{compactenum}
\label{thrm:nonlinearAfriat}
\end{theorem}
Theorem~\ref{thrm:nonlinearAfriat} says that a sequence of budget constraints and consumption bundles are rationalized by a utility function if and only if a set of linear inequalities~\eqref{eqn:Lagrange_nonlinearafriat} has a feasible solution. The GARP condition \eqref{eqn:GARP_nonlinear} is equivalent (due to \cite{VR82}) to the cyclical consistency condition proposed in \cite{AF67}. For completeness, we remark that constraining $\lambda_\dpiter$ in the feasibility test~\eqref{eqn:Lagrange_nonlinearafriat} to be a constant for all $\dpiter$, and assuming a linear budget $\nonlinb_\dpiter(\cdot)$ in Theorem~\ref{thrm:nonlinearAfriat} is equivalent to testing for quasi-linear utility maximization~\cite{BR07}, that is, the cyclical monotonicity condition of \cite[Theorem 2.2]{BR07} holds. Indeed, cyclical monotonicity for quasi-linear utility maximization is a stronger condition than GARP. 

Theorem~\ref{thrm:nonlinearAfriat} is a standard result in revealed preference literature. In Sec.\,\ref{sec:change-observables} below, we extend Theorem~\ref{thrm:nonlinearAfriat} to the scenario when the the analyst knows the decision maker's utility, and tests for the existence of budget constraints that rationalize the decision maker's actions.

\subsection{Modifying the revealed preference test: Observed utility, unobserved budget constraint}\label{sec:change-observables}
Our key aim in this paper is to establish a correspondence between revealed preference and revealed rational inattention. However, both results differ in what the analyst knows about the decision maker's decisions. Revealed rational inattention assumes the analyst knows the decision maker's utility function and tests for the existence of a rationalizing information acquisition cost. Hence, to establish the correspondence, we need to modify Theorem~\ref{thrm:nonlinearAfriat} to the case where the analyst knows the decision maker's utility function and tests for the existence of a rationalizing cost.\footnote{In classical revealed preference, the decision maker maximizes its utility (unknown to the analyst) subject to an upper bound on their budget (known to the analyst). Assumption~\ref{asmp:CRP} modifies the classical setup to the case where the decision maker minimizes a cost (unknown to the analyst) subject to a lower bound on their utility (known to the analyst). Under mild conditions on the decision maker's cost and utility, both optimization problems are equivalent. However, for notational convenience, we pose the decision problem in such a way that the revealed preference estimand is the decision maker's objective function.} We formalize this departure from the problem setting in Theorem~\ref{thrm:nonlinearAfriat} in Assumption~\ref{asmp:CRP} below.

\begin{assumption}\label{asmp:CRP} Consider the decision maker and analyst described in Theorem~\ref{thrm:nonlinearAfriat}. \\
(A1.1) The analyst's aim is to test if the decision maker chooses its consumption bundles optimally by solving the following optimization problem:
\begin{equation}\label{eqn:UM-nonlinear-alt-intro}
    \response_\dpiter \in \argmin_{\response\in\reals_+^\dim}~\nonlinb(\response),~\utilitysymbolagent{\dpiter}(\response)\geq \utilitysymbolagent{\dpiter}^\ast,~\forall\dpiter=1,2,\ldots,\numdp.
\end{equation}
In \eqref{eqn:UM-nonlinear-alt-intro}, the decision maker minimizes the cost of choosing bundle $\response$ subject to a lower bound on the utility.\\
(A1.2) The utility constraint in \eqref{eqn:UM-nonlinear-alt-intro} is active, that is, $\utilitysymbolagent{\dpiter}(\response_\dpiter) = \utilitysymbolagent{\dpiter}^\ast$ for all $\dpiter=1,2,\ldots,\numdp$.\\
(A1.3) The analyst knows the dataset $\dataset_{RP}$ defined as:
\begin{equation}\label{eqn:dataset-RP-alt}
    \dataset_{RP} = \{\utilitysymbolagent{\dpiter},\utilitysymbolagent{\dpiter}^*,\response_\dpiter\}_{\dpiter=1}^\numdp.
\end{equation}
(A1.4) The analyst's aim is to identify if there exists a cost $\nonlinb(\response)$ that rationalizes the analyst's dataset $\dataset_{RP}$~\eqref{eqn:dataset-RP-alt}, that is, the observed responses $\{\response_\dpiter\}_{\dpiter=1}^\numdp$ solve the optimization problem~\eqref{eqn:UM-nonlinear-alt-intro}.
\end{assumption}
In \eqref{eqn:UM-nonlinear-alt-intro}, we refer to $\nonlinb(\response)$ in \eqref{eqn:UM-nonlinear-alt-intro} as the `cost' of purchasing consumption bundle $\response$, in comparison to a budget {\em constraint} in classical revealed preference where the utility function is unknown. The optimization problem in \eqref{eqn:dataset-RP-alt} is a cost minimization problem subject to a lower bound on the utility. A related model is studied in \cite{VAR84-costminimization} where the decision maker at time $\dpiter=1,2,\ldots,\numdp$ minimizes a cost function but is constrained to choose its response from a specified compact set. In \cite{VAR84-costminimization}, analogous to WARP, the {\em Weak Axiom of Cost Minimization} (WACM) is proposed as a necessary and sufficient condition that rationalizes the dataset. In this paper, we focus on relating GARP from revealed preference theory to revealed rational inattention results. Hence, in Theorem~\ref{thrm:argmin} below, our necessary and sufficient condition for rationalizability is expressed in terms of the GARP condition~\eqref{eqn:GARP_nonlinear-alt} even though it is straightforward to express \eqref{eqn:GARP_nonlinear-alt} in the style of \cite[Theorem~1]{VAR84-costminimization}. We are now ready to state Theorem~\ref{thrm:argmin}. In complete analogy to Theorem~\ref{thrm:nonlinearAfriat}, Theorem~\ref{thrm:argmin} below states necessary and sufficient conditions for utility maximization when the analyst knows the decision maker's utility constraints~\eqref{eqn:UM-nonlinear-alt-intro}.

\begin{theorem}[Revealed Preference (Unknown Cost, Known Utility Constraints)] Consider a decision maker that, at time $\dpiter = 1,2,\ldots,\numdp$, chooses a consumption bundle $\response_\dpiter\in\reals_+^m$ subject to a utility constraint $\utilitysymbol_\dpiter(\response)\geq \utilitysymbol_\dpiter^\ast$. Suppose Assumption~\ref{asmp:CRP} holds and the set $\{\response\vert\utilitysymbol_{\dpiter}(\response)\geq \utilitysymbol_\dpiter^\ast\}$ of feasible bundles is compact for all $\dpiter$. Then, the following statements are equivalent:
\begin{compactenum}[1)]
\item There exists a monotone, continuous cost $\nonlinb:\reals_+^m\rightarrow\reals_+$ that rationalizes the dataset $\dataset_{RP}$, that is, \eqref{eqn:UM-nonlinear-alt-intro} holds.
\item The data set $\{\response_\dpiter,\utilitysymbol_\dpiter(\response_\dpiter)-\utilitysymbol_\dpiter(\cdot)\}_{\dpiter=1}^\numdp$ satisfies GARP~\eqref{eqn:GARP_nonlinear}:
\begin{align}\label{eqn:GARP_nonlinear-alt}
    \response_\dpiter\geq_H\response_\dpitertwo \Rightarrow \utilitysymbol_\dpitertwo(\response_\dpitertwo)\leq\utilitysymbol_\dpitertwo(\response_\dpiter),~\forall~\dpiter\neq\dpitertwo,
\end{align}
where the relation $\response_\dpiter\geq_H\response_\dpitertwo$ implies there exists indices $i_1,i_2,\ldots,i_L$ such that $\utilitysymbolagent{\dpiter}(\response_{i_1})\geq \utilitysymbolagent{\dpiter}(\response_\dpiter),~\utilitysymbolagent{i_1}(\response_{i_2})\geq \utilitysymbolagent{i_1}(\response_{i_1}),~\ldots,\utilitysymbolagent{i_L}(\response_{\dpitertwo})\geq \utilitysymbolagent{i_L}(\response_{i_L})$.
\item There exist positive scalars $\nonlinb_\dpiter,\lambda_\dpiter>0,~\dpiter=1,2,\ldots,\numdp$ such that the following inequalities hold:
\begin{equation}\label{eqn:Lagrange_nonlinearafriat_alt}
\begin{split}
     \nonlinb_\dpitertwo- \nonlinb_\dpiter-\lambda_\dpiter~ (\utilitysymbol_\dpiter(\response_\dpitertwo) - \utilitysymbol_\dpiter(\response_\dpiter)) \geq~0 &~\forall \dpiter,\dpitertwo\in\{1,2,\ldots,\numdp\},~\dpiter\neq\dpitertwo.\\
    \text{Or equivalently, }~ \lambda_\dpiter~\utilitysymbolagent{\dpiter}(\response_\dpiter) - \nonlinb_\dpiter \geq \lambda_\dpiter~\utilitysymbolagent{\dpiter}(\response_\dpitertwo) - \nonlinb_\dpitertwo,&~\forall \dpiter,\dpitertwo\in\{1,2,\ldots,\numdp\},~\dpiter\neq\dpitertwo
\end{split}
\end{equation}
We denote the inequalities~\eqref{eqn:Lagrange_nonlinearafriat_alt} in abstract notation as $\crp(\{\utilitysymbol_\dpiter,\response_\dpiter\})$.
The reconstructed cost $\nonlinbest$ defined in \eqref{eqn:construct_nonlinearafriat_alt} below in terms of the feasible variables in \eqref{eqn:Lagrange_nonlinearafriat_alt} is monotone, continuous and rationalizes $\{\response_\dpiter, \utilitysymbol_\dpiter(\cdot)\geq \utilitysymbol_\dpiter^\ast\}_{\dpiter=1}^\numdp$~\eqref{eqn:UM-nonlinear-alt-intro}:
\begin{equation}\label{eqn:construct_nonlinearafriat_alt}
    \nonlinbest(\response) = \max_{\dpiter\in\{1,2,\ldots \numdp\}} \{\nonlinb_\dpiter + \lambda_\dpiter~ (\utilitysymbol_{\dpiter}(\response)-\utilitysymbol_{\dpiter}(\response_\dpiter))\}.
\end{equation}
\end{compactenum}
\label{thrm:argmin}
\end{theorem}
Theorem~\ref{thrm:argmin} above yields necessary and sufficient conditions for utility maximization behavior when the decision maker's utility function is observed by the analyst; its budget (cost) is unobserved and must be reconstructed by the analyst by testing for feasibility of a set of Afriat-type inequalities~\eqref{eqn:Lagrange_nonlinearafriat_alt} for rationalizability. The proof of Theorem~\ref{thrm:argmin} is in the appendix. At first sight, \eqref{eqn:UM-nonlinear-alt-intro} in Theorem~\ref{thrm:argmin} appears to be a dual statement to the optimization problem~\eqref{eqn:utilitymaximization_nonlinear} in Theorem~\ref{thrm:nonlinearAfriat}. However, the proof does not use duality. Also, in comparison to the reconstruction procedure~\eqref{eqn:construct_nonlinearafriat} that yields a point-wise {\em minimum} of piece-wise monotone functions, notice the reconstructed cost $\nonlinb$ in~\eqref{eqn:construct_nonlinearafriat_alt} is a point-wise {\em maximum} of piece-wise monotone functions. Indeed, if $\utilitysymbolagent{\dpiter}(\cdot)$ is differentiable for all $\dpiter$, then the cost $\nonlinb(\response) = \max_{\dpiter\in\{1,\ldots,\numdp\}} \{\nonlinb_\dpiter + \lambda_\dpiter\nabla\utilitysymbolagent{\dpiter}(\response_\dpiter)'(\response-\response_\dpiter)\}$ also rationalizes the decision maker's actions. This result follows from \cite[Proposition 2]{FO09}. Notice how in comparison to a piece-wise linear, concave utility reconstruction in Afriat's theorem for linear budget constraints, we now have a piece-wise linear {\em convex} cost that rationalizes the decision maker's actions if $\utilitysymbolagent{\dpiter}$ is differentiable.

Recall that our key objective is to establish a correspondence between revealed preference and revealed rational inattention~\cite{CD15}. In revealed rational inattention, the analyst knows the agent's utility and tests for the existence of a rationalizing information acquisition cost. In Sec.\,\ref{sec:main_result}, we will show that the setup in \cite{CD15} is a Bayesian analog of Theorem~\ref{thrm:argmin} and relate the reconstructed cost in \eqref{eqn:construct_nonlinearafriat_alt} to the information acquisition cost. The key takeaway of Theorem~\ref{thrm:argmin} is that we have a revealed preference test for unobserved costs in terms of GARP. Authors in \cite{CH17-nonseparable} have generalized the forward optimization problem in revealed rational inattention~\cite{CD15} so that the inverse learner uses a Bayesian analog of GARP instead of NIAC to test for rational inattention. However, as will be discussed in Theorem~\ref{thrm:eq}, the variable map for relating NIAC and GARP in this paper is distinct from \cite{CH17-nonseparable}; we rely on the auxiliary revealed preference result of Theorem~\ref{thrm:argmin} to establish the one-to-one correspondence.


\subsection{Revealed Rational Inattention}\label{sec:vanilla-RRI}
Having introduced revealed preference results for non-linear budgets, we now turn our attention to the Bayesian utility maximization setup of \cite{CD15}. 
Since our aim is to relate Theorem~\ref{thrm:argmin} to revealed rational inattention, we review the key revealed rational inattention result of~\cite{CD15}.

Before we state the key result, we describe the decision model of the Bayesian agent. Suppose the decision maker acts in a sequence of decision problems $\dpiter=1,2,\ldots,\numdp$, where the decision problem parametrizes the decision maker's utility. The decision maker in \cite{CD15} is a Bayesian agent - it has a prior probability distribution $\prior$ over a finite set of states $\stateset$. In every decision problem $\dpiter$, the agent chooses an information structure $\attfunagent{\dpiter}(\cdot\vert\state)$ that maps every state $\state\in\stateset$ to a probability distribution over a finite set of subjective signals $\obsset$; the kernel $\attfunagent{\dpiter}$ is termed as the agent's {\em attention strategy} in decision problem $\dpiter$:
\begin{equation}\label{eqn:attention-strategy}
    \text{\em Attention Strategy}\hspace{0.3cm}\attfunagent{\dpiter}: \stateset\rightarrow\Delta(\obsset),
\end{equation}
where $\Delta(\obsset)$ denotes the space of probability distributions over the set $\obsset$.  The agent then observes a realized signal (random variable) $\obs\in\obsset$ (and not the ground truth $\state$) and updates its belief (posterior) of the unobserved state $\state$ using Bayes rule:
\begin{equation}\label{eqn:bayes-rule}
    \belief_\obs(\state) = \frac{\prior(\state)\attfunagent{\dpiter}(\obs\vert\state)}{\sum_{\state'\in\stateset} \prior(\state')\attfunagent{\dpiter}(\obs\vert\state')}
\end{equation}
After computing the belief $\belief_\obs$, the agent then chooses an action $\action$ from a finite set of actions $\actionset$ according the {\em conditional action policy}:
\begin{equation}\label{eqn:conditional-action-policy}
    \text{\em Conditional Action Policy}\hspace{0.3cm}\actpol_\dpiter: \Delta(\stateset)\rightarrow\actionset
\end{equation}
The agent's attention strategy $\attfunagent{\dpiter}$ and conditional action policy $\actpol_\dpiter$ induce the {\em action selection policy} $p_\dpiter(\action\vert\state)$ and is defined as the probability of choosing action $\action$ if the true state is $\state$:
\begin{equation}\label{eqn:action-selection}
    p_\dpiter(\action\vert\state) = \sum_{\obs\in\obsset}\attfunagent{\dpiter}(\obs\vert\state)~\actpol_\dpiter(\action\vert\belief_\obs),
\end{equation}
where $\belief_\obs$ is defined in \eqref{eqn:bayes-rule} and {\em conditional action selection policy} $\actpol_\dpiter(\cdot\vert\belief)$ is the probability the agent chooses action $\action$ given belief $\belief$ in decision problem $\dpiter$. In the revealed rational inattention problem, we assume the analyst knows the dataset:
\begin{equation}\label{eqn:dataset-Bayesian}
\dataset_{RRI} = \{\prior,\{p_\dpiter(\action\vert\state),\utilitysymbolbig_\dpiter\}_{\dpiter=1}^\numdp\},
\end{equation}
In \eqref{eqn:dataset-Bayesian}, $\prior$ is the prior probability distribution over the state space $\stateset$ and $p_\dpiter(\action\vert\state)$ is the agent's action selection policy defined in \eqref{eqn:action-selection}. The variable $\utilitysymbolbig_\dpiter$ is the agent's discrete-valued utility in decision problem $\dpiter$ that depends on the state $\state\in\stateset$ and action $\action\in\actionset$:
\begin{equation}\label{eqn:utility-bayesian}
    \utilitysymbolbig_\dpiter~\equiv \{\utilitysymbolbig_\dpiter(\state,\action)\in\reals,~\state\in\stateset,\action\in\actionset\}.
\end{equation}
The analyst's aim in revealed rational inattention is to test if the Bayesian agent acts optimally and maximizes its {\em expected} utility maximization less a non-negative information cost $\RIcost(\attfunagent{})$ that depends only the attention strategy $\attfunagent{}$. The forward optimization problem is termed as `rationally inattentive utility maximization' in the literature:
\begin{align}
&\text{\em Rationally Inattentive Utility Maximization:}\nonumber\\
(a)~&\actpol_\dpiter(\action\vert\belief_\obs) \in \underset{p\in\Delta(\actionset)}{\argmax}~\sum_{\state\in\stateset} \belief_\obs(\state)\utilitysymbolbig_\dpiter(\state,\action),~\forall\obs\in\obsset\label{eqn:optimal-terminal-action}\\  
(b)~&\attfunsymb_\dpiter \in \underset{\attfunsymb:\stateset\rightarrow\Delta(\obsset)}{\argmax}~ \exputil(\attfunsymb,\utilitysymbolbig_\dpiter) - \RIcost(\attfunsymb), \text{ where}\label{eqn:optimal-attention-strategy}\\
&\exputil(\attfunsymb,\utilitysymbolbig_\dpiter) = \sum_{\obs\in\obsset}~p_\dpiter(\obs)~\left(\max_{\action'\in\actionset}\sum_{\state}~\belief_\obs(\state) \utilitysymbolbig_\dpiter(\state,\action')\right),~ p_\dpiter(\obs) = \sum_{\state} \prior(\state)\attfunagent{\dpiter}(\obs\vert\state).~\label{eqn:def_J}
\end{align}
In \eqref{eqn:optimal-terminal-action}, $\belief_\obs$ denotes the agent's belief computed using Bayes rule in \eqref{eqn:bayes-rule} after observing signal $\obs\in\obsset$. In \eqref{eqn:optimal-attention-strategy}, $\Delta(S)$ denotes the space of probability distributions over a set $S$.  Eq.\,\ref{eqn:optimal-terminal-action} ensures that the agent maximizes its conditional expected utility given any posterior probability distribution $\pi$ computed using \eqref{eqn:bayes-rule}. Assuming \eqref{eqn:optimal-terminal-action} is true, \eqref{eqn:optimal-attention-strategy} ensures that agent's chosen attention strategy $\attfunagent{\dpiter}$ maximizes its objective function, namely, expected utility~\eqref{eqn:optimal-attention-strategy} minus an information acquisition cost. The analyst's aim is to test for the existence of an information cost $\RIcost$~\eqref{eqn:def_J} that rationalizes the dataset $\dataset_{RRI}$~\eqref{eqn:dataset-Bayesian}, that is, the {\em unobserved} agent variables $\{\attfunagent{\dpiter},\actpol_\dpiter\}_{\dpiter=1}^\numdp$ satisfy the optimality conditions~\eqref{eqn:optimal-terminal-action},~\eqref{eqn:optimal-attention-strategy}, which makes the revealed rational inattention result of \cite{CD15} stated below remarkable.

\begin{theorem}[Revealed rational inattention~\cite{CD15}]\label{thrm:BRP-vanilla} Consider a Bayesian agent that faces $\numdp$ decision problems, where decision problem  $\dpiter$ is parameterized by a finite set of states $\stateset$, signals $\obsset$, actions $\actionset$ and utility $\utilitysymbolbig_\dpiter$~\eqref{eqn:utility-bayesian}. Suppose an analyst knows the dataset $\dataset_{RRI}$~\eqref{eqn:dataset-Bayesian} that comprises the agent's action selection policies $\{\actselectagent{\dpiter}\}_{\dpiter=1}^\numdp$ and its utility functions $\{\utilitysymbolbig_\dpiter\}_{\dpiter=1}^\numdp$ in the $\numdp$ decision problems. Then, the following statements are equivalent:\\
1) There exists a monotone information acquisition cost $\RIcost(\attfunsymb)$ 
that rationalizes the dataset $\dataset_{RRI}$. That is, the agent's unobserved attention strategy $\attfunagent{\dpiter}$~\eqref{eqn:attention-strategy} and conditional action policy $\actpol_\dpiter$~\eqref{eqn:conditional-action-policy} solve the nested optimization problem~\eqref{eqn:optimal-terminal-action},~\eqref{eqn:optimal-attention-strategy} for all decision problems $\dpiter=1,2,\ldots,\numdp$.

\noindent 2) The dataset $\dataset_{RRI}$ satisfies the `No-Improving-Action-Switches' (NIAS) and the `No-Improving-Action-Cycles' (NIAC) conditions:
\begin{align}
&\text{NIAS:}~\sum_{\state\in\stateset}~p_{\dpiter}(\state\vert\action)~\left(~\utilitysymbolbig_\dpiter(\state,\action)-\utilitysymbolbig_\dpiter(\state,\actiontwo)~\right)\geq 0,~\forall~\action\neq\actiontwo,~\action,\actiontwo\in\actionset,~\dpiter=1,2,\ldots,\numdp\label{eqn:nias-vanilla}\\
&\text{NIAC:}~\text{For any sequence of distinct indices } i_1,i_2,\ldots,i_M\in\{1,2,\ldots,\numdp\}~(M\leq \numdp), \text{the following inequality holds}:\nonumber\\
& \quad\quad\quad\quad\sum_{m=1}^{M}~\left(\sum_{\state\in\stateset,\action\in\actionset}~\prior(\state)~p_{i_m}(\action|\state)~\utilitysymbolbig_{i_m}(\state,\action) - \grosspayoff(p_{i_{m+1}}(\action\vert\state),\utilitysymbolbig_{i_m}) \right) \geq 0,~\text{where }~i_{M+1}=i_1~\text{and}:\label{eqn:niac-vanilla}\\
& \quad\quad\quad\quad\grosspayoff(p_{j}(\action\vert\state),\utilitysymbolbig_{i}) = \sum_{\action}p_{j}(\action)~\max_{\actiontwo\in\actionset}\sum_{\state}p_{j}(\state\vert\action)~\utilitysymbolbig_{i}(\state,\actiontwo)\label{eqn:exputil-surrogate}
\end{align}
The variable $\grosspayoff(\cdot,\cdot)$ in \eqref{eqn:exputil-surrogate} is the Bayesian agent's surrogate expected utility. In \eqref{eqn:niac-vanilla}, the variable $p_{\dpiter}(\action)=\sum_{\state}\prior(\state)\actselectagent{\dpiter}$ is the marginal distribution of the action $\action$, the variable $p_{\dpiter}(\state\vert\action)=\prior(\state)\actselectagent{\dpiter}/p_\dpiter(a)$ is the posterior belief of the state when action $\action$ is realized.

\noindent 3) The dataset $\dataset_{RRI}$ satisfies the data-matching condition:\\
\begin{align}\label{eqn:data-matching}
\text{There exists }\actpol_\dpiter:\obsset\rightarrow\Delta(\actionset) \text{ s.t. }p_\dpiter(\action\vert\state) = \sum_{\obs} \actpol_\dpiter(\action\vert\obs)~\attfunagent{\dpiter}(\obs\vert\state),~\forall~\dpiter.
\end{align}
\end{theorem}
Theorem~\ref{thrm:BRP-vanilla} is well-known in the information economics literature~\cite{CD15}; see \cite[Sec.\,10.2]{CD15} and \cite[Appendix~C.2.2]{PK23} for the proof. The `No-Improving-Action-Switches' (NIAS)~\eqref{eqn:nias-vanilla} and `No-Improving-Action-Cycles' (NIAC)~\eqref{eqn:niac-vanilla} conditions in Theorem~\ref{thrm:BRP-vanilla} are necessary and sufficient for the existence of an information acquisition cost $\RIcost$ that rationalizes the dataset $\dataset_{RRI}$, that is, conditions  (a)~\eqref{eqn:optimal-terminal-action} and (b)~\eqref{eqn:optimal-attention-strategy} hold. The first term in the LHS in \eqref{eqn:niac-vanilla} is the expected utility of the agent in decision problem $\dpiter$. The second term in the LHS is the agent's {\em surrogate  expected utility} $\grosspayoff(p_{i_{m+1}}(\action\vert\state),\utilitysymbolbig_{i_m})$, surrogate since the expectation is in terms of the action selection policy of the agent, and not its attention strategy. It is straightforward to show that the attention strategy Blackwell dominates the action selection policy. We make the notion of Blackwell dominance precise in Lemma~\ref{lem:partialorder_relax} below. Due to Blackwell dominance and Lemma~\ref{lem:partialorder_relax}, we further have the following inequality:
\begin{equation}
\grosspayoff(p_{i_{m+1}}(\action\vert\state),\utilitysymbolbig_{i_m})\leq \grosspayoff(\attfunagent{i_{m+1}},\utilitysymbolbig_{i_m}) = \exputil(\attfunagent{i_{m}},\utilitysymbolbig_{i_m})~\eqref{eqn:def_J},
\end{equation}
where equality holds when $i_{m+1} = i_m$. A second observation that is crucial for the unification result of Theorem~\ref{thrm:eq} below is that $\grosspayoff(p_{i_{m}}(\action\vert\state),\utilitysymbolbig_{i_m}) = \exputil(\attfunagent{i_m},\utilitysymbolbig_{i_m}) =  \sum_{\state\in\stateset,\action\in\actionset}~\prior(\state)~p_{i_m}(\action|\state)~\utilitysymbolbig_{i_m}(\state,\action)$ if and only if NIAS holds. Proving this relation is straightforward and omitted for brevity. Intuitively, the term $\grosspayoff(p_{i_{m+1}}(\action\vert\state),\utilitysymbolbig_{i_m})$ is a surrogate for the agent's expected utility since the inverse learner does not know the agent's attention strategies. While necessity for optimality is straightforwardly determined, the sufficiency proof assumes $\actpol_\dpiter$ to be a one-to-one map; the surrogate  expected utility matches the true expected utility, that is, $\grosspayoff(p_{i_{m+1}}(\action\vert\state),\utilitysymbolbig_{i_m})= \grosspayoff(\attfunagent{i_{m+1}},\utilitysymbolbig_{i_m})$.

Abstractly, the NIAS condition is true if and only if condition (a)~\eqref{eqn:optimal-terminal-action} is true, and the NIAC condition is true if and only if condition (b)~\eqref{eqn:optimal-attention-strategy} is true. Finally, the data-matching condition~\eqref{eqn:data-matching} ensures that the dataset $\dataset_{RRI}$ is indeed generated from a {\em Bayesian} decision maker that makes an action based on its realized posterior belief.

{\em Discussion of Theorem~\ref{thrm:BRP-vanilla}.} 
\begin{compactenum}
\item Classical revealed rational inattention \cite{CD15} assumes that there exists a {\em single} utility function $\utilitysymbolbig(\state,\action),~\state\in\stateset,~\action\in\actionset$, and the Bayesian agent's action choice in decision problem $\dpiter$ is restricted to a subset $\actionset_\dpiter\subset\actionset$. This restriction can be equivalently modeled as the decision maker having a utility function $\utilitysymbolbig_\dpiter$ in decision problem $\dpiter$ without any restriction on the choice of actions.\footnote{The problem setting in \cite{CD15} is equivalent to setting $\utilitysymbolbig_\dpiter(\state,\action)=\utilitysymbolbig(\state,\action)$ if $\action\in\actionset_\dpiter$ and $-\infty$ otherwise, where $\utilitysymbolbig$ is the agent's fixed utility over decision problems.}
\item Abstractly, Theorem~\ref{thrm:BRP-vanilla} says that the analyst can test for rationally inattentive utility maximization~\eqref{eqn:optimal-terminal-action},~\eqref{eqn:optimal-attention-strategy} even if the analyst only has access to a stochastically garbled~\footnote{Indeed, $p_\dpiter(\action\vert\state) = \sum_{\obs\in\obsset}\actpol(\action\vert\belief_\obs)\attfunagent{\dpiter}(\obs\vert\state)$. Hence, the matrix $Q\in[0,1]^{\vert\actionset\vert\times\vert\obsset\vert}$ with elements $Q_{\action,\obs} = \actpol(\action\vert\belief_\obs)$ can be viewed as a noisy channel that takes as input the attention strategy $\attfunagent{\dpiter}$ and outputs the action selection policy $p_\dpiter(\action\vert\state)$. } version of the attention strategy, namely, the action selection policy. We discuss this concept of stochastic garbling in more detail later in the paper in the context of Blackwell~\cite{BW53} partial order on the space of probability distributions.
\end{compactenum}

\subsection{Revealed rational inattention and observability of Bayesian agent's decisions by the analyst}\label{sec:justification-actsel-attfun}
In revealed preference~(Theorem~\ref{thrm:argmin}), the analyst observes the agent decisions accurately. In revealed rational inattention (Theorem~\ref{thrm:BRP-vanilla}), the analyst observes a noisy version of the Bayesian agent's decisions. A key yet unusual takeaway of revealed rational inattention is that the analyst can test for rational inattention by treating the noisy measurement of the agent decision as the true decision. We justify this claim below.

In the rational inattention model~\eqref{eqn:optimal-terminal-action},~\eqref{eqn:optimal-attention-strategy}, the decision maker, in decision problem $\dpiter$, chooses its attention strategy $\attfunagent{\dpiter}$ and conditional action policy $\actpol_\dpiter$. The quantities of interest to the Bayesian decision maker, namely, the information acquisition cost~$\RIcost$ and expected utility $\exputilsymb$ only depend on $\attfunagent{\dpiter}$ and $\actpol_\dpiter$, in addition to the prior $\prior$ that is assumed known to the analyst. Hence, it is intuitive to expect that test for optimal Bayesian decision-making is possible only if the chosen attention strategies $\{\attfunagent{\dpiter}\}_{\dpiter=1}^\numdp$ and conditional action policies $\{\actpol_\dpiter\}_{\dpiter=1}^\numdp$ are known to the analyst performing revealed rational inattention.

However, unlike revealed preference, the analyst only observes the action selection policy $p_\dpiter(\action|\state) = \sum_{\obs\in\obsset}~\actpol_\dpiter(\obs)\attfunagent{\dpiter}(\obs|\state)$ that is a noisy (less informative) version of the attention strategy $\attfunagent{\dpiter}$. Theorem~\ref{thrm:BRP-vanilla} indicates that the knowledge of the action selection policies suffices for testing Bayesian rationality. In the context of our equivalence result stated in Theorem~\ref{thrm:eq} below, testing if rational inattention~\eqref{eqn:optimal-terminal-action},~\eqref{eqn:optimal-attention-strategy} holds is {\em equivalent} to testing if rational inattention~\eqref{eqn:optimal-terminal-action},~\eqref{eqn:optimal-attention-strategy} holds when the attention strategy $\attfunagent{\dpiter}$ is replaced by the action selection policy $p_\dpiter(\action|\state)$. Hence, for the purpose of our equivalence result, we can treat the effective Bayesian decision maker's `response' as simply its action selection policy $p_\dpiter(\action|\state)$ in decision problem $\dpiter=1,2,\ldots,\numdp$. Let us briefly elaborate on the above claims:
\begin{compactitem}
\item If the agent is Bayes rational~\eqref{eqn:optimal-terminal-action},~\eqref{eqn:optimal-attention-strategy}, then the optimality conditions~\eqref{eqn:optimal-terminal-action},~\eqref{eqn:optimal-attention-strategy} also  hold when the  attention strategy is replaced with the action selection policy, a noisy version of the attention strategy. Indeed, the proof of necessity of NIAS and NIAC for rational inattention shows that replacing the attention strategy in \eqref{eqn:optimal-terminal-action},~\eqref{eqn:optimal-attention-strategy} with the action selection policy, and testing for a weaker version of \eqref{eqn:optimal-attention-strategy} to ensure optimality over a finite number of strategies yields the NIAS and NIAC inequalities. The key component in the necessity of the NIAS and NIAC conditions for rational inattention is Blackwell dominance discussed in Lemma~\ref{lem:partialorder_relax}; at a deeper level, NIAS and NIAC is necessary for \eqref{eqn:optimal-terminal-action} and \eqref{eqn:optimal-attention-strategy} to hold since the attention strategy `Blackwell dominates' the action selection policy.
\item The sufficiency proof of NIAS and NIAC for rational inattention assumes a one-to-one map from the observation $\obs$ to the action $\action$. Since the attention strategy is not observed, the analyst can assume that the {\em observed} action selection policy is the same as the {\em unobserved} attention strategy without loss of generality. Afriat~\cite{AF67} computes a set-valued utility function that rationalizes the finite dataset $\dataset_{RP}$~\eqref{eqn:dataset-RP-alt}. In complete analogy, \cite{CD15} exploit a result from quadratic assignment problems~\cite{KM57} to construct a set-valued rational inattention cost that is non-zero at the observed finitely many action selection policies (or equivalently, the attention strategies) in the $\numdp$ environments, and $\infty$ elsewhere. Since the reconstruction of the rational inattention cost only occurs in the sufficiency part of the proof, it thus suffices to replace the attention strategy with the action selection policy and simply check for Bayesian rationality of the chosen action selection policies. For clarity, we also express the reconstructed information acquisition cost as a function of the action selection policy in the generalization of the revealed rational inattention test of \cite{CD15} stated in Corollary~\ref{corl:niac-to-garp} below and implicitly assume a one-to-one map from the observations to the actions.
\end{compactitem}

To summarize, in this section we justify how an analyst can test for Bayesian rationality~\eqref{eqn:optimal-terminal-action},~\eqref{eqn:optimal-attention-strategy} by {\em assuming the observed action selection policy is the same as the unobserved attention strategy}. The key idea is that since the attention strategy is not observed, the analyst can, without loss of generality, assume a one-to-one mapping from the observation $\obs$ to the action $\action$. As a result, in the equivalence result below, we will show that the Bayesian decision maker's equivalent response is the action selection policy, and not the unobserved attention strategy and conditional action policy. Also,  in Corollary~\ref{corl:niac-to-garp} (a generalization of Theorem~\ref{thrm:BRP-vanilla}), the reconstructed information acquisition cost from the revealed rational inattention test is expressed in terms of the action selection policy; it is assumed that the action selection policy is the same as the unobserved attention strategy.

\section{Main Result. Unification of  Revealed preference and Revealed rational inattention}\label{sec:main_result}
We present our first key result in this section that unifies revealed preference and revealed rational inattention. Our unification result is Theorem~\ref{thrm:eq} below. 
Informally, the key takeaway of Theorem~\ref{thrm:eq} is as follows:\\ {\em The NIAC condition of \cite{CD15} is a special case of GARP~\eqref{eqn:GARP_nonlinear-alt} if NIAS holds. If NIAC~\eqref{eqn:niac-vanilla} holds, then the GARP condition~\eqref{eqn:GARP_nonlinear-alt} for utility maximization is true under an equivalent variable map. If NIAS holds, and the Afriat-type feasibility inequalities \eqref{eqn:Lagrange_nonlinearafriat_alt} are feasible with the Lagrange multipliers $\lambda_\dpiter$ set to $1$ under the variable map, then the NIAC condition is true.}

Let us now formalize the above takeaways in Theorem~\ref{thrm:eq} below.
\begin{theorem}[Unification of revealed preference and revealed rational inattention] \label{thrm:eq}
Consider the revealed rational inattention result of Theorem~\ref{thrm:BRP-vanilla} and the revealed preference result of Theorem~\ref{thrm:argmin}. Recall that the analyst uses dataset $\dataset_{RRI}$~\eqref{eqn:dataset-Bayesian} to test for rational inattention~\eqref{eqn:optimal-terminal-action},~\eqref{eqn:optimal-attention-strategy} in Theorem~\ref{thrm:BRP-vanilla}, and uses dataset $\dataset_{RP}$~\eqref{eqn:dataset-RP-alt} to test for utility maximization~\eqref{eqn:UM-nonlinear-alt-intro}. Also, suppose the NIAS condition~\eqref{eqn:nias-vanilla} holds for the dataset $\dataset_{RRI}$. Then:
\begin{compactenum}
\item 
The NIAC condition \eqref{eqn:niac-vanilla} in Theorem~\ref{thrm:BRP-vanilla} is a special case of GARP~\eqref{eqn:GARP_nonlinear-alt} under the variable map below:
\begin{alignat*}{3}
    &\text{\underline{Revealed Preference}} & \quad \Leftrightarrow\quad && \text{\underline{Revealed Rational Inattention}} \\
    \bullet\quad &\text{Time step } \dpiter & \quad \Leftrightarrow\quad && \text{Decision problem }  \dpiter\\
    \bullet\quad &\text{Response } \response_\dpiter & \quad \Leftrightarrow\quad && \text{Action selection policy }p_\dpiter(\action|\state)\\
    \bullet\quad &\text{Utility Function } \utilitysymbolagent{\dpiter}(\response)  & \quad \Leftrightarrow\quad && \text{Expected Utility }~ \grosspayoff(p(\action|\state),\utilitysymbolbig_\dpiter)~\eqref{eqn:exputil-surrogate}\\
    \bullet\quad & \text{Utility Bound }\utilitysymbolagent{\dpiter}^\ast & \quad \Leftrightarrow\quad && \text{Expected Utility }~ \grosspayoff(p_\dpiter(\action|\state),\utilitysymbolbig_\dpiter)~\eqref{eqn:exputil-surrogate}\\
    \bullet\quad & \text{Cost } \nonlinb(\response) & \quad \Leftrightarrow \quad && \text{Information Acquisition Cost }~\RIcost~\eqref{eqn:def_J}\\
    \bullet\quad & \text{Element-wise partial order on the space} & \quad \Leftrightarrow\quad && \text{Blackwell partial order on the space}\\
    & \text{of consumption bundles (Euclidean space)} &  && \text{of attention strategies (pmfs)}
\end{alignat*}
\item The following modification of rationally inattentive utility maximization~\eqref{eqn:optimal-terminal-action},~\eqref{eqn:optimal-attention-strategy} generalizes NIAC in the revealed rational inattention test of Theorem~\ref{thrm:BRP-vanilla} to a GARP-type condition:\\
\underline{Modified Rationally Inattentive Utility Maximization} $~\equiv~$ \eqref{eqn:optimal-terminal-action} and the following modification of \eqref{eqn:optimal-attention-strategy} holds:
\begin{align}
    &~\attfunsymb_\dpiter \in \underset{\attfunsymb:\stateset\rightarrow\Delta(\obsset)}{\argmax}~ \lambda_\dpiter~\exputil(\attfunsymb,\utilitysymbolbig_\dpiter) - \RIcost(\attfunsymb),~\text{or equivalently,} \label{eqn:mod-optimal-attention-strategy}\\
    &~\attfunsymb_\dpiter \in \underset{\attfunsymb:\stateset\rightarrow\Delta(\obsset)}{\argmin}~\RIcost(\attfunsymb),~\text{subject to  }\exputil(\attfunsymb,\utilitysymbolbig_\dpiter) \geq \exputilsymb_\dpiter^\ast ,\label{eqn:mod-optimal-attention-strategy_alt}
\end{align}
where $\exputil(\cdot)$ is defined in \eqref{eqn:def_J}, and $\lambda_\dpiter>0$ in \eqref{eqn:mod-optimal-attention-strategy} is a utility multiplier.
We formalize the GARP-type generalization of NIAC in Corollary~\ref{corl:niac-to-garp} below. Also, the setup in \eqref{eqn:mod-optimal-attention-strategy} is the ``minimum'' modification needed wrt the forward decision model in \cite{CD15} specified by conditions \eqref{eqn:optimal-terminal-action},~\eqref{eqn:optimal-attention-strategy} that allows checking for optimality of the chosen attention strategy via a GARP-type condition.
\end{compactenum}
\end{theorem}

We prove Theorem~\ref{thrm:eq} in  Appendix~\ref{app:proof-outline} and briefly discuss the intuition behind the proof below. A key aspect of the unification result (statement (1)) in Theorem~\ref{thrm:eq} is that the equivalent variables in the Bayesian decision setup comprise only the variables observed by the external analyst. For example, although the analyst knows the Bayesian decision maker chooses the attention strategy $\attfunagent{\dpiter}$ and the action selection policy $\actpol_\dpiter$~\eqref{eqn:conditional-action-policy}, the equivalent response under the variable map is only the observed action selection policy $p_\dpiter(\action|\state)$ that depends on $\attfunagent{\dpiter}$ and $\actpol_\dpiter$. We justify this unusual claim in Sec.\,\ref{sec:justification-actsel-attfun}.

The key idea behind relating NIAC~\eqref{eqn:niac-vanilla} and GARP~\eqref{eqn:GARP_nonlinear-alt} is to first express the NIAC condition for revealed rational inattention test in Theorem~\ref{thrm:BRP-vanilla} as a feasibility inequality~\eqref{eqn:niac-feasibility} (see Appendix~\ref{app:proof-equivalence-Afriat-NIAC} for the proof), and then compare the feasibility inequality to the Afriat-type inequality~\eqref{eqn:Lagrange_nonlinearafriat_alt} for the modified revealed preference test in Theorem~\ref{thrm:nonlinearAfriat}. Under the variable map outlined in statement (1) in Theorem~\ref{thrm:eq} above, we observe in the proof that the inequality \eqref{eqn:niac-feasibility} is the same as \eqref{eqn:Lagrange_nonlinearafriat_alt} with the Lagrange multipliers $\lambda_\dpiter$ in \eqref{eqn:Lagrange_nonlinearafriat_alt} set to 1. As a result, we show that NIAC is a special case of GARP, when NIAS is true, under the variable map of Theorem~\ref{thrm:eq}.

\subsection{Discussion of Theorem~\ref{thrm:eq}} 
Theorem~\ref{thrm:eq} presents three key results on the unification of revealed preference and revealed rational inattention discussed in more detail below:
\begin{compactenum}
\item {\em Assuming NIAS holds for the unification result.} Recall from Theorem~\ref{thrm:BRP-vanilla} that the first term in the summation in the LHS of the NIAS feasibility condition is the expected utility under the joint distribution $\prior(\state)p_{i_m}(\action|\state)$. The variable map in statement (1) of Theorem~\ref{thrm:eq} requires that the surrogate expected utility $\grosspayoff(p_{i_m}(\action|\state),\utilitysymbolbig_{i_m})$ be equal to the expected utility $\sum_{\state,\action}~\prior(\state)p_{i_m}(\action|\state)\utilitysymbolbig_{i_m}(\state,\action)$ that holds only if NIAS is true. In words, the revealed rational inattention test of Theorem~\ref{thrm:BRP-vanilla} checks (a) if the decision maker chooses the optimal action given its posterior belief from a realized observation, and (b) if the decision maker's expected utility from the chosen attention strategy less the information acquisition cost exceeds that for any other attention strategy chosen in the remaining $\numdp-1$ decision problems. Assuming NIAS is true ensures the decision's expected utility $\sum_{\state,\action}~\prior(\state)p_{i_m}(\action|\state)\utilitysymbolbig_{i_m}(\state,\action)$ is the maximum possible utility for the decision maker, where the maximum is taken over all conditional action policies $\actpol_{i_m}$. Put differently, assuming NIAS to be true only requires the analyst to check NIAC for testing rational inattention, and hence, enables a one-to-one comparison with revealed preference.
\item {\em Treating the action selection policy as the effective response for the Bayesian agent.} The variable map in statement (1) in Theorem~\ref{thrm:eq} states the the equivalent response in the Bayesian setup is the action selection policy, a noisy version of the unobserved attention strategy chosen by the agent. Although unusual, it suffices for the analyst to treat the action selection policy to be the same as the attention strategy for testing Bayesian rationality; we discuss this in more detail in Sec.\,\ref{sec:justification-actsel-attfun}. 
\item {\em Relating NIAC and GARP via a variable map.} Statement (1) in Theorem~\ref{thrm:eq} establishes a one-to-one correspondence between revealed preference and revealed rational inattention and relates both approaches via a variable map. The key takeaway is that NIAC is a special case of GARP, when NIAS is true.
We see from the variable map in Theorem~\ref{thrm:eq} that in the Bayesian decision framework, the ``effective'' utility function from revealed preference is the surrogate expected utility $\grosspayoff$~\eqref{eqn:exputil-surrogate} that encodes both the prior pmf $\prior$ and utility $\utilitysymbolbig$, and depends on the observed action selection policy. The cost $\nonlinb$ in revealed preference translates to the information acquisition cost $\RIcost$ in revealed rational inattention.
\\
Statement (2) introduces a generalization of the forward optimization model considered in \cite{CD15} for which the NIAC condition generalizes to a GARP-type condition. We formalize the revealed rational inattention test for the generalized model in Corollary~\ref{corl:niac-to-garp} below. The generalization of NIAC, namely, GARRI defined in \eqref{eqn:GARRI} in Corollary~\ref{corl:niac-to-garp} is equivalent to a Bayesian analog of GARP, thus completely unifying revealed rational inattention and revealed preference.
\item {\em Generalizing the rational inattention model of \cite{CD15}.} The key distinction between the generalized model~\eqref{eqn:mod-optimal-attention-strategy} and the classical rational inattention model considered in~\cite{CD15} specified by~\eqref{eqn:optimal-terminal-action} and \eqref{eqn:optimal-attention-strategy} is the free variable $\lambda_\dpiter$. Analogous to Theorem~\ref{thrm:nonlinearAfriat} where the Lagrange multipliers~\eqref{eqn:Lagrange_nonlinearafriat} can be interpreted as the marginal utility of the decision maker, $\lambda_\dpiter$ in \eqref{eqn:mod-optimal-attention-strategy} can be interpreted as the marginal cost in the constrained cost minimization problem~\eqref{eqn:mod-optimal-attention-strategy_alt}. Eq.\,\ref{eqn:optimal-attention-strategy} is equivalent to \eqref{eqn:mod-optimal-attention-strategy} with $\lambda_\dpiter$ set to a constant. As a result, the revealed rational inattention test of Theorem~\ref{thrm:BRP-vanilla} yields the cyclic NIAC condition for checking Bayesian rationality~\eqref{eqn:optimal-attention-strategy}. However, the generalized model of \eqref{eqn:mod-optimal-attention-strategy} has $\lambda_\dpiter$ as a free variable and must be estimated by the external analyst in addition to testing for the existence of an information acquisition cost. The revealed rational inattention test for the generalized model yields the acyclic GARRI condition defined in \eqref{eqn:GARRI} in Corollary~\ref{corl:niac-to-garp} below, and is equivalent to GARP under the above variable map.
\item {\em Change of partial order from revealed preference to revealed rational inattention.} In the variable mapping of Theorem~\ref{thrm:eq}, the ``response'' $\attfunsymb_\dpiter$ in the Bayesian setup lies in the unit simplex of probability mass functions. More precisely, the response belongs to the space $\Delta(\obsset)^{\vert\stateset\vert}$, where $\Delta(\obsset)$ is the unit simplex of pmfs over the set of signals $\obsset$. Clearly, with respect to the natural element-wise partial order of Euclidean spaces, the expected utility $\exputil(\attfunsymb,\utilitysymbolbig)$ and information acquisition cost $\RIcost(\attfunsymb)$ are not monotonically increasing in $\attfunsymb$. Hence, the unification result of Theorem~\ref{thrm:eq} involves equipping the space of attention strategies with a different partial order, namely, the Blackwell partial order~\cite{BW53} for probability measures discussed in more detail below.
\end{compactenum}

\subsection{Change of partial order from revealed preference to revealed rational inattention}
The change of partial order from the natural element-wise partial ordering of positive vectors in the Euclidean space for revealed preference to the Blackwell~\cite{BW53} partial ordering of attention strategies is a key component in establishing the one-to-one correspondence in Theorem~\ref{thrm:eq} above. Let us discuss the Blackwell order in detail.
\begin{definition}[Blackwell order~\cite{BW53}]\label{def:BD} Consider two attention strategies $\attfunsymb,\bar{\attfunsymb}\in\Delta(\obsset)^{\vert\stateset\vert}$, where $\stateset$ and $\obsset$ denote the finite set of states and private signals, respectively, in the rationally inattentive utility maximization framework of Theorem~\ref{thrm:BRP-vanilla}. Then, $\attfunsymb$ Blackwell dominates $\bar{\attfunsymb}$ (denoted as $\attfunsymb\BD\bar{\attfunsymb}$) if there exists a row-stochastic matrix $\BDmat$ such that $\bar{\attfunsymb}=\attfunsymb~\BDmat$.
\end{definition}
The Blackwell order introduces the notion of monotonicity in the space of attention strategies (probability distributions). The Blackwell order is a partial order, since there exist attention strategy pairs that cannot be ordered via the Blackwell relation~(Definition~\ref{def:BD}). Intuitively, attention strategy $\attfunsymb$ Blackwell dominates $\bar{\attfunsymb}$ if $\bar{\attfunsymb}$ is a noisy (garbled)  version of $\attfunsymb$. In classical revealed preference results, the decision maker's response belongs to the Euclidean space. The standard assumption (and key to establishing revealed preferences results) is to impose a monotonicity condition on the decision maker's budget constraint with respect to the element-wise partial ordering for the Euclidean space. The Blackwell partial order can be viewed as a Bayesian analog of the element-wise ordering for rational inattention. Recall from the equivalence result of Theorem~\ref{thrm:eq} that a constraint on the expected utility is the rational inattention analog of the decision maker's budget constraint in revealed preference. For the reader's clarity, we show below the expected utility is monotone with respect to the Blackwell partial order.\footnote{That a convex functional in monotone with respect to the Blackwell partial order is well-known in the literature, and stated in the main text for completeness. In future work, we will investigate more general partial orders such as interval dominance and integral precision dominance and how they affect revealed rational inattention results.
}
 
\begin{lemma}\label{lem:partialorder_relax}
Consider the rationally inattentive utility maximization setup in Theorem~\ref{thrm:argmin}. Suppose NIAS holds, that is, the decision maker chooses the optimal action given its posterior belief. Also, suppose the space of attention strategies $\Delta(\obsset)^{\vert\stateset\vert}$ (probability simplices) are equipped with the Blackwell partial order $\BDsymb$ (Definition~\ref{def:BD}). Then, the decision maker's expected utility~\eqref{eqn:exputil-surrogate} is monotonically increasing and convex in the attention strategy.
\end{lemma} 
The proof of Lemma~\ref{lem:partialorder_relax} is in Appendix~\ref{appdx:proof-lemma}. Lemma~\ref{lem:partialorder_relax} facilitates the one-to-one correspondence between the revealed preference results of Theorem~\ref{thrm:argmin} (element-wise partial order over consumption vectors) and the revealed rational inattention result of Corollary~\ref{corl:niac-to-garp} (Blackwell partial order over attention strategies). 

\noindent{\em Remark.} Lemma~\ref{lem:partialorder_relax} states the expected utility is monotone is the Bayesian decision maker's `response', namely, the action selection policy under the Blackwell order. However, it is straightforward to show {\em any convex functional} is monotone wrt the Blackwell order. Indeed, the expected utility~\eqref{eqn:exputil-surrogate} is convex in the action selection strategy.

To summarize, in revealed preference, the cost of consumption is monotone with respect to the natural element-wise partial order on the Euclidean space. The reconstructed utility function~\cite{AF67} is a monotone function of the consumption cost, and hence, is also monotone with respect to the natural element-wise partial order. In complete analogy, in revealed rational inattention, the {\em expected} utility is monotone with respect to the Blackwell order on the space of attention strategies. Corollary~\ref{corl:niac-to-garp} below presents an Afriat-type reconstruction of a valid information acquisition cost. The reconstructed information acquisition cost is a monotone function of the expected utility, and hence, is also monotone with respect to the Blackwell order.

Finally, Lemma~\ref{lem:partialorder_relax} is particularly useful in justifying why testing for rational inattention (Theorem~\ref{thrm:BRP-vanilla}) is possible when only the action selection policies are known; hence, the analyst can treat the action selection policy $p_\dpiter(\action|\state)$ as the Bayesian decision maker's response in decision problem $\dpiter$.

\subsection{Generalizing Revealed Rational Inattention to Variable Attention Spans}
Having stated our equivalence result in Theorem~\ref{thrm:eq} and Lemma~\ref{lem:partialorder_relax} above, we now state our second theoretical result, Corollary~\ref{corl:niac-to-garp}. Recall from statement (1) in Theorem~\ref{thrm:eq} that NIAC is a special case of GARP. Corollary~\ref{corl:niac-to-garp} generalizes the revealed rational inattention result of \cite[Th.\,1]{CD15} to a decision model with added degrees of freedom to accommodate variable attention spans of the Bayesian decision maker in different decision problems, or equivalent, different marginal costs of information acquisition in the rational inattention setup of \eqref{eqn:optimal-attention-strategy}. The key idea is to introduce minimum additional degrees of freedom in the rationally inattentive utility maximization model of \cite{CD15} so that the rationalizability condition for the {\em inverse} task generalizes from NIAC to a GARP-type condition. We term the rational inattention analog of GARP as {\em GARRI}, defined in \eqref{eqn:GARRI} below.

\subsubsection*{Motivation to generalize the rational inattention model of \cite{CD15}}
Generalizing the revealed rational inattention test of \cite{CD15} is useful because empirical studies~\cite{YT-diff-1,YT-diff-2,YT-diff-3} on online user engagement data show that online multimedia users have {\em different} attention spans in different decision problems. In the rational inattention context, different attention spans translate to different marginal costs of information acquisition. The forward optimization model of \cite{CD15} is restrictive in that the decision maker has the same marginal cost in all decision problems (video categories in the YouTube context). In comparison, the generalized revealed rational inattention test proposed in Corollary~\ref{corl:niac-to-garp} below allows for non-constant attention spans of the decision maker in different decision problems. Also, cognitive psychology literature~\cite{diff-lambda-supp-1,diff-lambda-supp-2,diff-lambda-supp-3} suggests the human attention span (hence, the information acquisition cost) is task-dependent (decision problem-dependent), in contrast to the rational inattention model of \cite{CD15} where the expected utility and information acquisition cost are weighed equally for decision making. The above works serve as motivation to generalize the rational inattention model of \cite{CD15} to test for non-constant values for the margin cost of information acquisition in datasets aggregated from human decisions.

\begin{corollary}\label{corl:niac-to-garp} Consider the Bayesian decision maker in Theorem~\ref{thrm:BRP-vanilla}. Suppose the analyst knows the dataset $\dataset_{RRI}$~\eqref{eqn:dataset-Bayesian} and tests for generalized rationally inattentive utility maximization, namely, conditions \eqref{eqn:optimal-terminal-action} and \eqref{eqn:mod-optimal-attention-strategy}. In~\eqref{eqn:mod-optimal-attention-strategy}, the positive utility multiplier $\lambda_\dpiter$ is unknown to the analyst. Then, the following statements are equivalent:\\
1) There exists a monotone (wrt Blackwell order (Lemma~\ref{lem:partialorder_relax})) information acquisition cost $\RIcost(\attfunsymb)$ 
that rationalizes the dataset $\dataset_{RRI}$. That is, the dataset $\dataset_{RRI}$ satisfies the data-matching condition~\eqref{eqn:data-matching}, and the agent's unobserved attention strategy $\attfunagent{\dpiter}$~\eqref{eqn:attention-strategy} and conditional action policy $\actpol_\dpiter$~\eqref{eqn:conditional-action-policy} solve the nested optimization problem~\eqref{eqn:optimal-terminal-action} and \eqref{eqn:mod-optimal-attention-strategy} for all decision problems $\dpiter=1,2,\ldots,\numdp$.

\noindent 2) The dataset $\dataset_{RRI}$ satisfies the data-matching condition~\eqref{eqn:data-matching}, NIAS~\eqref{eqn:nias-vanilla} and the Generalized Axiom of Revealed Rational Inattention (GARRI) defined below. GARRI~\eqref{eqn:GARRI} is equivalent to GARP~\eqref{eqn:GARP_nonlinear-alt} under the variable map of statement (1) in Theorem~\ref{thrm:eq}.
\begin{align}\label{eqn:GARRI}
 \text{GARRI: } &  p_\dpiter(\action\vert\state)\geq_H p_\dpitertwo(\action\vert\state) \Rightarrow \grosspayoff(p_\dpiter(\action\vert\state),\utilitysymbolbig_\dpitertwo) \leq \grosspayoff(p_\dpitertwo(\action\vert\state),\utilitysymbolbig_\dpitertwo),
\end{align}
where the relation $p_\dpiter(\action\vert\state)\geq_H p_\dpitertwo(\action\vert\state)$ means there exists indices $i_1,i_2,\ldots,i_L$ s.t.\,  $\grosspayoff(p_{i_1}(\action\vert\state),\utilitysymbolbig_\dpiter)\geq  \grosspayoff(p_\dpiter(\action\vert\state),\utilitysymbolbig_\dpiter)$, $\grosspayoff(p_{i_2}(\action\vert\state),\utilitysymbolbig_{i_1})\geq \grosspayoff(p_{i_1}(\action\vert\state),\utilitysymbolbig_{i_1}),\ldots,\grosspayoff(p_\dpitertwo(\action\vert\state),\utilitysymbolbig_{i_L})\geq \grosspayoff(p_{i_L}(\action\vert\state),\utilitysymbolbig_{i_L})$. The surrogate expected utility $\grosspayoff$ in \eqref{eqn:GARRI} is defined in \eqref{eqn:exputil-surrogate}.

\noindent 3) The dataset $\dataset_{RRI}$ satisfies the data-matching condition~\eqref{eqn:data-matching} and NIAS~\eqref{eqn:nias-vanilla}. Also, there exist positive scalars $\lambda_\dpiter,~\runcostinst_{\dpiter},~\dpiter=1,2,\ldots,\numdp$ such that the following inequalities hold for all pairs of decision variables $\dpiter,~\dpitertwo,~\dpiter\neq\dpitertwo$:
\begin{align}
\lambda_\dpiter\sum_{\state\in\stateset,\action\in\actionset}~p_\dpiter(\action|\state)~\prior(\state)~\utilitysymbolbig_\dpiter(\state,\action)~ - ~\runcostinst_\dpiter~\geq~ \lambda_\dpiter~\sum_{\action\in\actionset}~p_\dpitertwo(a)~\left(\max_{\actiontwo\in\actionset}~\sum_{\state\in\stateset} p_\dpitertwo(\state|\action)~\utilitysymbolbig_\dpiter(\state,\actiontwo)\right)~ -~ \runcostinst_\dpitertwo,\label{eqn:niac-garri-feasibility}
\end{align}
where $p_\dpiter(\action)=\sum_{\state\in\stateset}\prior(\state)p_\dpiter(\action|\state)$ is the marginal distribution of the agent's action $\action$ in decision problem $\dpiter$, $p_\dpiter(\state|\action) = \prior(\state)p_\dpiter(\action|\state)/p_\dpiter(\action)$ is the posterior distribution of the agent's state in decision problem $\dpiter$.

\noindent 4) If NIAS and GARRI hold, the following reconstructed information acquisition cost $\RIcostest$ rationalizes the dataset $\dataset_{RRI}$:
\begin{equation}\label{eqn:reconstruct_RI}
\RIcostest(p(\action|\state))  = 
        \underset{\dpiter\in\{1,2,\ldots,\numdp\}}{\max}~\{ \runcostinst_\dpiter + \lambda_\dpiter(\grosspayoff(p(\action|\state),\utilitysymbolbig_\dpiter) - \grosspayoff(p_\dpiter(\action|\state),\utilitysymbolbig_\dpiter))\}
\end{equation}
where the positive scalars $\runcostinst_\dpiter,\lambda_\dpiter$ are feasible solutions of~\eqref{eqn:niac-garri-feasibility}, and $\grosspayoff(\cdot)$ is the decision maker's surrogate expected utility defined in \eqref{eqn:exputil-surrogate}.\\
The Afriat-type~\cite{AF67}  reconstructed cost $\RIcostest$ in \eqref{eqn:reconstruct_RI} is a point-wise maximum of monotone convex functions, monotone with respect to the Blackwell order on the space of action selection policies.
The reconstructed cost also satisfies the axiomatic properties of weak monotonicity and mixture feasibility as postulated in~\cite[Theorem 2]{CD15}.
\end{corollary}

The proof of Corollary~\ref{corl:niac-to-garp} is in Appendix~\ref{appdx:proof-corollary}.

{\em Equivalence between GARRI and GARP.} Corollary~\ref{corl:niac-to-garp} assumes the same problem setting as that of Theorem~\ref{thrm:BRP-vanilla}. The only difference in the problem setting between Theorem~\ref{thrm:BRP-vanilla} and Corollary~\ref{corl:niac-to-garp} is the additional scalar multiplier $\lambda_\dpiter$ in \eqref{eqn:mod-optimal-attention-strategy}. The key takeaway is that the NIAC condition for checking optimality of attention strategies across $\numdp$ decision problems is replaced by the generalized axiom of revealed rational inattention (GARRI)~\eqref{eqn:GARRI}. Indeed, NIAC is a special case of GARRI since \eqref{eqn:optimal-attention-strategy} is a special case of \eqref{eqn:mod-optimal-attention-strategy} with $\lambda_\dpiter=1$ for all $\dpiter$. In fact, the addition of the non-constant multiplier $\lambda_\dpiter$ is the minimum modification needed in the decision model of \cite{CD15} for checking the optimality of the chosen attention strategies via the cyclical consistency of condition of GARP, instead of the more restrictive cyclical monotonicity condition of NIAC.

{\em Afriat-type reconstruction of information acquisition cost.} Corollary~\ref{corl:niac-to-garp} builds on \cite{CD15} and reconstructs an Afriat-type monotone convex cost of information acquisition~\eqref{eqn:reconstruct_RI} that rationalizes the dataset $\dataset_{RRI}$~\eqref{eqn:dataset-Bayesian}. Afriat's~\cite{AF67} utility reconstruction involves `stitching' a piece-wise linear, concave utility function that rationalizes the agent's actions and is expressed in terms of the utility value earned by the agent at every time step and the budget constraints. In complete analogy, the piece-wise convex monotone cost~\eqref{eqn:reconstruct_RI} rationalizes the dataset $\dataset_{RRI}$~\eqref{eqn:dataset-Bayesian} and is expressed in terms of the information acquisition cost incurred by the agent in every decision problem and the surrogate expected utility functional $\grosspayoff$. Recall from Sec.\,\ref{sec:justification-actsel-attfun} that the analyst performing revealed rational inattention can treat the observed action selection policy as the unobserved attention strategy chosen by the Bayesian agent. Hence, the {\em reconstructed} cost of information acquisition is a function of the observed variable, namely, the action selection policy. The reconstructed utility function in \cite{AF67} is locally non-satiated, monotone and concave, and rationalizes the observed price and consumption bundles. In complete analogy, the reconstructed cost~\eqref{eqn:reconstruct_RI} is weakly monotonic (in information (Blackwell partial order~\cite{BW53})), mixture feasible (convex) and normalized, and rationalizes the observed utility functions and action selection strategies in $\dataset_{RRI}$~\eqref{eqn:dataset-Bayesian}.

{\em Remark.} The authors of \cite{CD15} generalize their results to posterior separable costs of information acquisition in \cite{CDL19rationalinattention}. Specifically, \cite[Theorem 2]{CDL19rationalinattention} provides a constructive procedure for recovering a posterior separable cost that satisfies the rationalizability axioms for optimality in decision making under posterior cost constraints. Apart from the Afriat-style of cost construction, the construction style in \cite[Theorem 2]{CDL19rationalinattention} differs from that in Theorem~\ref{thrm:eq} in one key aspect. Our reconstruction procedure assumes the inverse learner only has access to a finite set of agent utility functions from finitely many environments, that is, the completeness axiom in \cite[Axiom A4]{CDL19rationalinattention} is {\em not} assumed.

\section{Related Works} \label{sec:existing-works}
There are several works in the economics literature that generalize classical results of revealed preference and revealed rational inattention. In this section, we compare GARRI~\eqref{eqn:GARRI}, the Bayesian analog of GARP, in  Corollary~\ref{corl:niac-to-garp} with related existing works.

{\em GARRI and GAPP.} \cite{DEB18-GAPP} test for the existence a preference relation over {\em prices} set by the buyer, instead of a preference relation over consumption bundles. The key testable axiom is Generalized Axiom of Price Preference (GAPP). Corollary~\ref{corl:niac-to-garp} proposes a GARP-type condition, GARRI that generalizes NIAC. In complete analogy, GARP is a generalization of the GAPP condition of \cite{DEB18-GAPP}. A key point is that GAPP generalizes (and is not equivalent to) the test for quasi-linear utility maximization, even though the above relations might suggest drawing this conclusion. We discuss the relation between GARI and revealed preference test for quasi-linear utility maximization below.

{\em GARRI, NIAC and Strong law of demand.} We now relate revealed rational inattention~(Corollary~\ref{corl:niac-to-garp}) and revealed preference for quasi-linear utility maximization~\cite{BR07}. Theorem 2.2 in \cite{BR07} proposes testable conditions for quasilinear utility maximization and term this rationalization as strong law of demand. Test for quasilinear utility maximization checks for the existence of a utility function $\utility$ such that the following condition holds for $\dpiter=1,2,\ldots,\numdp$:
\begin{equation}\label{eqn:quasilinear-UM}
    \response_\dpiter \in\argmax_{\response:\price_\dpiter'\response\leq 1} \utilitysymbolagent{}(\response) - \price_\dpiter'\response-1,~\price_\dpiter'\response_\dpiter=1.
\end{equation}
The objective function of \eqref{eqn:quasilinear-UM} is a non-Bayesian analog of the decision framework of \cite{CD15}, with the response $\response_\dpiter$ replaced with $p_\dpiter(\action\vert\state)$, utility $\utilitysymbolagent{\dpiter}(\cdot)$ replaced with $\exputil(\cdot,\utilitysymbolbig_\dpiter)$, and cost $\price_\dpiter'(\cdot)$ replaced with the information acquisition cost $\RIcost(\cdot)$. Under the variable map of Theorem~\ref{thrm:eq}, the cyclical monotonicity condition of  \cite[Definition 3]{BR07}, a special case of GARP~\eqref{eqn:GARP_nonlinear-alt}, is equivalent to NIAC~\eqref{eqn:niac-vanilla}, a special case of GARRI~\eqref{eqn:GARRI}. Testing for quasilinear utility maximization~\cite[Theorem 2.2]{BR07} is equivalent to checking for the feasibility of Afriat's inequalities with constant Lagrange multipliers. In complete analogy, testing for the classical rational inattention setup of \cite{CD15} is equivalent to checking for the feasibility of \eqref{eqn:niac-garri-feasibility} with the Lagrange multipliers set to a constant.


{\em GARRI and GACI.} 
\cite{CH17-nonseparable,REH20-rational} generalize \cite{CD15} to the case where the decision maker maximizes a non-separable objective function. The key axiom that generalizes NIAC to the non-separable case is the {\em Generalized Axiom of Costly Information} (GACI)~\cite[Condition~1]{CH17-nonseparable}, that possesses the cyclical monotonicity structure of GARP. However, on careful examination, we observed that the variable map from the Bayesian to non-Bayesian decision framework in \cite{CH17-nonseparable} is distinct from the one proposed in Theorem~\ref{thrm:eq} in spite of the GARP flavor in both GACI and our proposed generalization of NIAC, namely, GARRI~\eqref{eqn:GARRI} in Corollary~\ref{corl:niac-to-garp}. \cite{CH17-nonseparable} relate the revealed attention strategy $p_\dpiter(\action\vert\state)$ to the price of a good, and the 
expected utility functional $\exputil(\cdot,\utilitysymbolbig_\dpiter)$ to the response of the decision maker in revealed preference. However, in our equivalence result, the expected utility functional is analogous to the utility constraint~\eqref{eqn:UM-nonlinear-alt-intro}, and attention strategy is analogous to the decision maker's response in the modified revealed preference setup in Sec.\,\ref{sec:change-observables}. The difference in the variable map between GARRI and \cite{CH17-nonseparable} to GARP can be attributed to Theorem~\ref{thrm:argmin} that yields a GARP-type condition for testing if the decision maker minimizes an unobserved cost subject to a lower bound on its utility. We remark that the decision model of \cite{CH17-nonseparable} accommodates both models of \cite{CD15} and Corollary~\ref{corl:niac-to-garp} as special cases. However, the addition of the scalar multiplier in \eqref{eqn:mod-optimal-attention-strategy} is the {\em minimum} modification required in the decision framework of \cite{CD15} for the dataset to be rationalized by a GARP-type condition.
\vspace{0.15cm}\\
 

\section{Extending Robustness Measures in Revealed Preference to Revealed Rational Inattention}\label{sec:extension-robustness}
We now exploit the equivalence result of Theorem~\ref{thrm:eq} and the generalized revealed rational inattention result of Corollary~\ref{corl:niac-to-garp} to construct robustness measures for the revealed rational inattention test. The key idea is to compute the minimum perturbation needed for a dataset to pass the revealed rational inattention test, namely, the feasibility of NIAS and GARRI conditions in Corollary~\ref{corl:niac-to-garp}. There are several works in the revealed preference literature that characterize how far a sequence of budget constraints and consumption bundles is from satisfying GARP~\eqref{eqn:GARP_nonlinear}. To the best of our knowledge, there is no formal approach in the literature to measure how well a dataset $\dataset_{RRI}$~\eqref{eqn:dataset-Bayesian} fits the rational inattention model.

Abstractly, the key idea behind the robustness measures in revealed preference is to minimally perturb the observed dataset so that GARP holds. A few notable robustness measures include:\\
1. The `Afriat Efficiency Index (AEI)'~\cite{AFT72-efficiency} that yields the minimum relaxation (expenditure wastage) needed in the budget constraints to rationalize the data.\\
2. The chi-squared `Minimal Perturbation Test (MPT)'~\cite{VAR85-measurement} where the analyst assumes an additive measurement error in the observed response, and performs a chi-squared test on the minimum $\mathcal{L}_2$-deviation from the observed responses such that the perturbed responses rationalize the dataset.\\
3. The `Money Pump Index (MPI)'~\cite{ECH11-moneypump} that yield the maximum profit a seller can make from a dataset violating GARP.\\
4. The `Minimum Cost Index (MCI)'~\cite{DEA16-mincost} that yields the lowest normalized cost of breaking all revealed preference cycles from a dataset.\\
5. The `Houtman-Maks Index (HMI)'~\cite{HTM85-consistency} that, for a specified rationalizability tolerance, outputs the largest subset that satisfies GARP.

In this section, we exploit the equivalence result of Theorem~\ref{thrm:eq} and extend the robustness measures described above to revealed rational inattention. For brevity, we only discuss the rational inattention analogs of robustness measures 1-3 above, namely, the Afriat Efficiency Index (AEI), Minimal Perturbation test (MPT) and the Money Pump Index (MPI). However, we emphasize that {\em any} robustness measure from the revealed preference literature can be extended to the revealed rational inattention test via the unification result of Theorem~\ref{thrm:eq}.

\subsection{Afriat Efficiency Index for Rational Inattention (RI-AEI)}
In the classical revealed preference setup with linear budget constraints, the Afriat Efficiency Index (AEI)~\cite{AFT72-efficiency} is a uniform lower bound on the scalar multiplier $e$ so that GARP holds for a dataset $\{\price_\dpiter,\response_\dpiter\}_{\dpiter=1}^\numdp$ The variables $\price_\dpiter$ and $\response_\dpiter$ denote the price vector and consumption bundle at time $\dpiter$, the consumer's budget constraint is given by $\price_\dpiter'\response_\dpiter\leq 1$. AEI is defined as:
\begin{equation}\label{eqn:AEI}
    \text{Afriat Efficiency Index }(\operatorname{AEI}) = \argmin_{e\in\geq 0}~e,~\text{such that GARP($e$) holds},
\end{equation}
where GARP($e$) is a generalization of GARP defined as:
\begin{equation}\label{eqn:garp-e}
    \mathbf{1}\{e~\price_\dpiter'\response_\dpiter\geq\price_\dpiter'\response_\dpitertwo\}~\price_\dpitertwo'(e~\response_\dpitertwo-\response_\dpiter)\leq 0,~\forall~\dpitertwo,\dpiter,~\dpitertwo\neq\dpiter,~e\geq 0
\end{equation}
In \eqref{eqn:garp-e} above, $\mathbf{1}\{\cdot\}$ is the indicator function. In words, the constraint in \eqref{eqn:garp-e} says that, for a relaxation level $e$, if $\response_\dpitertwo$ is e-affordable at time $\dpiter$, then it must be that $\response_\dpiter$ must not be e-affordable at time $\dpitertwo$. Clearly, setting $e=1$ in \eqref{eqn:garp-e} yields the classical GARP condition~\eqref{eqn:GARP_nonlinear} for linear budget constraints. The parameter $e$ can be viewed as a relaxation of the GARP condition. Hence, AEI measures the {\em minimum} relaxation in budget constraints needed for the dataset to satisfy utility maximization behavior. 

It is straightforward to show that, for a fixed value of $e$, checking if GARP($e$)~\eqref{eqn:garp-e} holds is equivalent to checking for the feasibility of the following set of linear inequalities:
\begin{equation}
\begin{split}
    &\text{Find non-negative scalars }\lambda_\dpiter,\utilitysymbol_\dpiter~\text{ such that } \utilitysymbol_\dpitertwo - \utilitysymbol_\dpiter - \lambda_\dpiter~\price_\dpiter'(\response_\dpitertwo - e~\response_\dpiter) \leq 0,~\text{for all index pairs }\dpitertwo,\dpiter,~\dpitertwo\neq\dpiter.
\end{split}    
\label{eqn:AEI-feasibility}
\end{equation}
Hence, AEI for the dataset $\{\price_\dpiter,\response_\dpiter\}_{\dpiter=1}^\numdp$ can be computed as:
\begin{equation}\label{eqn:AEI-compute}
    \boxed{\operatorname{AEI} = \argmin_{e,\{\lambda_\dpiter,\utilitysymbol_\dpiter\}_{\dpiter=1}^\numdp\geq 0}~e,~\text{ such that } \utilitysymbol_\dpitertwo - \utilitysymbol_\dpiter - \lambda_\dpiter~\price_\dpiter'(\response_\dpitertwo - e~\response_\dpiter) \leq 0},~\text{for all index pairs }\dpitertwo,\dpiter,~\dpitertwo\neq\dpiter.  
\end{equation}
If the dataset $\{\price_\dpiter,\response_\dpiter\}_{\dpiter=1}^\numdp$ satisfies Afriat's inequalities for utility maximization behavior, then AEI$\geq 1$ when computed via~\eqref{eqn:AEI-compute}.\footnote{Indeed, GARP($e$) is equivalent to the square matrix $A(e) = [\price_\dpiter'(\response_\dpitertwo-e~\dpiter)]_{\dpiter,\dpitertwo=1}^\numdp$ satisfying GARP. Using the relation between the elements of the square GARP matrix to the Afriat inequalities~\cite[Th.\,2]{FOS04}, the constraint in \eqref{eqn:AEI-compute} results as an equivalent formulation for GARP($e$).} We now extend AEI to the revealed rational inattention setup of Corollary~\ref{corl:niac-to-garp} by invoking the equivalence result of Theorem~\ref{thrm:eq}. For clarity, we term AEI for the revealed rational inattention case as {\em Rationally Inattentive}-AEI (RI-AEI).

\begin{definition}[Rationally Inattentive Afriat Efficiency Index (RI-AEI)] \label{def:RI-AEI} Consider an external analyst with the stochastic choice dataset $\dataset_{RRI}$~\eqref{eqn:dataset-Bayesian}. The rationally inattentive Afriat efficiency index (RI-AEI) is the {\em minimum} relaxation in expected utility constraints~\eqref{eqn:mod-optimal-attention-strategy_alt} required for the dataset to be consistent with rationally inattentive utility maximization behavior. RI-AEI is defined as:
\begin{equation}
\begin{split}\label{eqn:RI-AEI}
    &\operatorname{RI-AEI} = \underset{e\geq 1,\{\lambda_\dpiter,\runcostinst_\dpiter\}_{\dpiter=1}^\numdp\geq 0}{\argmin}~e,~\text{ such that NIAS~\eqref{eqn:niac-garri-feasibility} and the following set of inequalities hold:}\\
    &\runcostinst_\dpitertwo - \runcostinst_\dpiter -\lambda_\dpiter\left(\sum_{\action\in\actionset}~p_\dpitertwo(a)~\left(\max_{\actiontwo\in\actionset}~\sum_{\state\in\stateset} p_\dpitertwo(\state|\action)~\utilitysymbolbig_\dpiter(\state,\actiontwo)\right)~ -~e~\sum_{\state\in\stateset,\action\in\actionset}~p_\dpiter(\action|\state)~\prior(\state)~\utilitysymbolbig_\dpiter(\state,\action)\right)\geq 0,\\
    &\text{for all index pairs }\dpitertwo,\dpiter\in\{1,2,\ldots,\numdp\},~\dpitertwo\neq\dpiter.
    \end{split}
\end{equation}
\end{definition}
If $e=1$ satisfies the feasibility inequality in \eqref{eqn:RI-AEI} above, then the dataset is consistent with rationally inattentive utility maximization. If not, then the minimum value of $e$ for which \eqref{eqn:RI-AEI} has a feasible solution is bounded from below by $1$, in contrast to AEI~\eqref{eqn:AEI}, where the minimum perturbation needed for a dataset to satisfy GARP is less than unity. This difference arises due to the decision maker's constraint in the rationally inattentive case. Recall from \eqref{eqn:utilitymaximization_nonlinear} that for the non-Bayesian decision model in revealed preference, the decision maker faces an {\em upper} bound on its budget, whereas in the rational inattention setup of Corollary~\ref{corl:niac-to-garp}, the decision maker faces a {\em lower} bound on the expected utility~\eqref{eqn:mod-optimal-attention-strategy_alt}.

\subsection{Minimum Perturbation Test for Rational Inattention (RI-MPT)}
The minimum perturbation test (MPT) introduced in~\cite{VAR85-measurement} assumes the decision maker's chosen consumption bundles are measured in noise, and computes the {\em minimum} perturbation needed in the consumption bundles for the dataset to be consistent with utility maximization behavior.

Suppose the analyst has a noisy dataset $\ndataset_{RP} = \{\price_\dpiter,\nresponse_\dpiter\}_{\dpiter=1}^\numdp$, where $\nresponse_\dpiter = \response_\dpiter + \noise_\dpiter$ is a noisy version of the true response $\response_\dpiter$ unobserved by the analyst, and the measurement error $ \noise_\dpiter\sim f_\noise$ is an i.i.d.\ random variable with pdf $f_\noise$. Assume, WLOG, that the decision maker's budget constraint at time $\dpiter$ is given by $\price_\dpiter'\response_\dpiter\leq 1$.
Let us now introduce the null and alternate hypotheses $H_0$ and $H_1$:\\
$H_0$: The true (noiseless) dataset $\dataset_{RP}=\{\price_\dpiter,\response_\dpiter\}_{\dpiter=1}^\numdp$ satisfies GARP~\eqref{eqn:GARP_nonlinear}, $H_1$: The true dataset $\dataset_{RP}$ does NOT satisfy GARP~\eqref{eqn:GARP_nonlinear}.\\
The analyst then performs MPT on the noisy dataset $\ndataset_{RP}$, namely, computes a test statistic defined below and performs a hypothesis test to reject or accept the null hypothesis $H_0$:
\begin{equation}\label{eqn:mpt-compute}
\begin{split}
    &~\operatorname{MPT}: \phi(\ndataset_{RP})~\lessgtr_{H_1}^{H_0}~\eta_f,~\text{where the test statistic $\phi(\cdot)$ is defined as:}\\
    & \phi(\ndataset_{RP}) = \min_{\eps_{1,2,\ldots,\numdp}\in\reals^\dim,\{\lambda_\dpiter,\utilitysymbol_\dpiter\}_{\dpiter=1}^\numdp\geq 0}~\sum_{\dpiter=1}^\numdp \|\eps_\dpiter\|_2^2,~\text{such that}\\
    &(i)~\response_\dpiter + \eps_\dpiter\geq \boldsymbol{0},~\price_\dpiter'(\response_\dpiter+\eps_\dpiter)=1,~(ii)~\utilitysymbol_\dpitertwo - \utilitysymbol_\dpiter - \lambda_\dpiter~\price_\dpiter'(\response_\dpitertwo + \eps_\dpitertwo - \response_\dpiter - \eps_\dpiter) \leq 0,~\text{for all index pairs }\dpitertwo,\dpiter,~\dpitertwo\neq\dpiter,
\end{split}
\end{equation}
where $\eta$ is a parameter that bounds the detector's Type-I error probability $\prob(H_1|H_0)$. The rationale behind~\eqref{eqn:mpt-compute} is that if $H_0$ holds, then $\phi(\ndataset_{RP})$ is a lower bound on the measurement error $\sum_{\dpiter=1}^\numdp \|\response_\dpiter - \nresponse_\dpiter\|_2^2$. This observation further implies~(see \cite{KH17,PKB22-jrnl} for details) that the Type-I error probability of the hypothesis test is upper bounded by $F_w^{-1}(\eta_f)$, where $F_w(\cdot)$ is the cdf of the noise pdf $f_w$.

We now extend MPT~\eqref{eqn:mpt-compute} to the rationally inattentive utility maximization setup of \cite{CD15} by exploiting the equivalence result of Theorem~\ref{thrm:eq}. Suppose the analyst has a noisy dataset $\ndataset_{RRI} = \{\prior,\{\hat{p}_\dpiter(\action\vert\state),\utilitysymbolbig_\dpiter\}_{\dpiter=1}^\numdp\}$, where $\hat{p}_\dpiter(\action|\state)$ is a noisy version of the true action selection policy $p_\dpiter(\action|\state)$.\footnote{Noise in probability mass functions may arise due to multiple factors such as misspecification error, or computing the action selection policy empirically from a finite number of samples; see \cite{PK23} for a finite sample analysis of the revealed preference test.} For clarity, we term MPT for the revealed rational inattention case as {\em Rationally Inattentive}-MPT (RI-MPT).
\begin{definition}[Rationally Inattentive Minimum Perturbation Test (RI-MPT)] \label{def:RI-MPT} Consider an external analyst with the stochastic choice dataset $\dataset_{RRI}$~\eqref{eqn:dataset-Bayesian}. The rationally inattentive minimum perturbation test (RI-MPT) is the {\em minimum} perturbation in the observed action selection policies required for the dataset to be consistent with rationally inattentive utility maximization behavior. The RI-MPT is defined as:
\begin{equation}\label{eqn:RI-MPT}
\begin{split}
    &~\operatorname{RI-MPT}: \phi(\ndataset_{RRI})~\lessgtr_{H_1}^{H_0}~\eta_f,~\text{where the test statistic $\phi(\cdot)$ is defined below:}\\
    & \phi(\ndataset_{RRI}) = \min_{\{\actselectagentnoise{\agent},\lambda_\dpiter,\utilitysymbol_\dpiter\}_{\dpiter=1}^\numdp\geq 0}~\sum_{\dpiter=1}^\numdp\sum_{\state\in\stateset,\action\in\actionset}\|\actselectagentnoise{\agent} - \actselectagent{\agent}\|_2^2,~\text{such that}\\
    (i)&~\sum_{\action\in\actionset}~\actselectagentnoise{\dpiter}=1~\forall,\state,\dpiter~(\text{Valid pmf})\\
    (ii)&~NIAS:\sum_{\state\in\stateset}~\tilde{p}_\dpiter(\state|\action)(\utilitysymbolbig_\dpiter(\state,\action)-\utilitysymbolbig_\dpiter(\state,\actiontwo))\geq 0~\forall~\dpiter,\action,\actiontwo,\\\
    (iii)&~GARRI:\runcostinst_\dpitertwo - \runcostinst_\dpiter -\lambda_\dpiter\left(\sum_{\action\in\actionset}~\tilde{p}_\dpitertwo(a)~\max_{\actiontwo\in\actionset}~\sum_{\state\in\stateset} \tilde{p}_\dpitertwo(\state|\action)~\utilitysymbolbig_\dpiter(\state,\actiontwo) ~-~\sum_{\state\in\stateset,\action\in\actionset}~\tilde{p}_\dpiter(\action|\state)~\prior(\state)~\utilitysymbolbig_\dpiter(\state,\action)\right)\geq 0,\\
    &\quad\quad\quad\quad\text{for all index pairs }\dpitertwo,\dpiter,~\dpitertwo\neq\dpiter.
\end{split}
\end{equation}
In \eqref{eqn:RI-MPT} above, $\tilde{p}_\dpiter(\action) = \sum_{\state\in\stateset}~\actselectagentnoise{\agent}\prior(\state)$ is the marginal action probability, and $\tilde{p}_\dpiter(\state|\action) = \frac{\actselectagentnoise{\dpiter}\prior(\state)}{\sum_{\state'\in\stateset}~\actselectagentnoisesymb{\dpiter}(\action|\state')\prior(\state')}$ is the posterior state distribution given action selection policy $\actselectagentnoise{\dpiter}$.
\end{definition}
RI-MPT defined in \eqref{eqn:RI-MPT} above is a hypothesis test that considers the minimum perturbation needed in the action selection policies for the feasibility of NIAS and GARRI conditions as the sufficient statistic for the test. In complete analogy to \eqref{eqn:mpt-compute}, the variable $\eta_f$ in \eqref{eqn:RI-MPT} controls the Type-I error probability of detecting Bayesian rationality, that is, NIAS and GARRI conditions hold for the noise-less dataset $\dataset_{RRI}$~\eqref{eqn:dataset-Bayesian}.

\subsection{Money Pump Index for Rational Inattention (RI-MPI)}
The money pump index (MPI) introduced in \cite{ECH11-moneypump} quantifies the severity of violations of GARP. If the decision maker's choices are observed in noise, the computed value of MPI can be used to test if the decision maker is rational or not. In this section, we consider the case where the decision maker's responses are measured accurately. MPI is defined for a sequence of tuples of prices and consumption bundles that violate GARP. Consider the classical revealed preference setup in Theorem~\ref{thrm:nonlinearAfriat} with linear budget constraints $\price_\dpiter'\response_\dpiter\leq1$. Suppose the sequence $\{\price_{\dpiter_i},\response_{\dpiter_i}\}_{i=1}^l$ violates GARP ($l\leq \numdp$). Then MPI for the violating sequence is defined as:
\begin{equation}\label{eqn:mpi-compute}
    \boxed{\operatorname{MPI}_{\{\price_{\dpiter_i},\response_{\dpiter_i}\}_{i=1}^l} = \frac{1}{l}\sum_{i=1}^l(\price_{\dpiter_{i+1}} - \price_{\dpiter_i})'(\response_{\dpiter_{i+1}} - \response_{\dpiter_i})}\quad\quad\quad(\dpiter_{l+1}\equiv\dpiter_{1})
\end{equation}
If GARP fails for the sequence $\{\price_{\dpiter_i},\response_{\dpiter_i}\}_{i=1}^l$, it is straightforward to show that MPI$(\{\price_{\dpiter_i},\response_{\dpiter_i}\}_{i=1}^l)>0$ in \eqref{eqn:mpi-compute}. Intuitively, MPI measures the profit a malicious arbitrager can make by buying the consumption bundles in the GARP-violating sequence at a lower price, and selling the same bundles to the non-rational decision maker at a higher price.

We now extend MPI~\eqref{eqn:mpi-compute} to the rationally inattentive utility maximization setup of \cite{CD15} by exploiting the unification result of Theorem~\ref{thrm:eq}. We term MPI for the revealed rational inattention case as {\em Rationally Inattentive}-MPI (RI-MPI).
\begin{definition}[Rationally Inattentive Money Pump Index (RI-MPI)]\label{def:RI-MPI} Consider an external analyst with the stochastic choice dataset $\dataset_{RRI}$~\eqref{eqn:dataset-Bayesian}. The rationally inattentive money pump index RI-MPI is defined as:
\begin{equation}\label{eqn:RI-MPI}
    \operatorname{RI-MPI} = \max_{\{\price_{\dpiter_{1:l}},\response_{\dpiter_{1:l}},~l\leq \numdp\}}~ \frac{\sum_{i=1}^l
    \bar{\exputilsymb}_{\prior,\dpiter_{i},\dpiter_{i+1}}(\actselectagent{\dpiter_{i}},\actselectagent{\dpiter_{i+1}})}{\sum_{i=1}^l \exputil(\actselectagent{\dpiter_i},\utilitysymbolbig_{\dpiter_i}) } 
\end{equation}
where $\dpiter_{l+1}\equiv\dpiter_{1}$ and $\bar{\exputilsymb}_{\prior,\dpiter_{i},\dpiter_{i+1}}(\actselectagent{\dpiter_{i}},\actselectagent{\dpiter_{i+1}})$ defined below is the net expected utility a malicious arbitrager can gain by exploiting the fact that the Bayesian decision maker's choices fail the GARRI~\eqref{eqn:GARRI} condition:
\begin{equation}\label{eqn:exputil-diff}
\bar{\exputilsymb}_{\prior,\dpiter_{i},\dpiter_{i+1}}(\actselectagent{\dpiter_{i}},\actselectagent{\dpiter_{i+1}}) = \exputil(\actselectagent{\dpiter_{i+1}},\utilitysymbolbig_{\dpiter_i}) - \exputil(\actselectagent{\dpiter_{i}},\utilitysymbolbig_{\dpiter_i}) + \exputil(\actselectagent{\dpiter_{i}},\utilitysymbolbig_{\dpiter_{i+1}}) - \exputil(\actselectagent{\dpiter_{i+1}},\utilitysymbolbig_{\dpiter_{i+1}})
\end{equation}
In \eqref{eqn:exputil-diff}, $\exputil(\cdot)$ is the expected utility functional defined in \eqref{eqn:def_J}.
\end{definition}
In \cite{ECH11-moneypump}, the money pump index is defined for a sequence of indices for which GARP fails. In complete analogy, \eqref{eqn:RI-MPI} in Definition~\ref{def:RI-MPI} computes the {\em maximum} `profit' a malicious arbitrager can make over all possible sequences of decision problems combinations, normalized by the sum of expected utilities of the Bayesian decision maker in the decision problems. The extent of irrationality of the Bayesian maker that facilitates arbitrage is captured by the variable $\bar{\exputilsymb}_{\prior,\dpiter_{i},\dpiter_{i+1}}(\actselectagent{1},\actselectagent{2})$ in \eqref{eqn:RI-MPI} and discussed below in more detail.

Let us briefly discuss the intuition behind RI-MPI \eqref{eqn:RI-MPI}. Without loss of generality, suppose GARRI fails for indices $1,2$ (sequence of length 2), which implies the following set of inequalities hold:
\begin{equation}\label{eqn:RI-MPI-2}
\begin{split}
&\exputil(\actselectagent{2},\utilitysymbolbig_{1})~\geq~\exputil(\actselectagent{1},\utilitysymbolbig_{1})~\text{ and } \exputil(\actselectagent{1},\utilitysymbolbig_2)~\geq~\exputil(\actselectagent{2},\utilitysymbolbig_2)\\
\implies & \underbrace{\exputil(\actselectagent{2},\utilitysymbolbig_{1}) - \exputil(\actselectagent{1},\utilitysymbolbig_{1})}_{\geq 0} + \underbrace{\exputil(\actselectagent{1},\utilitysymbolbig_2)~\geq~\exputil(\actselectagent{2},\utilitysymbolbig_2)}_{\geq 0} \geq 0\\
\implies & \bar{\exputilsymb}_{\prior,1,2}(\actselectagent{1},\actselectagent{2}) \geq 0
\end{split}
\end{equation}
The term $\bar{\exputilsymb}_{\prior,1,2}(\actselectagent{1},\actselectagent{2})$ measures the excess expected utility a malicious arbitrager can gain by `buying' choices $\actselectagent{2},~\actselectagent{1}$ when presented with utilities $\utilitysymbolbig_1,~\utilitysymbolbig_2$ in decision problems $1,2$,~respectively, and selling choices $\actselectagent{2},~\actselectagent{1}$ to the Bayesian decision maker in decision problems $2,1$, respectively.

{\em Summary.} In this section, we extended three robustness measures from revealed preference to the revealed rational inattention result of Corollary~\ref{corl:niac-to-garp}. Specifically, we extended the Afriat Efficiency Index (AEI)~\cite{AFT72-efficiency}, Varian's~\cite{VAR85-measurement} Minimum Perturbation Test (MPT) and the Money Pump Index (MPI)~\cite{ECH11-moneypump} to the Bayesian case. We now illustrate the Bayesian analogs of AEI, MPT and MPI on a real-world YouTube metadata comprising user engagement from approximately 140,000 videos. We characterize, using the robustness measures for rational inattention defined above, the goodness-of-fit to the YouTube dataset to the rationally inattentive utility maximization model.

\section{Example. Testing YouTube metadata for rational inattention}\label{sec:real-world}
The first part of the paper thus far comprised theoretical results on unification of tests for revealed preference and revealed rational inattention, and extension of measures for goodness-of-fit in revealed preference to revealed rational inattention.
The second part of the paper, namely, this section, focuses on computational aspects of the results presented in the first part. In this section, we illustrate the robustness metrics introduced in Sec.\,\ref{sec:extension-robustness}, namely, RI-AEI, RI-MPT and RI-MPI on a real-world YouTube dataset.\footnote{Although our past works~\citep{HKP20,PK23} use the same dataset for numerical experiments, the dataset pre-processing steps and experimental results in this paper are new.}

For our numerical experiment on a real world dataset, we consider a YouTube dataset comprising approximately 140,000 videos across 25,000 channels spanning 18 video categories and over 9 millions users from April 2007 to May 2015. The diversity of videos in YouTube is immense; YouTube users engage differently with YouTube videos in different video categories~\cite{YT-commenting}. There are several works in the literature~\cite{PK23,HKP20,YT-content} that perform YouTube metadata analysis using tools from microeconomics and machine learning. However, to the best of our knowledge, a principled approach to characterizing the goodness-of-fit of YouTube user engagement to revealed rational inattention tests have not been addressed in the literature. A revealed rational inattention-based analysis is especially suitable for the YouTube dataset since the dataset does not comprise the visual cues (private signals) perceived by the online user from the video webpage before choosing to engage on the YouTube platform. In complete analogy, the analyst performing revealed rational inattention has no knowledge of the decision maker's private signals,and yet, due to Blackwell dominance, can reconstruct feasible utility functions and information acquisition costs that rationalize the decision maker's actions.

Our aim in the second part of the paper is to analyze groups of YouTube users in different video categories and test using revealed rational inattention if YouTube user engagement is consistent with rationally inattentive utility maximization. The authors in \cite{YT-cognition, YT-cognition-2, YT-cognition-3} analyze how YouTube video content affects user behavior, emotion and cognitive engagement. Our aim in this paper is to study YouTube user engagement from an information economics perspective -  characterize YouTube user engagement in different video categories with {\em different} utility functions, but the {\em same} cost of information acquisition. Formally, our aim is to:\\
(1) Transform the Youtube video metadata in a form amenable to the rationally inattentive utility maximization framework of \eqref{eqn:optimal-terminal-action} and \eqref{eqn:mod-optimal-attention-strategy} in order to perform revealed rational inattention analysis.\\
(2) Identify if YouTube user engagement is rationally inattentive (checking the feasibility of NIAS and GARRI conditions of Corollary~\ref{corl:niac-to-garp}), and if so, construct the YouTube users' utility functions in different video categories, and their cost of information acquisition. In the YouTube context, the utility function is the online social reward a typical YouTube user earns by engaging with a particular YouTube video. The information acquisition cost abstracts the cognitive perception cost expended by the online user on the YouTube video.\\
(3) Compute the goodness-of-fit of YouTube user engagement to the revealed rational inattention test of Corollary~\ref{corl:niac-to-garp} by computing the robustness metrics in Sec.\,\ref{sec:extension-robustness}. The goodness-of-fit for rational inattention quantifies the maximum perturbation needed in the YouTube dataset so that the revealed rational inattention test of Corollary~\ref{corl:niac-to-garp} fails. Goodness-of-fit tests for rational inattention are novel and are facilitated primarily due to the unification result of Theorem~\ref{thrm:eq}, and, to the best of our knowledge, have not been previously explored in the literature.

\subsection{Context. YouTube user engagement and rational inattention}
YouTube is a social multimedia platform where human users interact with video content on YouTube channels by posting comments and rating videos. Empirical studies (\cite{KH17,HG17,ABCH15,AK17}) show that the comments and ratings from users are  influenced by the thumbnail, title, category, and perceived popularity of each video. Models for human decision making in the context of online multimedia platforms have been studied extensively in the literature. Two widely-used classes of models that motivate us to understand YouTube user engagement from the lens of rational inattention are `parallel constraint satisfaction models' and `evidence accumulation models'.\\
{\em Parallel constraint satisfaction models}~(\cite{GCK08,MC89}) assume that information is screened sequentially to highlight salient alternatives and final choice is made when the decision maker reaches sufficient internal coherence.\\
{\em Evidence accumulation models}~(\cite{KRA10,RC04}) model consumers' attention by drift-diffusion models that accumulate evidence based on whether they are fixating their gaze on either the product or its price. The decision is taken when any of the alternatives' evidence threshold level is achieved.

Both classes of models described above have one aspect in common - the decision maker makes a final choice {\em after} sequentially accumulating information, and naturally fits our rational inattention framework. In particular, we refer the reader to \cite[Sec.\,5.1]{CD15},~\cite{PK23} where sequential information accumulation frameworks are expressed as a one-shot decision framework~\eqref{eqn:optimal-attention-strategy}. The key idea is to map the sequence of realized observations $\obs_{1:\tau}$ to a single meta-observation, where $\tau$ is a stopping time. The main takeaway is that NIAS and NIAC are still necessary and sufficient for rational inattention in a sequential information accumulation framework. Hence, we are justified in performing the revealed rational inattention test on YouTube metadata.

In terms of YouTube webpage parameters, we hypothesize the YouTube user is a Bayesian agent that sequentially consumes webpage cues such as thumbnail and title and incurs a cost of attention, followed by engaging on the YouTube platform and gaining a social utility~\cite{YT-social-utility}. Our revealed rational inattention aim is to embed YouTube user engagement in the rational inattention framework, and construct utility functions identify using the YouTube dataset, if online users engage `optimally' on the YouTube multimedia platform in a rationally inattentive utility maximization sense.

\subsection{YouTube Dataset and Model Parameters}\label{sec:YT-parameters}
\begin{figure}[ht]
	\centering
	\vspace{-9pt}
	\includegraphics[width=0.75\textwidth]{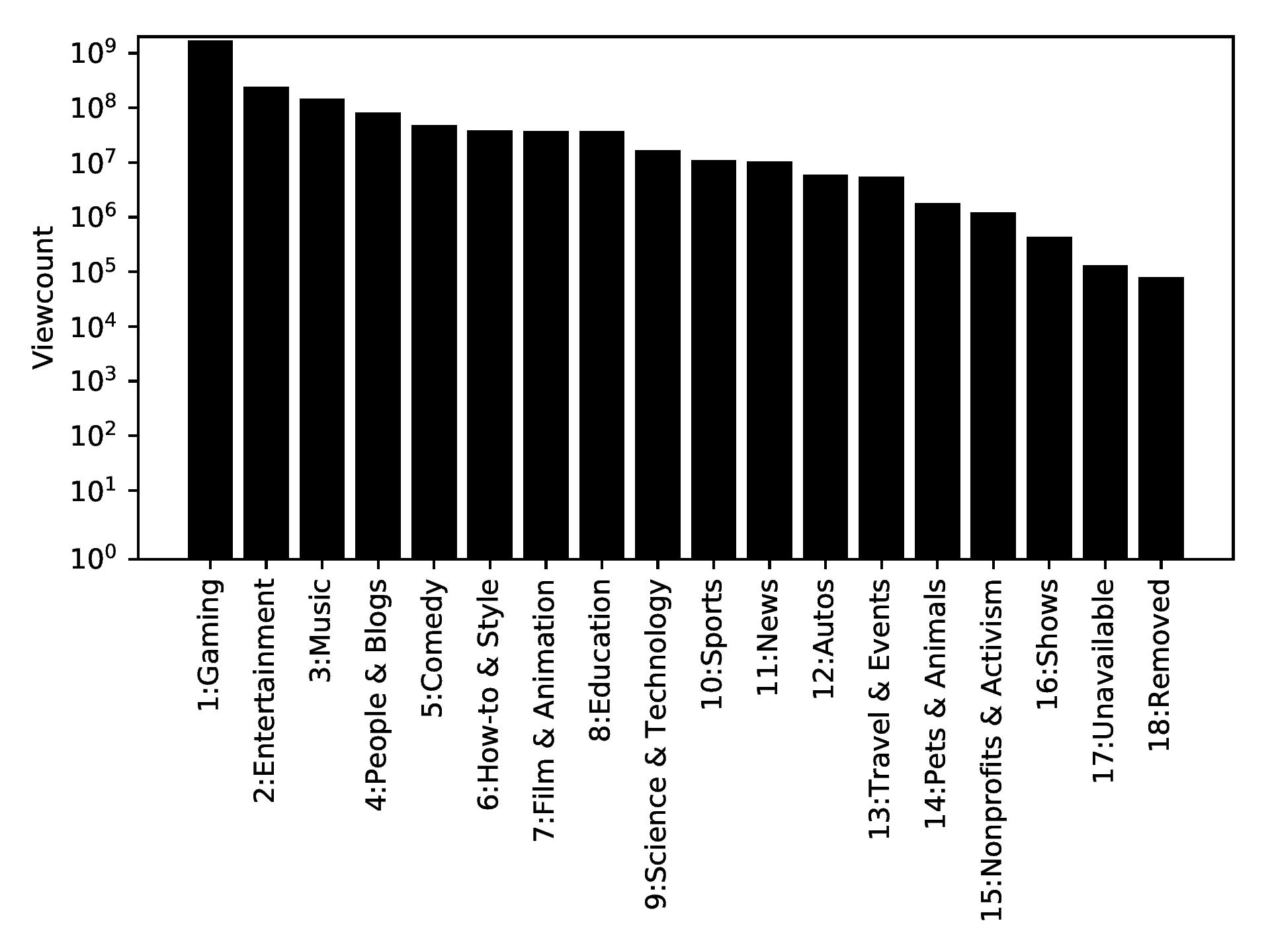}
	\vspace{-10pt}
	\caption{YouTube Dataset Overview. Viewcount summed over all videos (vertical axis) of $\numagents = 18$ video categories. The $18$ categories are listed on the horizontal axis.}
	\label{fig:categoryandviewcount}
\end{figure}

Categories in YouTube (e.g.\, News, Gaming, Music etc.) are numbered from $1 - 18$ (See Fig.~\ref{fig:categoryandviewcount} for the full listing). Recall from the rationally inattentive utility maximization setup of \eqref{eqn:optimal-terminal-action} and \eqref{eqn:mod-optimal-attention-strategy} that the decision maker observes the ground truth (state $\state$) in noise (private measurement $\obs$), and then takes an action $\action$. In our numerical experiments on the YouTube dataset, we assume the user's state $\state$ depends on the YouTube video's thumbnail (image) and title (text), since the thumbnail and title influence whether or not a user clicks on the video thumbnail and interacts with the video (through comments and likes/dislikes). The user's private measurement $\obs$ are the perceived visual cues abstracted away in the dataset. Finally, we assume the user's action $\action$ depends on (i) video viewcount (did the user watch the video or not), (ii) total comments on the video (user inclination to comment on a video), (iii) number of likes and dislikes on the video (user inclination to like/dislike a video) and (iv) difference in the number of likes and dislikes (indicates user engagement polarity). We formalize this abstraction of YouTube metadata to rational inattention framework's variables below.

\begin{figure}[h]
    \centering
    \includegraphics[width = 0.65\columnwidth]{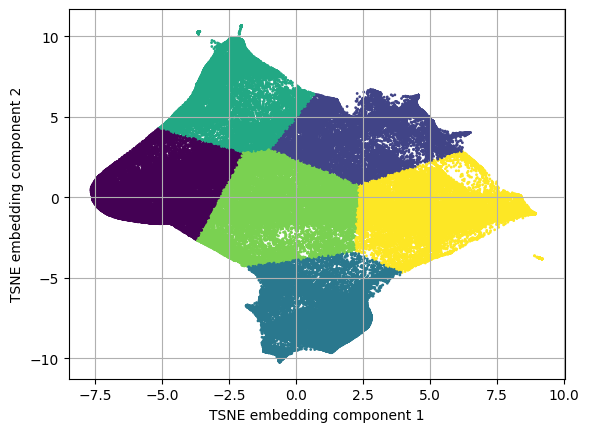}
    \caption{Plot of 2-dimensional embedding of 140,000 YouTube video thumbnails obtained from a deep auto-encoder. Every video in the YouTube dataset is associated with a cluster index (distinct color), where the clustering on the embedding space is performed using the k-means clustering algorithm~\cite{kmeans,scikit-learn}.}
    \label{fig:yt_thumbnails}
\end{figure}

\begin{figure}
    \centering
    \includegraphics[scale = 0.65]{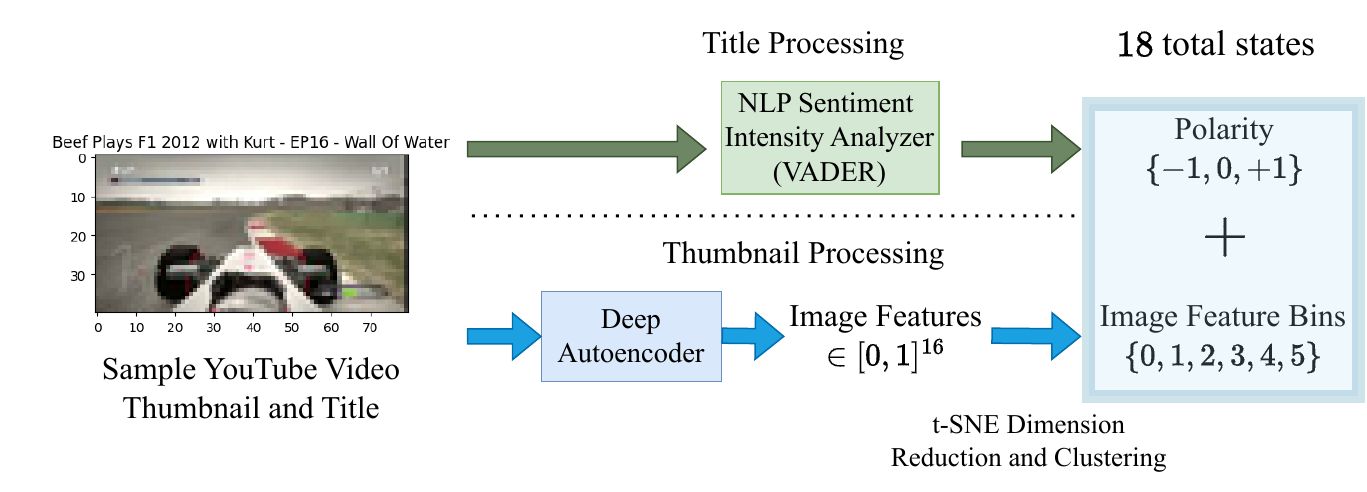}
    \caption{Deep auto-encoder and natural language processing (NLP) based data processing pipeline for converting YouTube video thumbnail and title to categorical state variables. The video title (description) is passed through NLP-based VADER sentiment analyzer~\cite{Vader} that outputs the polarity of the text description. The video thumbnail undergoes two sequential transformations: first, the image is passed through a deep auto-encoder that performs feature extraction; second, the extracted features' dimensions are converted to 2-dimensional embeddings using the t-SNE dimension reduction technique~\cite{tsne} followed by k-means clustering~\cite{scikit-learn,kmeans} that partitions the embedding space into 6 clusters. The thumbnail embeddings of all 140,000 videos in the YouTube dataset are illustrated in Fig.\,\ref{fig:yt_thumbnails}. The tuple comprising the title polarity and the thumbnail feature cluster forms the state for the rationally inattentive user interacting with the YouTube video.}
    \label{fig:state-processing}
\end{figure}

\begin{figure*}
        \centering
        \begin{subfigure}[b]{0.475\textwidth}
            \centering
            \includegraphics[width=\textwidth]{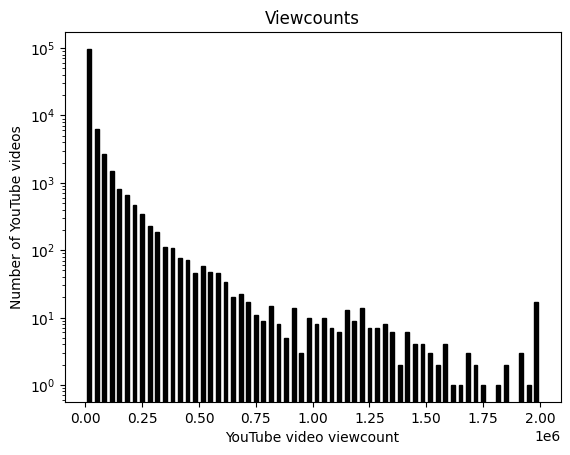}
            \caption{{\small Histogram of YouTube video view counts}}   
        \end{subfigure}
        \hfill
        \begin{subfigure}[b]{0.475\textwidth}  
            \centering 
            \includegraphics[width=\textwidth]{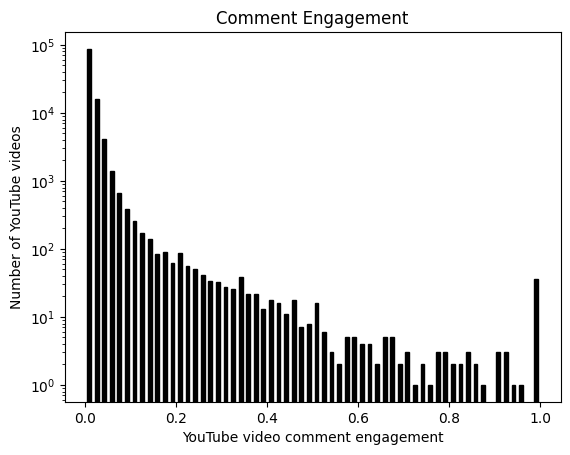}
            \caption{{\small Histogram of YouTube video comment engagement}}    
        \end{subfigure}
        \vskip\baselineskip
        \begin{subfigure}[b]{0.475\textwidth}   
            \centering 
            \includegraphics[width=\textwidth]{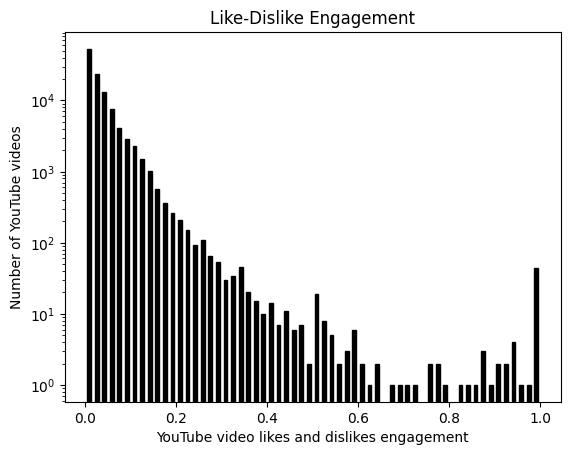}
            \caption{{\small Histogram of YouTube video like/dislike engagement}}    
        \end{subfigure}
        \hfill
        \begin{subfigure}[b]{0.475\textwidth}   
            \centering 
            \includegraphics[width=\textwidth]{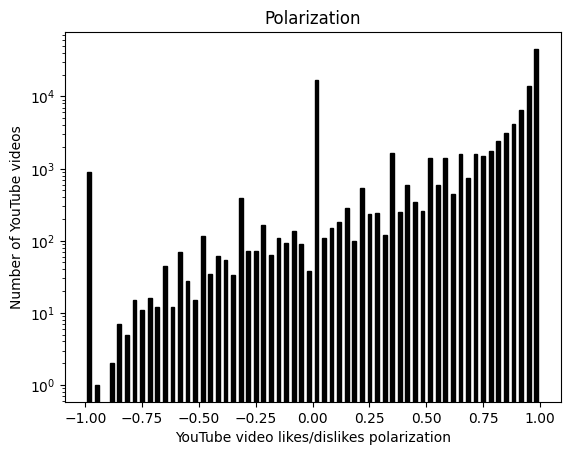}
            \caption{{\small Histogram of YouTube video polarization}} 
        \end{subfigure}
        \caption{\small Histograms of user engagement metrics in the YouTube video metadata dataset comprising 140,000 videos. The illustrated user engagement metrics are aggregated into a categorical action variable via the procedure described in Table~\ref{tab:action-discretization}. The key idea is to embed the YouTube dataset into a rational inattention setup where the decision maker receives visual cues from the YouTube video's thumbnail and title, and interacts with the video by taking an action that accounts for viewing, commenting on and liking/disliking the video. Our aim is to compute the goodness-of-fit of the YouTube metadata to the rationally inattentive utility maximization model by embedding the illustrated as the `action' taken by the Bayesian agent in the rational inattention model. 
        } 
        \label{fig:metadata-illustration}
    \end{figure*}

The video categories have mean numbers of users ranging from $149$ to $4596$  for high viewcount (greater than $10000$) videos and $8$ to $1801$ for low viewcount videos (less than $10000$). Figure~\ref{fig:categoryandviewcount} lists each video category along with the total number of views. Note that the video categories ``Unavailable'' or ``Removed'' are videos flagged by YouTube as being suspected of violating YouTube's video policies.
The YouTube dataset contains the view counts, comment counts, likes, dislikes, thumbnail, title, and category of each video. Figure~\ref{fig:metadata-illustration} shows the histograms for various user engagement metadata in the YouTube dataset across 140,000 videos, namely, video viewcount, video comment engagement (number of comments normalized by the viewcount), video like/dislike engagement (number of likes and dislikes normalized by the viewcount) and polarization in video engagement (difference in likes and dislikes normalized by the total number of likes and dislikes). As discussed formally below, the YouTube user's action $\action$ in the rational inattention context is a function of the above metadata. To relate to the revealed rational inattention result of Corollary~\ref{corl:niac-to-garp}, we define the following:\newline
\textit{$1$. Decision Maker}: Group of users interacting with videos in each video segment. User engagement in different video categories can be interpreted as the decision maker acting in different decision problems, namely, video categories. Recall that the utility function of the decision maker is parametrized by the decision problem.\newline 
\textit{$2$. Decision problem ($\dpiter$)}: Environment $\dpiter$ corresponds to the video category in our YouTube dataset, excluding the categories ``Unavailable'' or ``Removed'' that comprise videos that are flagged by YouTube as being suspected of violating YouTube's video policies\footnote{Refer to~\url{https://www.youtube.com/yt/about/policies/\#community-guidelines} for details}. Fig.~\ref{fig:categoryandviewcount} lists the cumulative viewcount on all videos belonging to each video category. For our numerical experiments, we only consider $\numdp = 8$ video categories with more than 500 videos. 
\newline
\textit{$3$. State ($\state$)}: In the YouTube dataset, the state comprises aggregated information from: (i) the video thumbnail, and (ii) the sentiment of the video title. Figure~\ref{fig:state-processing} illustrates how the video thumbnail and title are processed to yield a categorical variable that takes values in $\stateset = \{1,2,\ldots,18\}$. The video thumbnail is first passed through an auto-encoder that yields a 16 dimensional representation of the thumbnail. We further reduce the dimension of the auto-encoder output to 2 using t-SNE~(t-distributed stochastic neighbor embedding)~\cite{tsne}, a widely used method for dimension reduction, and obtain our {\em thumbnail embeddings}. The 2-dimensional representation of all video thumbnails in the YouTube dataset is illustrated in Fig.\,\ref{fig:yt_thumbnails}. Finally, we cluster the thumbnail embeddings into 6 partitions using the k-means clustering algorithm~\cite{scikit-learn,kmeans}\footnote{The number of clusters as a hyperparameter was optimized using the widely used `elbow' heuristic~\cite{elbow-kmeans}.}. To summarize, the video thumbnail is now quantized into 6 embedding clusters. The video title is passed through VADER sentiment analyzer~\cite{Vader} that outputs $-1,0,+1$ if the title has a negative, neutral or positive sentiment, respectively. The two-tuple comprising the 2-dimensional thumbnail embedding cluster and title sentiment comprises the video's state. In YouTube, the video viewcount and thumbnail are the independent quantities that govern the user engagement on the video since videos need to be viewed first before users can comment on or like/dislike the video. \newline
\textit{$4$. (Unobserved) Private Observation ($\obs$)}: The observation $\obs$ for a YouTube user abstracts the visual cues a user perceives that depends on video metadata; in our numerical experiment, we assume the YouTube user takes into account visual cues from the video thumbnail, sentiment cures from the video title, and the video category the video belongs to. The observation likelihood is indicative of the attention expended by the user on a video. We note that although neither the observations $\obs$ nor the observation likelihood $p(\obs|\state)$ are contained in the YouTube dataset, our IRL algorithm abstracts away these unobserved model parameters, and still yields necessary and sufficient conditions for Bayes optimality.\newline
\textit{$5.$ Action ($\action$)}: In the YouTube dataset, the action $\action$ is related to the overall user engagement with the YouTube videos. YouTube user engagement comprises:\\
(i) the video viewcount,\\
(ii) the comment rate defined as the number of comments on a video normalized by the video viewcount,\\
(iii) like/dislike rate defined as the total number of likes and dislikes normalized by the video viewcount, and\\
(iv) engagement polarity defined as the difference between the number of likes and dislikes on a video normalized by the the total number of likes and dislikes on the video.

The 4 engagement metrics listed above were recorded 20 days after the video was posted on YouTube for all YouTube videos in the YouTube dataset. The procedure for converting the engagement metrics to a finite set of actions in described in Table~\ref{tab:action-discretization}. The key takeaway is that the 4 engagement metrics above are converted to 18 possible actions via the discretization scheme of Table~\ref{tab:action-discretization} into a categorical variable that takes one of 24 values. For our revealed rational inattention analysis on the YouTube dataset, we only consider 8 of the 24 possible actions for which the video count exceeds 100 videos; hence, the action set is $\actionset = \{1,2,\ldots,8\}$.

To summarize, the above description embeds YouTube user engagement into the rational inattention framework. The decision problem is parametrized by the video category. For our data analysis, we consider only $\numdp=8$ out of $18$ video categories for which the number of videos exceeds 500 videos; hence the set of decision problems is $\dpset = \{1,2,\ldots,8\}$. The ground truth (state) and is a function of the video thumbnail and title as illustrated in Fig.\,\ref{fig:state-processing} and belongs to the set $\stateset =\{1,2\ldots,18\}$. The user's private signals comprise of visual cues that are abstracted away in the YouTube dataset. Finally, the user's engagement with the YouTube videos is measured via the video viewcount, the total number of comments, and likes and dislikes on the video. These engagement metrics are converted to categorical action variables via a discretization scheme illustrated in Table~\ref{tab:action-discretization}; the user's action thus belongs to the set $\actionset=\{1,2,\ldots,8\}$.


Recall from Corollary~\ref{corl:niac-to-garp} that the inverse learner requires knowledge of the dataset $\dataset_{RRI} = \{\prior,\{\actselectagent{\dpiter},\utilitysymbolbig_\dpiter\}_{\dpiter=1}^\numdp\} $~\eqref{eqn:dataset-Bayesian} for checking if the NIAS~\eqref{eqn:nias-vanilla} and GARRI~\eqref{eqn:GARRI} conditions hold for Bayesian rationality. In the YouTube context, the variables $\prior,\{\actselectagent{\dpiter}\}_{\dpiter=1}^\numdp$ in dataset $\dataset_{RRI}$ can be constructed as:\\
\begin{equation}
\begin{split}
\prior(\state) &= \frac{1}{I}\sum_{i=1}^I \mathbbm{1}\{\state_{i} = \state\},~~\actselectagent{\dpiter} s= \frac{\sum_{i=1}^I\mathbbm{1}\{\state_i = \state, \action_i = \action, \operatorname{category}_i=\dpiter\}}{\sum_{i=1}^{I}\mathbbm{1}\{\state_i=\state, \operatorname{category}_i = \dpiter\}},
\end{split}
\label{eqn:YT_dataset}
\end{equation}
where $\mathbbm{1}\{\cdot\}$ is the indicator function, variable $i$ indexes the YouTube videos, $I=140,000$ is the total number of YouTube videos in the dataset, and environment $\dpiter\in\{1,2,\ldots,8\}$ indexes the video categories with more than 500 videos. Also, $\state_i, \action_i, \operatorname{category}_i$ denote the state, action and category of the YouTube video indexed by $i$, where the state and action interpretations for the YouTube videos are discussed above.

\begin{table}[ht]
    \centering
    \begin{tabular}{|l|l|}\hline
    & \\
        Viewcount indicator ($vi$) & 0, if viewcount $\leq$ viewcount median, and 1 otherwise
        \\
        & \\\hline
        & \\
         Comment Rate indicator ($cri$) & 0, if comment rate <= comment count median given $vi$, and 1 otherwise\\&\\\hline
         & \\
         Like/Dislike rate indicator ($ldri$) & 0, if like/dislike rate <= like/dislike median given $vi$, and 1 otherwise \\&\\\hline & \\
         Engagement polarity indicator ($epi$) & -1, if engagement polarity $\leq$ $33.33$ percentile of engagement polarities given $vi$,\\
         & ~0, if engagement polarity $\in [33.33~\text{percentile},~66.67~\text{percentile}]$ given $vi$,\\
         & and 1 otherwise. \\&\\\hline
    \end{tabular}
    \caption{User engagement in YouTube videos is quantized into $|\actionset| = 24$ categories. The user engagement depends on 4 categorical variables $vi,cri,ldri,epi$ described above. The variables $vi,cri,ldri$ are binary ($0$ for low value, and $1$ for high), whereas the polarity variables $epi$ is ternary to accommodate negative (-1), neutral (0) and positive (+1) sentiments. Our exploratory data analysis on the YouTube dataset showed that the video count for $16$ out of $24$ possible actions is less than $100$. Hence, we only consider the remaining $8$ actions for our revealed rational inattention analysis.
    }
    \label{tab:action-discretization}
\end{table}

\subsection{YouTube Dataset Analysis}
In this section, we discuss our experimental findings from performing revealed rational inattention tests on the YouTube dataset.\footnote{All our numerical results are completely reproducible and can be accessed from the GitHub repository \url{https://github.com/KunalP117/YouTube-Commenting-Analysis}.} Our aim is to perform the revealed rational inattention on the YouTube metadata, and characterize its goodness-of-fit to rationally inattentive utility maximization. Recall from Sec.\,\ref{sec:extension-robustness} that the robustness metrics RI-AEI~\eqref{eqn:RI-AEI}, RI-MPT~\eqref{eqn:RI-MPT} and RI-MPI~\eqref{eqn:RI-MPI} measure the goodness-of-fit of a stochastic choice dataset $\dataset_{RRI}$~\eqref{eqn:dataset-Bayesian}  to the rationally inattentive utility maximization model of \eqref{eqn:optimal-terminal-action} and \eqref{eqn:mod-optimal-attention-strategy_alt} when the decision maker's utility is known to the external analyst. However, in the YouTube metadata context, the utility functions in $\dataset_{RRI}$~\eqref{eqn:YT_dataset} are not available to the analyst, and must be reconstructed via the revealed rational inattention test of Corollary~\ref{corl:niac-to-garp}. The reconstructed utility functions, by definition, trivially satisfy the revealed rational inattention test of Corollary~\ref{corl:niac-to-garp}. Hence, instead of quantifying how close YouTube metadata is to passing the revealed rational inattention test, we adapt the notion of goodness-of-fit for YouTube to quantify {\em how well YouTube metadata fits the rational inattention model specified by~\eqref{eqn:optimal-terminal-action},~\eqref{eqn:mod-optimal-attention-strategy}}.  
Formally, our numerical experiments thus comprise two steps:\\
(i) Construct two feasible point estimates of utility functions and the corresponding cost of information acquisition using the revealed rational inattention test of Corollary~\ref{corl:niac-to-garp} for the dataset $\dataset_{RRI}$~\eqref{eqn:YT_dataset} computed from the raw YouTube metadata.\\
(ii) Adapt the robustness metrics for revealed rational inattention in Sec.\,\ref{sec:extension-robustness} to compute how far the two utility point estimates computed in step (i) with the prior and action selection policies computed in \eqref{eqn:YT_dataset} are from {\em failing} the revealed rational inattention test of Corollary~\ref{corl:niac-to-garp} given $\dataset_{RRI}$~\eqref{eqn:YT_dataset}.

\subsection{Construction of feasible utilities and information acquisition costs from YouTube metadata}\label{sec:util-youtube}
We construct two sets of utility functions for YouTube metadata, namely, (a) max-margin utility that maximizes the extent of feasibility of NIAS and GARRI conditions, and (b) sparse utility that minimizes the number of non-zero values for the utility function. The max-margin utility and sparse utility are defined below:
\begin{definition}[Utility reconstruction for YouTube metadata] \label{def:util-reconstruction} Consider the stochastic choice dataset $\dataset_{RRI}$~\eqref{eqn:YT_dataset}. The reconstructed max-margin and sparse utility functions for $\dataset_{RRI}$ that satisfy the NIAS and GARRI conditions of Corollary~\ref{corl:niac-to-garp} for Bayesian rationality are defined as:
\begin{align}
    \underline{\text{\em Max-margin utility}}: & \nonumber \\
    \{\{\utilitysymbolbig_\dpiter,\runcostinst_\dpiter,\lambda_\dpiter\}_{\dpiter=1}^\numdp,\eps\}_{\maxmargin} &~= \underset{\{\utilitysymbolbig_\dpiter,\runcostinst_\dpiter,\lambda_\dpiter\}_{\dpiter=1}^\numdp,\eps\geq 0}{\argmax}~\eps,~\text{such that:}\label{eqn:max-margin-utility}\\
    \text{NIAS:}~& \sum_{\state\in\stateset}~p_\dpiter(\state|\action)(\utilitysymbolbig_\dpiter(\state,\actiontwo)-\utilitysymbolbig_\dpiter(\state,\action)) + \eps \leq 0 \nonumber\\
    \text{GARRI:}~&\lambda_\dpiter~\sum_{\action\in\actionset}\left(~p_\dpitertwo(a)~\max_{\actiontwo\in\actionset}~\sum_{\state\in\stateset} p_\dpitertwo(\state|\action)~\utilitysymbolbig_\dpiter(\state,\actiontwo)- \sum_{\state\in\stateset}~p_\dpiter(\action|\state)~\prior(\state)~\utilitysymbolbig_\dpiter(\state,\action)\right) -~ (\runcostinst_\dpitertwo - \runcostinst_\dpiter) + \eps \leq 0\nonumber\\
    \text{Normalization:}~&\utilitysymbolbig_\dpiter(\state,\action)\in[10^{-3},1]~\forall\state,\action,\dpiter\nonumber\\
    \underline{\text{\em Sparse utility}}: & \nonumber\\
    \{\{\utilitysymbolbig_\dpiter,\runcostinst_\dpiter,\lambda_\dpiter\}_{\dpiter=1}^\numdp\}_{\sparse}&~= \underset{\{\utilitysymbolbig_\dpiter,\runcostinst_\dpiter,\lambda_\dpiter\}_{\dpiter=1}^\numdp\geq 0}{\argmin}~\sum_{\dpiter=1}^\numdp~\sum_{\state\in\stateset,\action\in\actionset}|\utilitysymbolbig_\dpiter(\state,\action)|,~\text{such that:}\label{eqn:sparse-utility}\\
    \text{NIAS:}~& \sum_{\state\in\stateset}~p_\dpiter(\state|\action)(\utilitysymbolbig_\dpiter(\state,\actiontwo)-\utilitysymbolbig_\dpiter(\state,\action)) \leq 0 \nonumber\\
    \text{GARRI:}~&\lambda_\dpiter~\sum_{\action\in\actionset}\left(~p_\dpitertwo(a)~\max_{\actiontwo\in\actionset}~\sum_{\state\in\stateset} p_\dpitertwo(\state|\action)~\utilitysymbolbig_\dpiter(\state,\actiontwo)- \sum_{\state\in\stateset}~p_\dpiter(\action|\state)~\prior(\state)~\utilitysymbolbig_\dpiter(\state,\action)\right) -~ (\runcostinst_\dpitertwo - \runcostinst_\dpiter) \leq 0\nonumber\\
    \text{Normalization:}~&\utilitysymbolbig_\dpiter(\state,\action)\in[10^{-3},1]~\forall\state,\action,\dpiter\nonumber
\end{align}
The two sets of utility functions computed in \eqref{eqn:max-margin-utility} and \eqref{eqn:sparse-utility} above facilitates construction of two stochastic choice datasets $\dataset_{RRI,\maxmargin}$ and $\dataset_{RRI,\sparse}$, respectively, defined as:
\begin{equation}\label{eqn:datasets-maxmargin-sparse}
    \begin{split}
        \text{Max-margin dataset: }&\dataset_{RRI,\maxmargin} = \{\prior,\{\actselectagent{\dpiter},\utilitysymbolbig_{\dpiter,\maxmargin}\}_{\dpiter=1}^\numdp\},\\
        \text{Sparse dataset: }&\dataset_{RRI,\sparse}  = \{\prior,\{\actselectagent{\dpiter},\utilitysymbolbig_{\dpiter,\sparse}\}_{\dpiter=1}^\numdp\},\\
    \end{split}
\end{equation}
where $\{\utilitysymbolbig_{\dpiter,\maxmargin}\}_{\dpiter=1}^\numdp$ and $\{\utilitysymbolbig_{\dpiter,\sparse}\}_{\dpiter=1}^\numdp$ are computed in \eqref{eqn:max-margin-utility} and \eqref{eqn:sparse-utility}, respectively, and prior pdf $\prior$ and action selection policies $\{\actselectagent{\dpiter}\}_{\dpiter=1}^\numdp$ are computed in \eqref{eqn:YT_dataset}.
\end{definition}

    \begin{figure}
        \centering
        \begin{subfigure}[b]{\textwidth}  
            \centering 
            \includegraphics[width=0.75\textwidth]{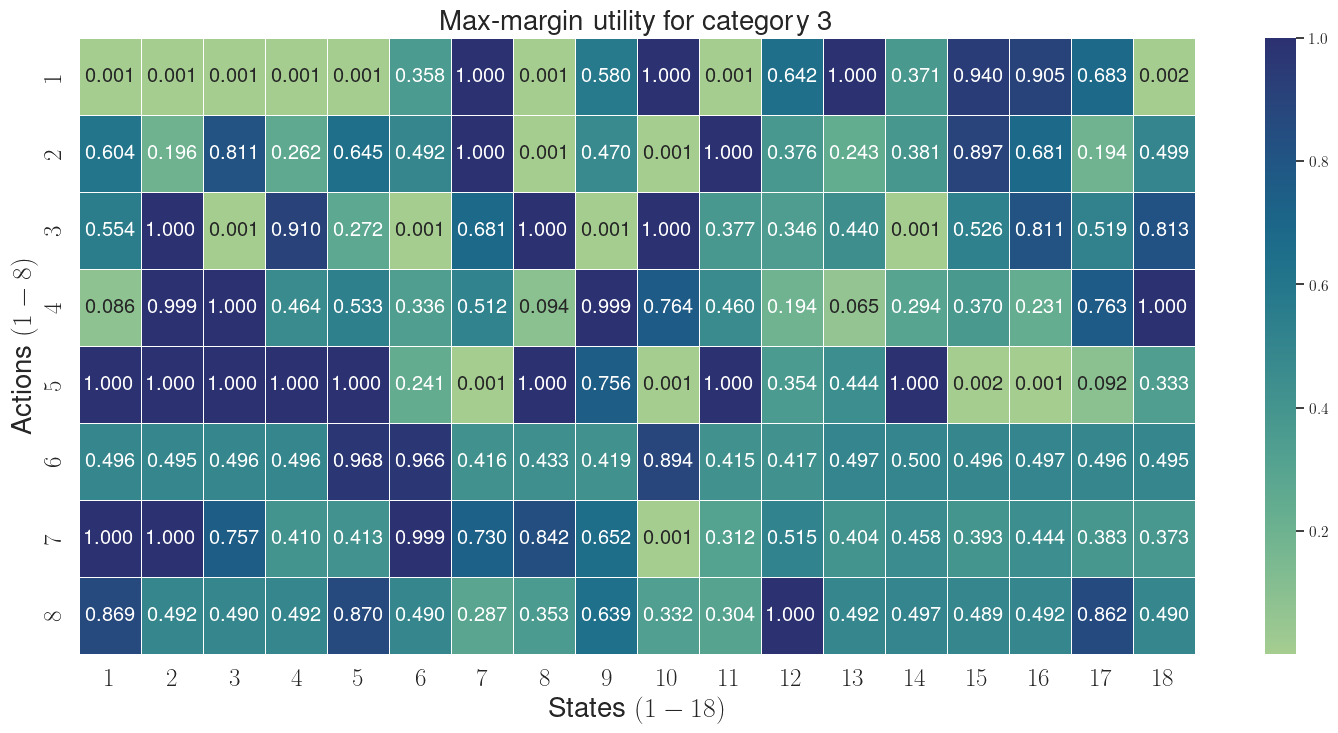}
            \caption{{\small Visualization of the max-margin reconstructed utility for video category 3.}}    
            \label{fig:maxmarginutil-cat-3}
        \end{subfigure}
        \vskip\baselineskip
            \begin{subfigure}[b]{\textwidth}
            \centering
            \includegraphics[width=0.75\textwidth]{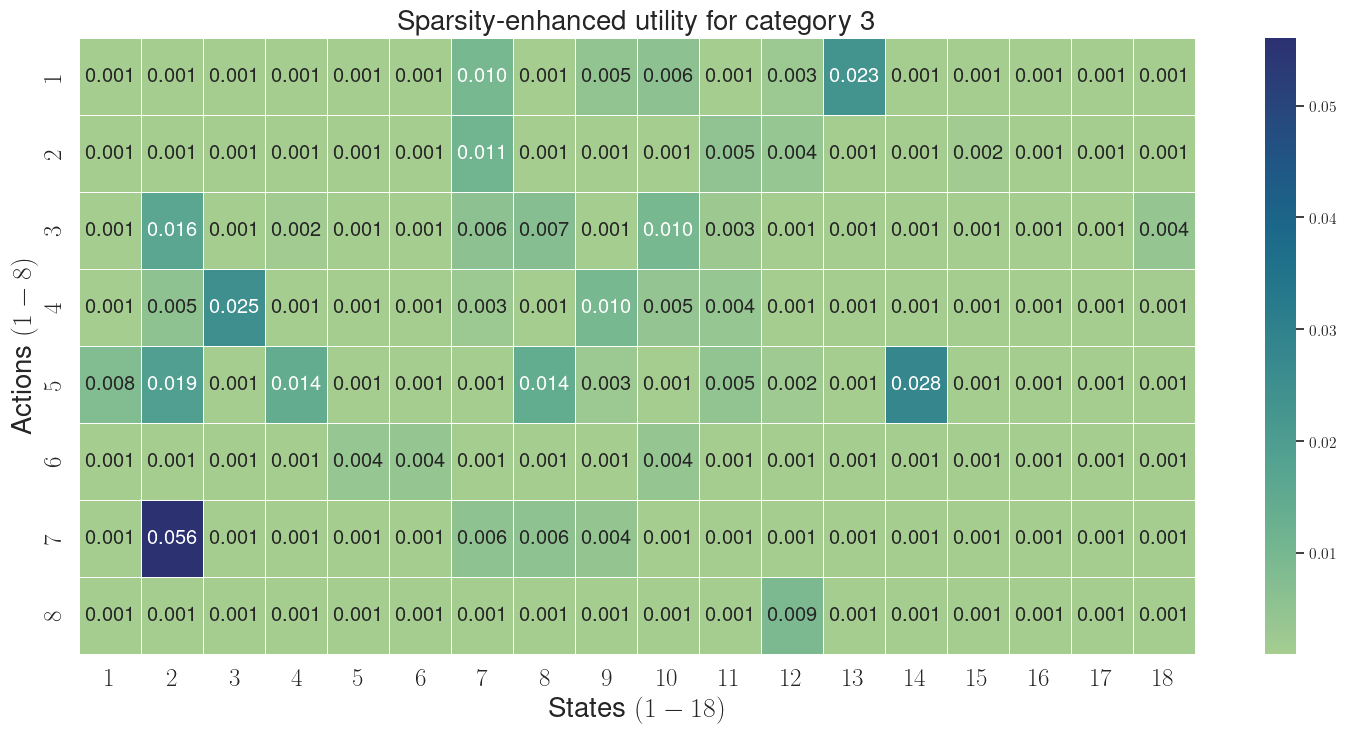}
            \caption{{\small Visualization of the sparsest reconstructed utility for video category 3.}}    
            \label{fig:sparseutil-cat-3}
        \end{subfigure}
        \caption{Visualization of the reconstructed max-margin and sparse utility function for YouTube metadata for video category 3 that satisfy the NIAS~\eqref{eqn:nias-vanilla} and GARRI~\eqref{eqn:GARRI} conditions in Corollary~\ref{corl:niac-to-garp} reconstructed using \eqref{eqn:max-margin-utility},~\eqref{eqn:sparse-utility}, respectively; the utility visualizations of other categories are illustrated in the Appendix.}
        \label{fig:util-visualization}
    \end{figure}

Definition~\ref{def:util-reconstruction} constructs two sets of stochastic choice data, $\dataset_{RRI,\maxmargin}$ and $\dataset_{RRI,\sparse}$ from YouTube metadata to perform a goodness-of-fit test to the rationally inattentive utility maximization model. The max-margin utility defined in \eqref{eqn:max-margin-utility} outputs the interior-most point of the feasibility polytope corresponding to the NIAS~\eqref{eqn:nias-vanilla} and GARRI~\eqref{eqn:GARRI} conditions of Corollary~\ref{corl:niac-to-garp}. Intuitively, the max-margin utility is a {\em robust point estimate}, robust to failing the revealed rational inattention test of Corollary~\ref{corl:niac-to-garp} against errors such as measurement error, misspecification error etc. The sparse utility, on the other hand, is a {\em parsimonious point estimate}, and provides the most compact characterization of a YouTube user using the fewest number of non-zero variables. Figure~\ref{fig:util-visualization} illustrates the max-margin (sub-figure (a)) and sparse utility (sub-figure (b)) reconstructed using \eqref{eqn:max-margin-utility} and \eqref{eqn:sparse-utility}, respectively, for video category\footnote{Recall from Sec.\,\ref{sec:YT-parameters} that the decision problem for the revealed rational inattention test is indexed by the video category.} number 3. The online GitHub repository contains the visualizations of max-margin and sparse utility functions for all 8 video categories. As expected, we observe from Fig.\,\ref{fig:util-visualization} that the max-margin utility has a higher variation in utility values compared to the sparse utility as the state and action indices are varied. Also, as expected, we observed that the the utility value exceeds the lower limit of 0.001 for 26\% of state, action pairs, and exceeds 0.005 for 15\% of state, action pairs for the sparse utility.

\subsection{Goodness-of-fit of YouTube metadata to rationally inattentive utility maximization}
Having constructed two sets of stochastic choice data in Definition~\ref{def:util-reconstruction} above, we now compute how far are the two datasets $\dataset_{RRI,\maxmargin}$ and $\dataset_{RRI,\sparse}$ from failing the revealed rational inattention test of Corollary~\ref{corl:niac-to-garp}. We adapt the robustness metrics in Sec.\,\ref{sec:extension-robustness} to compute how far a stochastic choice dataset is from failing the revealed rational inattention test, instead of computing how close the dataset is to passing the test in Definition~\ref{def:adaptation-robustness} below:
\begin{definition}[Adaptation of Robustness Metrics for Rational Inattention]\label{def:adaptation-robustness} Consider a dataset $\dataset_{RRI}$~\eqref{eqn:dataset-Bayesian} aggregated from the actions of a Bayesian decision maker. Suppose the dataset $\dataset_{RRI}$ satisfies the NIAS and GARRI conditions of Corollary~\ref{corl:niac-to-garp}. The goodness-of-fit measures below adapted from Definitions~\ref{def:RI-AEI},~\ref{def:RI-MPT} and \ref{def:RI-MPI} in Sec.\,\ref{sec:extension-robustness} that calculate the maximum admissible perturbation for $\dataset_{RRI}$ to fail the revealed rational inattention test of Corollary~\ref{corl:niac-to-garp}:
\begin{align}
    &\text{Adapted }\operatorname{RI-AEI}: \underset{e\in[0,1),\{\lambda_\dpiter,\runcostinst_\dpiter\}_{\dpiter=1}^\numdp\geq 0}{\argmin}~e,~\text{ such that NIAS~\eqref{eqn:niac-garri-feasibility} and the set of inequalities ~\eqref{eqn:RI-AEI} are feasible.}\label{eqn:adapted-RI-AEI}\\
    &\text{Adapted }\operatorname{RI-MPT}: 
\phi(\ndataset_{RRI})=
    \max_{\{\actselectagentnoise{\agent},\lambda_\dpiter,\utilitysymbol_\dpiter\}_{\dpiter=1}^\numdp\geq 0}~\sum_{\dpiter=1}^\numdp\sum_{\state\in\stateset,\action\in\actionset}\|\actselectagentnoise{\agent} - \actselectagent{\agent}\|_2^2,~\text{such that}\label{eqn:adapted-RI-MPT}\\
    &(i)~\sum_{\action\in\actionset}~\actselectagentnoise{\dpiter}=1~\forall,\state,\dpiter~(\text{Valid pmf}),\nonumber\\
    &(ii)~NIAS:\sum_{\state\in\stateset}~\tilde{p}_\dpiter(\state|\action)(\utilitysymbolbig_\dpiter(\state,\action)-\utilitysymbolbig_\dpiter(\state,\actiontwo))\geq 0~\forall~\dpiter,\action,\actiontwo,\nonumber\\
    &(iii)~GARRI:\runcostinst_\dpitertwo - \runcostinst_\dpiter -\lambda_\dpiter\left(\sum_{\action\in\actionset}~\tilde{p}_\dpitertwo(a)~\max_{\actiontwo\in\actionset}~\sum_{\state\in\stateset} \tilde{p}_\dpitertwo(\state|\action)~\utilitysymbolbig_\dpiter(\state,\actiontwo) ~-~\sum_{\state\in\stateset,\action\in\actionset}~\tilde{p}_\dpiter(\action|\state)~\prior(\state)~\utilitysymbolbig_\dpiter(\state,\action)\right)\geq 0,\nonumber\\
    &\quad\quad\quad\quad\text{for all index pairs }\dpitertwo,\dpiter,~\dpitertwo\neq\dpiter.\nonumber\\
    &\text{Adapted }\operatorname{RI-MPI}:~ -\max_{\{\price_{\dpiter_{1:l}},\response_{\dpiter_{1:l}},~l\leq \numdp\}}~\frac{\sum_{i=1}^l
    \bar{\exputilsymb}_{\prior,\dpiter_{i},\dpiter_{i+1}}(\actselectagent{\dpiter_{i}},\actselectagent{\dpiter_{i+1}})}{\sum_{i=1}^l \exputil(\actselectagent{\dpiter_i},\utilitysymbolbig_{\dpiter_i}) }\label{eqn:adapted-RI-MPI},
\end{align}
where $\dpiter_{l+1}\equiv\dpiter_{1}$ and $\bar{\exputilsymb}_{\prior,\dpiter_{i},\dpiter_{i+1}}(\actselectagent{\dpiter_{i}},\actselectagent{\dpiter_{i+1}})$ is defined in \eqref{eqn:exputil-diff}.
\end{definition}

A few words on how the adapted robustness metrics defined in Definition~\ref{def:adaptation-robustness} above differ from those defined in Sec.\,\ref{sec:extension-robustness}. Let us first discuss the adapted rationally inattentive Afriat efficiency index~(RI-AEI). Since Definition~\ref{def:adaptation-robustness} assumes the dataset $\dataset_{RRI}$ is consistent with rational inattention, we first note that setting $e=1$ satisfies  inequality in \eqref{eqn:RI-AEI}. Decreasing the value of $e$ decreases the LHS in the feasibility inequality~\eqref{eqn:RI-AEI} and hence, pushes the inequality closer to infeasibility. Hence, the smallest value of $e<1$ that satisfies \eqref{eqn:RI-AEI} can be interpreted as the goodness-of-fit of a dataset aggregated from a rational Bayesian agent to the NIAS~\eqref{eqn:nias-vanilla} and GARRI~\eqref{eqn:GARRI} conditions for Bayesian rationality. In comparison, observed that if $\dataset_{RRI}$ is not consistent with rational inattention, the inequalities~\eqref{eqn:RI-AEI} are infeasible for $e=1$. Hence, RI-AEI in \eqref{eqn:RI-AEI} computes the minimum relaxation needed for the dataset to satisfy NIAS and GARRI. The adapted rationally inattentive minimum perturbation test (RI-MPT) in \eqref{eqn:adapted-RI-MPT} computes the {\em maximum} permissible perturbation in the dataset $\dataset_{RRI}$~\eqref{eqn:dataset-Bayesian} so that the NIAS and GARRI conditions hold. In comparison, \eqref{eqn:RI-MPT} computes the minimum perturbation needed for NIAS and GARRI to hold, where the unperturbed dataset does not satisfy NIAS and GARRI. Finally, the adapted rationally inattentive money pump index (MPI) defined in \eqref{eqn:adapted-RI-MPI} is simply the negative of RI-MPI defined in \eqref{eqn:RI-MPI}. It is straightforward to verify that the value of RI-MPI will be larger for a dataset that is inconsistent with rational inattention compared to a dataset that satisfies NIAS and GARRI conditions. Hence, the negative of RI-MPI computes how well a dataset is consistent with rational inattention assuming the dataset satisfies NIAS and GARRI conditions for Bayesian rationality.

\subsubsection*{Robustness analysis of YouTube metadata}
\begin{figure*}
        \centering
        \begin{subfigure}[b]{0.48\textwidth}
            \centering
            \includegraphics[height = 0.724\textwidth]{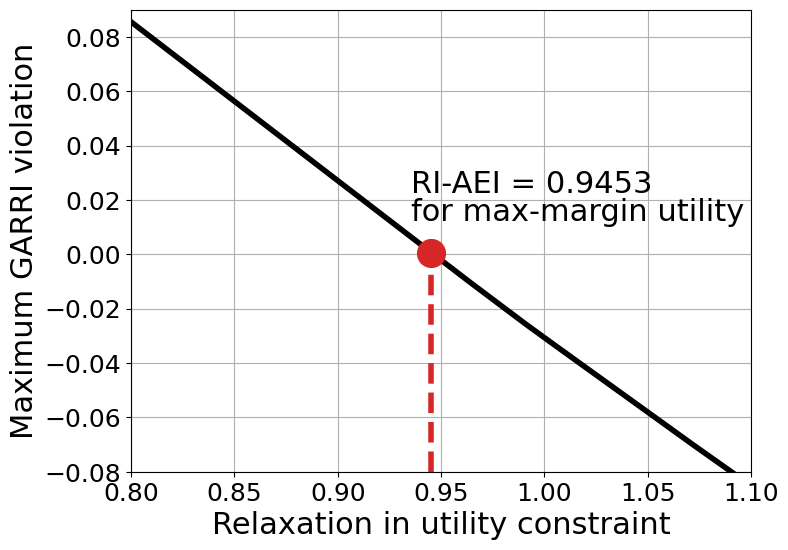}
            \caption{RI-AEI~\eqref{eqn:RI-AEI} for max-margin utility~\eqref{eqn:max-margin-utility}}
            \label{fig:ri-aei-maxmargin}
        \end{subfigure}
        \hspace{0.05cm}
        \begin{subfigure}[b]{0.48\textwidth}
            \centering 
            \includegraphics[height = 0.732\textwidth]{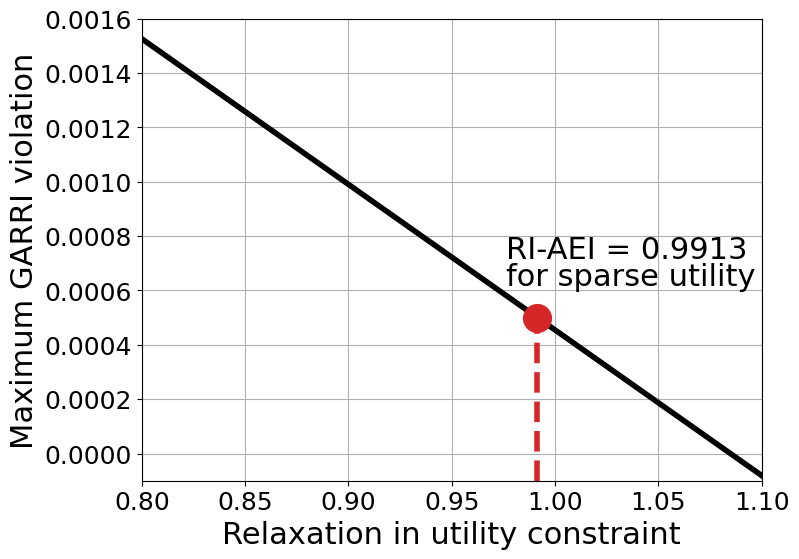}
            \caption{RI-AEI~\eqref{eqn:RI-AEI} for sparse utility~\eqref{eqn:sparse-utility}}    
            \label{fig:ri-aei-sparse}
        \end{subfigure}
        \caption{Rationally Inattentive Afriat Efficiency Index (RI-AEI) defined in \eqref{eqn:RI-AEI} computed for the YouTube dataset for two choices of utility functions: (a) the max-margin utility computed via \eqref{eqn:max-margin-utility} and (b) the sparse utility computed via \eqref{eqn:sparse-utility}. For context, the zero relaxation case (NIAS and GARRI conditions) corresponds to $x=1$ in the above plots. If the dataset is not consistent with rational inattention, then the minimum relaxation~\eqref{eqn:RI-AEI} required for the dataset to be consistent with rational inattention is greater than unity. In our case, the max-margin and sparse utility functions and the corresponding rational inattention costs satisfy NIAS and GARRI by definition. Hence, the red stem line in the above plots indicate the maximum relaxation~\eqref{eqn:RI-AEI} less than unity for which NIAS and GARRI conditions for rational inattention fail to hold for a specified inequality tolerance of 0.005. Our numerical results show that the RI-AEI computed for the max-margin utility is $\sim$6\% greater (hence, better fit to the dataset) than that computed for the sparse utility function. This experimental finding agrees with intuition, since the max-margin utility maximizes the goodness-of-fit of NIAS and GARRI constraints over all utility functions.} 
        \label{fig:ri-aei}
    \end{figure*}

\begin{figure}
    \centering
    \includegraphics[width = 0.65\textwidth]{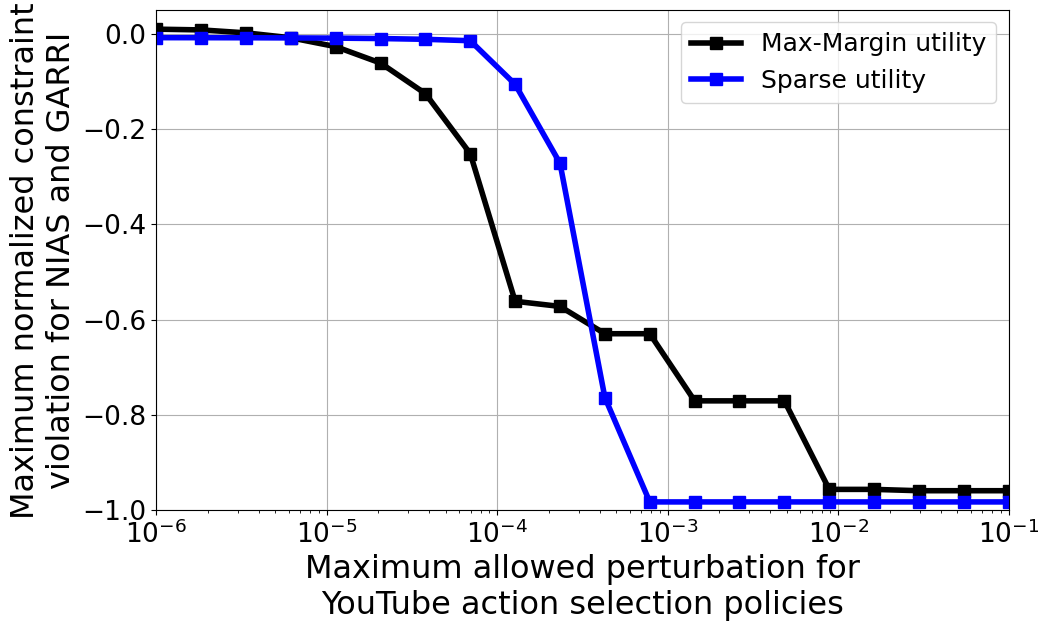}
    \caption{Maximum violation in NIAS and GARRI conditions versus maximum perturbation in action selection policies~\eqref{eqn:YT_dataset} aggregated from YouTube metadata. The key takeaway is that for low perturbations ($\leq$ 0.002), the sparse utility has a lower constraint (NIAS, GARRI) violation compared to the max-margin utility. However, for perturbations exceeding 0.002, the max-margin utility has a lower constraint violation.}
    \label{fig:ri-mpt}
\end{figure}

Figures~\ref{fig:ri-aei},~\ref{fig:ri-mpt} and Table~\ref{tab:RI-MPI} display the results of our numerical experiments on the robustness of YouTube metadata to rational inattention for the max-margin utility and sparse utility computed via \eqref{eqn:max-margin-utility} and \eqref{eqn:sparse-utility}, respectively. Our main observation is that the max-margin utility rationalizes the YouTube metadata better than the sparse utility, and has better goodness-of-fit values to the revealed rational inattention test obtained by the robustness metrics defined in \eqref{eqn:adapted-RI-AEI},~\eqref{eqn:adapted-RI-MPT} and \eqref{eqn:RI-MPI}. Our numerical results also agree with intuition since the max-margin utility, by definition, maximizes the margin of feasibility of the NIAS and GARRI conditions for rational inattention. 

{\em Adapted RI-AEI:} Figures~\ref{fig:ri-aei-maxmargin} and \ref{fig:ri-aei-sparse} display the maximum constraint violation of the GARRI condition for the max-margin utility~\eqref{eqn:max-margin-utility} and sparse utility~\eqref{eqn:sparse-utility}, respectively, as the relaxation parameter $e$ is varied in the inequality~\eqref{eqn:adapted-RI-AEI}. We observed that the goodness-of-fit (adapted RI-AEI) for the max-margin utility exceeds that for the sparse utility by 5\%, hence showing the max-margin utility rationalizes YouTube metadata better than the sparse utility.

{\em Adapted RI-MPT:} Figure~\ref{fig:ri-mpt} plots the maximum violation of NIAS and GARRI conditions as the maximum $\mathcal{L}_2$-perturbations in the action selection policies of YouTube metadata is varied, for both the max-margin and spare utility computed via \eqref{eqn:max-margin-utility} and \eqref{eqn:sparse-utility}. We observed that for a maximum perturbation less than 0.0002, the maximum NIAS and GARRI constraint violation for the sparse utility exceeds that for the max-margin utility. However, for a maximum perturbation exceeding 0.0002, the max-margin utility has a lower constraint violation than that for the sparse utility function.

{\em Adapted RI-MPI:} Table~\ref{tab:RI-MPI} displays the mean, standard deviation and maximum value of the normalized excess expected utility a malicious arbitrager can make (defined in \eqref{eqn:exputil-diff}) for all subsets of decision problems (video categories) in $\dpset = \{1,2,\ldots,8\}$ as the subset size is varied from 2 to 8. Recall from \eqref{eqn:adapted-RI-MPI} that the normalized excess expected utility is inversely proportional to the goodness-of-fit of a Bayesian decision maker's actions to rational inattention. We observe from Table~\ref{tab:RI-MPI} that the maximum arbitrage possible for the max-margin and sparse utilities is 0.201 and 0.215, respectively. From \eqref{eqn:adapted-RI-AEI}, the adapted RI-AEI for the max-margin and sparse utilities is simply the negative of the maximum excess expected utility obtained by arbitrage, Hence, the adapted RI-AEI is -0.201 for the max-margin utility, and -0.215 for the sparse utility. The adapted RI-AEI robustness value is approximately 7\% lower than that for the sparse utility. Hence, the max-margin utility rationalizes the YouTube metadata better than the sparse utility.

To summarize, we conducted a principled analysis of goodness-of-fit of YouTube metadata for two point utility estimates, namely, the max-margin utility~\eqref{eqn:max-margin-utility} and the sparse utility~\eqref{eqn:sparse-utility}. The max-margin utility maximizes the extent of feasibility of the NIAS and GARRI conditions for Bayesian rationality, while the sparse utility yields the most parsimonious utility representation of YouTube user engagement that rationalizes YouTube metadata. For our goodness-of-fit analysis, we first adapted the robustness measures for the revealed rational inattention test introduced in Sec.\,\ref{sec:extension-robustness} to measure how close a dataset is to failing the revealed rational inattention test, and computed the adapted robustness measures for the YouTube metadata. Our main takeaway is that the max-margin utility rationalizes the YouTube metadata substantially better than the sparse utility, and outperforms the sparse utility for all adapted robustness measures. To conclude, the max-margin utility and sparse utility estimates can be viewed as robust and parsimonious representations, respectively, of YouTube user engagement that rationalizes YouTube metadata.

\begin{table}
    \centering
    \begin{tabular}{|c|c|c|c|}\hline
    \multicolumn{4}{|c|}{RI-MPI~\eqref{eqn:RI-MPI} for Max-Margin Utility~\eqref{eqn:max-margin-utility}} \\ \hline
        Seq.\ length & Mean & St.\ dev.\ & Max. value  \\ \hline
        2 & -0.0455 & 0.0243 & 0.0201\\ \hline
        3 & -0.0456 & 0.0186 & -0.0021\\ \hline
        \blue{\bf 4} & -0.0453 & 0.0164 & {\bf \blue{0.0201}}\\ \hline
        5 & -0.0451 & 0.0145 & 0.0067\\ \hline
         6 & -0.0447 & 0.0131 & 0.0198\\ \hline
        7 & -0.0447 & 0.0122 & 0.0104\\ \hline
        8 & -0.0435 & 0.0114 & 0.0168\\ \hline
    \end{tabular}
\hspace{1cm}
        \begin{tabular}{|c|c|c|c|}\hline
        \multicolumn{4}{|c|}{RI-MPI~\eqref{eqn:RI-MPI} for Sparse Utility~\eqref{eqn:sparse-utility}} \\ \hline
        Seq.\ length & Mean & St.\ dev.\ & Max. value  \\ \hline
        2 & -0.0394 & 0.0195 & 0.0176\\ \hline
        3 & -0.0431 & 0.0157 & -0.0059\\ \hline
        4 & -0.0415 & 0.0135 & 0.0185\\ \hline
        5 & -0.0455 & 0.0132 & 0.0043\\ \hline
        \blue{\bf 6} & -0.0480 & 0.0127 & {\bf \blue{0.0215}}\\ \hline
        7 & -0.0496 & 0.0121 & 0.0098\\ \hline
        8 & -0.0495 & 0.0117 & 0.0201\\ \hline
    \end{tabular}
    \caption{Rationally inattentive Money Pump Index (RI-MPI) defined in \eqref{eqn:RI-MPI} computed for the YouTube dataset for two choices of utility functions: (a) max-margin utility computed via \eqref{eqn:max-margin-utility} and (b) sparse utility computed via \eqref{eqn:sparse-utility}. Our numerical experiments show that the YouTube dataset is prone to larger maximum arbitrage loss~\eqref{eqn:RI-MPI-2} for the sparse utility function compared to the max-margin utility (approximately $8\%$ greater arbitrage). For every index tuple of size $i\in\{2,3,\ldots,8\}$, we computed the arbitrage~\eqref{eqn:RI-MPI-2} for all $8!/(8-i)!$ index permutations, and report the mean, standard deviation and maximum value of the arbitrage. Our numerical results also agree with intuition, since the max-margin utility maximizes the goodness-of-fit of the NIAS and GARRI constraints over all utility functions.}
    \label{tab:RI-MPI}
\end{table}

\section{Conclusion and Future Work}
In this paper, we established the connection between  revealed preference and revealed rational inattention. Our main finding is that the NIAC condition~\cite{CD15} in revealed rational inattention is a special case of GARP~\cite{VR82} in revealed preference under a different partial order and a different state space (probability simplex). We exploit this result to construct a monotone convex information acquisition cost in revealed rational inattention. The construction procedure resembles that of the utility function reconstructed from consumer data in \cite{AF67}. Due to the equivalence result, we adapt goodness-of-fit measures from revealed preference to the revealed rational inattention case and characterize how well a dataset fits the rational inattention model. Finally, we characterize the goodness-of-fit of a massive dataset of YouTube metadata from 140,000 videos using the adapted robustness measures. To the best of our knowledge, our numerical experiments are a novel exercise on systematically analyzing the goodness-of-fit of decision data from Bayesian decision makers to the rational inattention model.

In future work it is worthwhile exploiting this unification to study revealed preference in Bayesian versions of potential games building on \cite{DEB08,DEB09}, market games building on \cite{FO09}, inverse reinforcement learning building on \cite{NG00,PK23}, and dynamic revealed preference building on \cite{CR10}.

\begin{acks}
This research was funded in part by the National Science Foundation grant CCF-2112457 and Army Research Office grant W911NF-21-1-0093.
\end{acks}

\bibliographystyle{unsrt_abbrv_custom}
\bibliography{bib-file}

\appendix

\section{Proof of Theorem~\ref{thrm:argmin}}\label{app:proof_argmin}
{\bf Statement (1)$\Rightarrow$(2).}\\
Fix indices $\dpitertwo,\dpiter$ and assume $\response_\dpiter\geq_{H}\response_\dpitertwo$. From the definition of the relation `$\geq_{H}$'~\eqref{eqn:GARP_nonlinear-alt}, there exist indices $i_1,i_2,\ldots,i_L$ such that such that $\utilitysymbolagent{\dpiter}(\response_{i_1})\geq \utilitysymbolagent{\dpiter}(\response_\dpiter),~\utilitysymbolagent{i_1}(\response_{i_2})\geq \utilitysymbolagent{i_1}(\response_{i_1}),~\ldots,\utilitysymbolagent{i_L}(\response_{\dpitertwo})\geq \utilitysymbolagent{i_L}(\response_{i_L})$. Since $\nonlinb(\cdot)$ rationalizes the decision maker's dataset, we must have $\nonlinb(\response_\dpiter)\leq \nonlinb(\response_{i_1})\leq \nonlinb(\response_{i_2}) \leq \ldots\leq \nonlinb(\response_{i_L})\leq \nonlinb(\response_{\dpitertwo})$ which, in turn, implies $\nonlinb(\response_\dpiter)\leq \nonlinb(\response_\dpitertwo)$.

Our aim is to show $\utilitysymbolagent{\dpitertwo}(\response_\dpitertwo)\geq\utilitysymbolagent{\dpitertwo}(\response_\dpiter)$. We prove this by contradiction. Suppose $\utilitysymbolagent{\dpitertwo}(\response_\dpitertwo)<\utilitysymbolagent{\dpitertwo}(\response_\dpiter)$. Then, from the continuity of $\utilitysymbolagent{\dpitertwo}(\cdot)$ and monotonicity of $\nonlinb(\cdot)$, we could find a $\response\in\reals_+^m$ in the neighborhood of $\response_\dpiter$ such that the $\response$ satisfies the two conditions below:\\
(a) $\utilitysymbolagent{\dpitertwo}(\response_\dpitertwo)<\utilitysymbolagent{\dpitertwo}(\response)\leq\utilitysymbolagent{\dpitertwo}(\response_\dpiter)$ ($\response$ satisfies utility constraint~\eqref{eqn:UM-nonlinear-alt-intro} for time step $\dpitertwo$) and\\
(b) $\nonlinb(\response)<\nonlinb(\response_\dpiter)\leq \nonlinb(\response_\dpitertwo)$, or equivalently, $\nonlinb(\response)\leq \nonlinb(\response_\dpitertwo)$ ($\response$ {\em strictly} costs lesser compared to $\response_\dpitertwo$).\\
Clearly, if both (a) and (b) hold, then $\response_\dpitertwo$ does not rationalize the decision maker's action at time $\dpitertwo$, i.e.\,, $\nonlinb$ does not rationalize the analyst's dataset, hence our assumption is false. Therefore, it must be the case that $\boxed{\utilitysymbolagent{\dpitertwo}(\response_\dpitertwo)\geq\utilitysymbolagent{\dpitertwo}(\response_\dpiter)}$.\vspace{0.2cm}\\
{\bf Statement (2)$\Rightarrow$(3).} Construct a matrix $A\in\reals^{\numdp\times\numdp}$ with elements $A_{j,k} = \utilitysymbolagent{\dpiter}(\response_\dpiter) - \utilitysymbolagent{\dpiter}(\response_\dpitertwo)$. Since GARP~\eqref{eqn:GARP_nonlinear-alt} holds, it is trivial to show the matrix A is cyclically consistent. From \cite[Lemma 2]{FO09} and \cite[Sections 2 and 3]{FOS04}, there exist scalars $\bar{\nonlinb}_\dpiter$ and $\lambda_\dpiter>0$ that satisfy the following set of Afriat-type~\eqref{eqn:Lagrange_nonlinearafriat} feasibility inequalities:
\begin{equation}\label{eqn:proof_th_2}
    \bar{\nonlinb}_\dpitertwo- \bar{\nonlinb}_\dpiter - \lambda_\dpiter(\utilitysymbol_\dpiter(\response_\dpiter)-\utilitysymbol_\dpiter(\response_\dpitertwo))\leq 0~\quad \forall~\dpiter,\dpitertwo. 
\end{equation}
Although \eqref{eqn:proof_th_2} resembles the Afriat inequalities in Theorem~\ref{thrm:argmin}, we cannot reconstruct the decision maker's cost that rationalizes its choices~\eqref{eqn:UM-nonlinear-alt-intro} without modifying the feasible variables $\bar{\nonlinb}_\dpiter,~\lambda_\dpiter$. However, if we were to reconstruct the cost $\nonlinb$ using $\bar{\nonlinb}_\dpiter$ in \eqref{eqn:proof_th_2}, we would obtain a decreasing function that violates the properties of the cost. To alleviate this issue and have a monotone reconstruction of the cost, we perform the following change of variables: \\
Without loss of generality, we restrict $\bar{\nonlinb}_\dpiter$ to be finite for all $\dpiter=1,2,\ldots,\numdp$. Let $M<\infty$ denote an arbitrary positive scalar that uniformly bounds $\{\bar{\nonlinb}_\dpiter\}_{\dpiter=1}^\numdp$ from above. Note that since $\bar{\nonlinb}_\dpiter$ is bounded for all $\dpiter$, such an $M$ exists and $\bar{\nonlinb}_\dpiter> 0$. We now define the variable $\nonlinb_\dpiter = M - \bar{\nonlinb}_\dpiter$ for all $\dpiter$ and rewrite \eqref{eqn:proof_th_2} in terms of the new variables $\{\nonlinb_\dpiter\}_{\dpiter=1}^\numdp$:
\begin{equation}\label{eqn:final-manipulation-cost}
\begin{split}
    & \bar{\nonlinb}_\dpitertwo- \bar{\nonlinb}_\dpiter - \lambda_\dpiter(\utilitysymbol_\dpiter(\response_\dpiter)-\utilitysymbol_\dpiter(\response_\dpitertwo))\leq 0,~\forall~\dpiter,\dpitertwo\\
    \Leftrightarrow & - \bar{\nonlinb}_\dpitertwo- (-\bar{\nonlinb}_\dpiter) - \lambda_\dpiter(\utilitysymbol_\dpiter(\response_\dpitertwo)-\utilitysymbol_\dpiter(\response_\dpiter))\geq 0,~\forall~\dpiter,\dpitertwo\\
    \Leftrightarrow & (M-\bar{\nonlinb}_\dpitertwo)- (M-\bar{\nonlinb}_\dpiter) - \lambda_\dpiter(\utilitysymbol_\dpiter(\response_\dpitertwo)-\utilitysymbol_\dpiter(\response_\dpiter))\geq 0,~\forall~\dpiter,\dpitertwo\\
    \Leftrightarrow & \boxed{\nonlinb_\dpitertwo- \nonlinb_\dpiter - \lambda_\dpiter(\utilitysymbol_\dpiter(\response_\dpitertwo)-\utilitysymbol_\dpiter(\response_\dpiter))\geq 0,~\forall~\dpiter,\dpitertwo \equiv \eqref{eqn:Lagrange_nonlinearafriat_alt}}.
    \end{split}
\end{equation}
Consider the reconstructed cost $\nonlinb(\response)=\max_{\dpiter} \{\nonlinb_\dpiter + \lambda_{\dpiter}(\utilitysymbol_\dpiter(\response)-\utilitysymbol_\dpiter(\response_\dpiter))\}$. Clearly, $\nonlinb$ is monotone and continuous since it is a point-wise maximum of monotone continuous functions. We now show that $\nonlinb_{recon}(\response) = \max_{\dpiter} \{\nonlinb_\dpiter + \lambda_{\dpiter}(\utilitysymbol_\dpiter(\response)-\utilitysymbol_\dpiter(\response_\dpiter))\}$ rationalizes the dataset $\{\response_\dpiter,\utilitysymbolagent{\dpiter}(\cdot)\geq\utilitysymbolagent{\dpiter}^\ast\}_{\dpiter=1}^\numdp$:\\
Fix index $\dpiter$. Then, $\nonlinb_{recon}(\response_\dpiter)\geq \nonlinb_\dpiter$ by definition. Also, from \eqref{eqn:final-manipulation-cost}, we have that:
\begin{equation*}
    \nonlinb_\dpiter \geq \nonlinb_\dpitertwo + \lambda_\dpitertwo (\utilitysymbolagent{\dpitertwo}(\response_\dpiter) - \utilitysymbolagent{\dpitertwo}(\response_\dpitertwo))~\forall \dpitertwo\neq\dpiter
\end{equation*}
Therefore, it is clear that $\nonlinb_{recon}(\response_\dpiter) = \nonlinb_\dpiter$~\eqref{eqn:final-manipulation-cost}. Now, let $\response$ denote any feasible consumption bundle at time $\dpiter$, that is, $\utilitysymbolagent{\dpiter}(\response)\geq\utilitysymbolagent{\dpiter}^\ast = \utilitysymbolagent{\dpiter}(\response_\dpiter)$. Then:\\
\begin{equation}\label{eqn:rationalizability}
\begin{split}
\nonlinb_{recon}(\response)& = \max_{\dpiter} \{\nonlinb_\dpiter + \lambda_{\dpiter}(\utilitysymbol_\dpiter(\response)-\utilitysymbol_\dpiter(\response_\dpiter))\}\\
& \geq \underbrace{\nonlinb_\dpiter}_{=\nonlinb_{recon}(\response_\dpiter)} +~ \underbrace{\lambda_{\dpiter}(\utilitysymbol_\dpiter(\response)-\utilitysymbol_\dpiter(\response_\dpiter))}_{\geq 0} \geq \nonlinb_{recon}(\response_\dpiter)\\
\Rightarrow~&\boxed{\response_\dpiter =\argmin_{\response}~\nonlinb_{recon}(\response)~\text{s.t. }~\utilitysymbolagent{\dpiter}(\response)\geq \utilitysymbolagent{\dpiter}^\ast}
\end{split}
\end{equation}
Since \eqref{eqn:rationalizability} holds for all $\dpiter=1,2,\ldots,\numdp$, the reconstructed cost $\nonlinb_{recon}(\response) = \max_{\dpiter} \{\nonlinb_\dpiter + \lambda_{\dpiter}(\utilitysymbol_\dpiter(\response)-\utilitysymbol_\dpiter(\response_\dpiter))\}$ rationalizes the analyst's dataset $\{\response_\dpiter,\utilitysymbolagent{\dpiter}(\cdot)\geq\utilitysymbolagent{\dpiter}^\ast\}_{\dpiter=1}^\numdp$.\vspace{0.2cm}\\
{\bf Statement (3)$\Rightarrow$(1).} Consider the reconstructed cost $\nonlinb_{recon}(\response)=\max_{\dpiter} \{\nonlinb_\dpiter + \lambda_{\dpiter}(\utilitysymbol_\dpiter(\response)-\utilitysymbol_\dpiter(\response_\dpiter))\}$. By construction, the cost $\nonlinb_{recon}$ is monotone, continuous (since it is a point-wise maximum of finitely many monotone continuous segments) and rationalizes the decision maker's actions~\eqref{eqn:rationalizability}.\hfill\qedsymbol

\section{Proof of Theorem~\ref{thrm:eq}} \label{app:proof-outline}
The proof of our unification result, namely,  Theorem~\ref{thrm:eq}, comprises two steps. First, we show the NIAC condition can be expressed as an equivalent feasibility inequality. Second, under the variable map of statement 1 of Theorem 1, we compare the equivalent inequality with the Afriat-type inequality~\eqref{eqn:Lagrange_nonlinearafriat_alt} in Theorem~\ref{thrm:argmin} for testing if non-Bayesian cost minimization~\eqref{eqn:UM-nonlinear-alt-intro} holds. 
Finally, statement (2) in Theorem~\ref{thrm:eq} is proved in Appendix~\ref{appdx:proof-corollary}.

\subsection{Expressing the NIAC condition as an equivalent feasibility inequality}\label{app:proof-equivalence-Afriat-NIAC}
Suppose NIAC~\eqref{eqn:niac-vanilla} is true. Then, one can show using the concept of KKT conditions from duality theory~\cite[Sec.\,5.5]{BD04} that 
there exist non-negative scalars $\{\runcostinst_\dpiter\}_{\dpiter=1}^\numdp$ that satisfy the following inequalities:
\begin{equation}\label{eqn:niac-feasibility}
     \boxed{\grosspayoff(p_\dpiter(\action|\state),\utilitysymbolbig_{\dpiter}) - \runcostinst_{\dpiter} \geq  \grosspayoff(p_{\dpitertwo}(\action|\state),\utilitysymbolbig_{\dpiter}) - \runcostinst_{\dpitertwo},~\forall~\dpitertwo,\dpiter=1,2,\ldots,\numdp}.
\end{equation}
In \eqref{eqn:niac-feasibility}, the scalars $\{\runcostinst_\dpiter\}_{\dpiter=1}^\numdp$ are Lagrange multipliers corresponding to an equivalent linear assignment problem~\cite{KM57} that is solved by the identity map if the NIAC condition is true. We refer the reader to \cite[Appendix~C.2.2]{PK23} and \cite[Sec.\,10.2]{CD15} for a more elaborate discussion on the existence of the scalars $\{\runcostinst_\dpiter\}_{\dpiter=1}^\numdp$ that satisfy \eqref{eqn:niac-feasibility} if NIAC is true.

To prove equivalence between \eqref{eqn:niac-feasibility} and NIAC, it suffices to show that if there exist non-negative scalars $\{\runcostinst_\dpiter\}_{\dpiter=1}^\numdp$ that satisfy \eqref{eqn:niac-feasibility}, then NIAC holds.
Fix a sequence of indices $\dpiter_1,\dpiter_2,\ldots,\dpiter_m,~m\leq\numdp$, where $\dpiter_i\in\{1,2,\ldots,\numdp\}$, $\forall i=1,2,\ldots,m$. Since there exists a feasible solution $\{\runcostinst_\dpiter\}_{\dpiter=1}^\numdp$ to the inequality \eqref{eqn:niac-feasibility}, the following inequalities result:
\begin{align*}
     \grosspayoff(p_{\dpiter_1}(\action|\state),\utilitysymbolbig_{\dpiter_1}) - \runcostinst_{\dpiter_1} &\geq \grosspayoff(p_{\dpiter_2}(\action|\state),\utilitysymbolbig_{\dpiter_1}) - \runcostinst_{\dpiter_2}\\
     \grosspayoff(p_{\dpiter_2}(\action|\state),\utilitysymbolbig_{\dpiter_2}) - \runcostinst_{\dpiter_2}& \geq \grosspayoff(p_{\dpiter_3}(\action|\state),\utilitysymbolbig_{\dpiter_2}) - \runcostinst_{\dpiter_3}\\
    & \ldots\\
     \grosspayoff(p_{\dpiter_{m-1}}(\action|\state),\utilitysymbolbig_{\dpiter_{m-1}}) - \runcostinst_{\dpiter_{m-1}} & \geq \grosspayoff(p_{\dpiter_m}(\action|\state),\utilitysymbolbig_{\dpiter_{m-1}}) - \runcostinst_{\dpiter_{m}}\\
     \grosspayoff(p_{\dpiter_m}(\action|\state),\utilitysymbolbig_{\dpiter_{m}}) - \runcostinst_{\dpiter_{m}} &\geq \grosspayoff(p_{\dpiter_1}(\action|\state),\utilitysymbolbig_{\dpiter_{m}}) - \runcostinst_{\dpiter_{1}}\\
     \Rightarrow \sum_{i=1}^{m} \bigg(\grosspayoff(p_{\dpiter_i}(\action|\state),\utilitysymbolbig_{\dpiter_i}) - \grosspayoff(p_{\dpiter_{i+1}}(\action|\state),&\utilitysymbolbig_{\dpiter_i})  \bigg)  \geq 0,~\text{where } \dpiter_{m+1} = \dpiter_1 \\
     \underset{\text{\footnotesize (if NIAS holds)}}{\Leftrightarrow} \sum_{i=1}^{m} \bigg(\sum_{\state,\action}\prior(\state)p_{\dpiter_i}(\action|\state)\utilitysymbolbig_{\dpiter_i}(\state,\action) - \grosspayoff(p_{\dpiter_{i+1}}(\action|\state),&\utilitysymbolbig_{\dpiter_i})  \bigg)  \geq 0,~\text{where } \dpiter_{m+1} = \dpiter_1 \equiv \operatorname{NIAC}~\eqref{eqn:niac-vanilla}
\end{align*}
Hence, $\boxed{\operatorname{NIAC}~\eqref{eqn:niac-vanilla}\equiv\text{ there exists feasible non-negative scalars }\{\runcostinst_\dpiter\}_{\dpiter=1}^\numdp\text{ that satisfy }\eqref{eqn:niac-feasibility}}$.\hfill\qedsymbol

\subsection{Relating the feasibility inequalities for NIAC~(\ref{eqn:niac-feasibility}) and GARP~(\ref{eqn:Lagrange_nonlinearafriat_alt}) under the variable map of Theorem~\ref{thrm:eq}}
We now relate the feasibility inequality \eqref{eqn:niac-feasibility} to the Afriat-type revealed preference inequality \eqref{eqn:Lagrange_nonlinearafriat_alt}:
\begin{equation}\label{eqn:niac-garp-compare}
\begin{split}
    \operatorname{NIAC}\underset{\text{\footnotesize if NIAS holds}}{\Leftrightarrow}\eqref{eqn:niac-feasibility}:~&\text{ given dataset $\dataset_{RRI}$~\eqref{eqn:dataset-Bayesian}, there exist } \{\runcostinst_\dpiter\}_{\dpiter=1}^\numdp\geq 0\text{ such that \eqref{eqn:niac-feasibility} is feasible:}\\
    &~\grosspayoff(p_\dpiter(\action|\state),\utilitysymbolbig_{\dpiter}) - \runcostinst_{\dpiter} \geq  \grosspayoff(p_{\dpitertwo}(\action|\state),\utilitysymbolbig_{\dpiter}) - \runcostinst_{\dpitertwo},~\forall~\dpitertwo,\dpiter\in\{1,2,\ldots,\numdp\},~\dpiter\neq\dpitertwo.\\
   \operatorname{GARP}\Leftrightarrow\eqref{eqn:Lagrange_nonlinearafriat_alt}:~&\text{ given dataset $\dataset_{RP}$~\eqref{eqn:dataset-RP-alt}, there exist } \{\lambda_\dpiter,\nonlinb_\dpiter\}_{\dpiter=1}^\numdp\geq 0\text{ such that \eqref{eqn:Lagrange_nonlinearafriat_alt} is feasible:}\\
   &~\lambda_\dpiter~\utilitysymbolagent{\dpiter}(\response_\dpiter) - \nonlinb_\dpiter \geq \lambda_\dpiter~\utilitysymbolagent{\dpiter}(\response_\dpitertwo) - \nonlinb_\dpitertwo,~\forall~\dpitertwo,\dpiter\in\{1,2,\ldots,\numdp\},~\dpiter\neq\dpitertwo.
   \end{split}
\end{equation}
Clearly, from \eqref{eqn:niac-garp-compare}, we observe that the feasibility of \eqref{eqn:niac-feasibility} is equivalent to the feasibility of
\eqref{eqn:Lagrange_nonlinearafriat_alt} with $\lambda_\dpiter=1$ for all $\dpiter$ under the variable map of statement (1) in Theorem~\ref{thrm:eq} . Hence, NIAC is a special case of GARP if NIAS holds, under the variable map of statement (1) of Theorem~\ref{thrm:eq}. This concludes the proof of Theorem~\ref{thrm:eq}.\hfill \qedsymbol

\section{Proof of Lemma~\ref{lem:partialorder_relax}}\label{appdx:proof-lemma}
We first show the expected utility~\eqref{eqn:exputil-diff} is monotone wrt the Blackwell partial order. Consider two attention strategies $\attfunagent{1},\attfunagent{2}\in\Delta(\obsset)^{|\stateset|}$, where $\attfunagent{1}\BD\attfunagent{2}$ without loss of generality\footnote{Although implicitly assumed in the proof, showing monotonicity of the expected utility wrt Blackwell dominance does not require both attention strategies $\attfunagent{1},~\attfunagent{2}$ to be defined on the same space of observations. Adapting the proof to the case where $\attfunagent{1}\in\Delta(\obsset_1)^{|\stateset|},~\attfunagent{2}\in\Delta(\obsset_2)^{|\stateset|},~\obsset_1\neq\obsset_2$ is straightforward, and hence, omitted.}. From the definition of Blackwell dominance~\cite{BD04}, there exists a matrix $Q\in[0,1]^{|\obsset|\times|\obsset|}$ such that $\attfunagent{2}(\obs|\state) = \sum_{\obs'\in\obsset}~\BDmat_{\obs',\obs}~\attfunagent{1}(\obs'|\state)$ and $\sum_{\obs\in\obsset}~\BDmat_{\obs',\obs} = 1$ for all $\obs,\obs'\in\obsset,~\state\in\stateset$. We now prove that the expected utility functional $\exputil(\cdot,\utilitysymbolbig)$ for a utility function $\utilitysymbolbig$ is monotone with respect to the Blackwell partial order, that is, $\exputil(\attfunagent{1},\utilitysymbolbig)\geq \exputil(\attfunagent{2},\utilitysymbolbig)$:
\begin{equation}\label{eqn:exputil-monotone}
    \begin{split}
        & \exputil(\attfunagent{2},\utilitysymbolbig) = \sum_{\obs\in\obsset}~p_2(\obs)~\max_{\action\in\actionset} \belief_{\obs,2}'\utilitysymbolbig(\cdot,\action) = \sum_{\obs\in\obsset} \max_{\action\in\actionset}\sum_{\state\in\stateset}~p_2(\obs)~\belief_{\obs,2}(\state)~\utilitysymbolbig(\state,\action) \\
        = & \sum_{\obs\in\obsset} \max_{\action\in\actionset}\sum_{\state\in\stateset}~\attfunagent{2}(\obs|\state)~\prior(\state)\utilitysymbolbig(\state,\action) = \sum_{\obs\in\obsset} \max_{\action\in\actionset}~\sum_{\obs'\in\obsset}Q_{\obs',\obs}~\left(\sum_{\state}\attfunagent{1}(\obs'|\state)~\prior(\state)\utilitysymbolbig(\state,\action)\right)\\
        \leq &  \sum_{\obs\in\obsset}\left(\sum_{\obs'\in\obsset}Q_{\obs',\obs}~ \max_{\action\in\actionset}~\left(\sum_{\state}\attfunagent{1}(\obs'|\state)~\prior(\state)\utilitysymbolbig(\state,\action)\right)\right) = \sum_{\obs'\in\obsset} \underbrace{\left(\sum_{\obs\in\obsset}~Q_{\obs',\obs}\right)}_{=1\forall~\obs'\in\obsset}~\max_{\action\in\actionset}~\sum_{\state}\attfunagent{1}(\obs'|\state)~\prior(\state)\utilitysymbolbig(\state,\action) = \exputil(\attfunagent{1},\utilitysymbolbig)\\
        &\text{Therefore, }\boxed{\attfunagent{1}\BD\attfunagent{2}\implies \exputil(\attfunagent{1},\utilitysymbolbig)\geq \exputil(\attfunagent{2},\utilitysymbolbig)~\text{for any utility function }\utilitysymbolbig}\qed
    \end{split}
\end{equation}
Eq.\,\ref{eqn:exputil-monotone} shows that the expected utility functional $\exputil(\cdot,\utilitysymbolbig)$ is monotone in the Blackwell partial order. We now prove the expected utility is convex in the attention strategy. Fix a scalar $\theta\in[0,1]$. Define $\attfunagent{\theta} = \theta~\attfunagent{1} + (1-\theta)~\attfunagent{2}$. Then:
\begin{equation}\label{eqn:exputil-convex}
\begin{split}
& \exputil(\attfunagent{\theta},\utilitysymbolbig) = \sum_{\obs\in\obsset} p_\theta(\obs) \max_{\action\in\actionset} \belief_{\obs,\theta}'\utilitysymbolbig(\cdot,\action) = \sum_{\obs\in\obsset} \max_{\action\in\actionset} \sum_{\state\in\stateset} p_\theta(\obs)\belief_{\obs,\theta}(\state)\utilitysymbolbig(\state,\action)\\
= & \sum_{\obs\in\obsset} \max_{\action\in\actionset} \sum_{\state\in\stateset} \attfunagent{\theta}(\obs|\state)\prior(\state)\utilitysymbolbig(\state,\action) = \sum_{\obs\in\obsset} \max_{\action\in\actionset} \sum_{\state\in\stateset} (\theta\attfunagent{1}(\obs|\state) + (1-\theta)\attfunagent{2}(\obs|\state))~\prior(\state)\utilitysymbolbig(\state,\action) \\
\leq & ~\theta~\sum_{\obs\in\obsset} \max_{\action\in\actionset} \sum_{\state\in\stateset} \attfunagent{1}(\obs|\state)~\prior(\state)~\utilitysymbolbig(\state,\action) + (1-\theta)~\sum_{\obs\in\obsset} \max_{\action\in\actionset} \sum_{\state\in\stateset} \attfunagent{2}(\obs|\state)~\prior(\state)~\utilitysymbolbig(\state,\action)\\
= & ~\theta~\exputil(\attfunagent{1},\utilitysymbolbig) + (1-\theta)~\exputil(\attfunagent{2},\utilitysymbolbig)
\end{split}
\end{equation}
Therefore, $\boxed{\exputil(\attfunagent{\theta},\utilitysymbolbig) \leq \theta~\exputil(\attfunagent{1},\utilitysymbolbig) + (1-\theta)~\exputil(\attfunagent{2},\utilitysymbolbig)~\forall~\theta\in[0,1],\attfunagent{1},\attfunagent{2},\utilitysymbolbig.}$\hfill \qedsymbol

{\em Remark.}
Lemma~\ref{lem:partialorder_relax} is crucial in the proof of the revealed rational inattention result of Theorem~\ref{thrm:BRP-vanilla} since the attention strategy $\attfunagent{\dpiter}$ in decision problem $\dpiter$ Blackwell dominates the action selection policy $p_\dpiter(\action|\state)$ observed by the analyst:
\begin{equation}
    p_\dpiter(\action|\state) = \sum_{\obs\in\obsset} p_\dpiter(\action,\obs|\state) = \sum_{\obs\in\obsset} p_\dpiter(\action|\obs,\state)\attfunagent{\dpiter} = \sum_{\obs\in\obsset} \eta_\dpiter(\action|\obs)\attfunagent{\dpiter}
\end{equation}
Hence, $\grosspayoff(p_\dpitertwo(\action|\state),\utilitysymbolbig_\dpiter)) = \exputil(p_\dpitertwo(\action|\state),\utilitysymbolbig_\dpiter)\leq \exputil(\attfunagent{\dpitertwo},\utilitysymbolbig_\dpiter)$, where equality holds when $\dpiter=\dpitertwo$, where the surrogate expected utility $\grosspayoff(\cdot)$ is defined in \eqref{eqn:exputil-surrogate}.

\section{Proof of Corollary~\ref{corl:niac-to-garp}}\label{appdx:proof-corollary}
The proof of Corollary~\ref{corl:niac-to-garp} is identical to that of Theorem~\ref{thrm:argmin} (identical via the variable map of statement (1) of Theorem~\ref{thrm:eq}) and exploits the unification result of Theorem~\ref{thrm:eq}. Since the external analyst does not observe the attention strategy $\attfunsymb$, we can assume WLOG that the mapping from observation $\obs$ to action $\action$ is one-to-one\footnote{This assumption is also used by \cite{CD15} to prove the sufficiency of NIAS and NIAC for rationally inattentive utility maximization}. Hence, $\attfunagent{\dpiter}$ can be replaced with $p_\dpiter(\action|\state)$. 

Statement (1)$\Rightarrow$(2):
Fix indices $\dpitertwo,\dpiter$ and assume $p_\dpiter(\action|\state)\geq_{H}p_\dpitertwo(\action|\state)$. From the definition of the relation `$\geq_{H}$'~\eqref{eqn:GARRI}, there exist indices $i_1,i_2,\ldots,i_L$ such that such that $\grosspayoff(p_{i_1}(\action|\state),\utilitysymbolbig_{\dpiter})\geq \grosspayoff(p_{\dpiter}(\action|\state),\utilitysymbolbig_{\dpiter}),~\grosspayoff(p_{i_2}(\action|\state),\utilitysymbolbig_{i_1})\geq \grosspayoff(p_{i_1}(\action|\state),\utilitysymbolbig_{i_1}),$
$\ldots,\grosspayoff(p_{i_L}(\action|\state),\utilitysymbolbig_{i_2})\geq \grosspayoff(p_{i_L}(\action|\state),\utilitysymbolbig_{i_L})$. Since there exists an information acquisition cost $\RIcost$ that rationalizes the decision maker's dataset, we must have $\RIcost(\attfunagent{\dpiter})\leq \RIcost(\attfunagent{i_1})\leq \RIcost(\attfunagent{i_2}) \leq \ldots\leq \RIcost(\attfunagent{i_L})\leq \RIcost(\attfunagent{\dpitertwo})$ which, in turn, implies $\RIcost(\attfunagent{\dpiter})\leq \RIcost(\attfunagent{\dpitertwo})$.

Our aim is to show $\grosspayoff(p_{\dpitertwo}(\action|\state),\utilitysymbolbig_{\dpitertwo})\geq\grosspayoff(p_{\dpiter}(\action|\state),\utilitysymbolbig_{\dpitertwo})$. We prove this by contradiction. Suppose $\grosspayoff(p_{\dpitertwo}(\action|\state),\utilitysymbolbig_{\dpitertwo})<\grosspayoff(p_{\dpiter}(\action|\state),\utilitysymbolbig_{\dpitertwo})$. We note here that the surrogate expected utility $\grosspayoff$~\eqref{eqn:exputil-surrogate} is continuous in its first argument, namely, the action selection policy. Hence, from the continuity of $\grosspayoff(\cdot,\utilitysymbolbig_{\dpitertwo})$ and monotonicity of the cost $\RIcost$, we could find an action selection policy $p(\action|\state)$ in the neighborhood of $p_\dpiter(\action|\state)$ such that:\\
(a) $\grosspayoff(p_{\dpitertwo}(\action|\state),\utilitysymbolbig_{\dpitertwo})<\grosspayoff(p(\action|\state),\utilitysymbolbig_{\dpitertwo})\leq \grosspayoff(p_\dpiter(\action|\state),\utilitysymbolbig_{\dpitertwo})$ ($p(\action|\state)$ satisfies utility constraint~\eqref{eqn:mod-optimal-attention-strategy_alt} for decision problem $\dpitertwo$) and\\
(b) $\RIcost(p(\action|\state))<\RIcost(\attfunagent{\dpiter})\leq \RIcost(\attfunagent{\dpitertwo})$, or equivalently, $\RIcost(p(\action|\state))\leq \RIcost(\attfunagent{\dpitertwo})$ ($p(\action|\state)$ {\em strictly} costs lesser compared to $p_\dpitertwo(\action|\state)$).\\
Clearly, if both (a) and (b) hold, then $p_\dpitertwo(\action|\state)$ does not rationalize the Bayesian decision maker's response in decision problem $\dpitertwo$, i.e.\,, $\RIcost$ does not rationalize the analyst's dataset, hence our assumption is false. Therefore, it must be the case that $\boxed{\grosspayoff(p_{\dpitertwo}(\action|\state),\utilitysymbolbig_{\dpitertwo})\geq\grosspayoff(p_\dpiter(\action|\state),\utilitysymbolbig_{\dpitertwo})}$, that is, GARRI~\eqref{eqn:GARRI} holds.

\subsubsection*{Equivalence of GARRI and GARP}
On closely examining GARRI~\eqref{eqn:GARRI} and GARP~\eqref{eqn:GARP_nonlinear-alt} together, we observe that they are both equivalent under the variable map of statement (1) in Theorem~\ref{thrm:eq}, if NIAS holds. We require the NIAS condition to be true for the equivalence between GARRI and GARP since the `effective' utility in the revealed rational inattention case via the variable map is the surrogate expected utility $\grosspayoff$~\eqref{eqn:exputil-surrogate} that, by definition, is the `maximum' expected utility generated from an action selection policy. The maximum is attained when the Bayesian decision maker chooses the optimal action given the posterior belief, or in other words, NIAS holds.
\vspace{0.2cm}\\
{\bf Statement (2)$\Rightarrow$(3).} Construct a matrix $A\in\reals^{\numdp\times\numdp}$ with elements $A_{j,k} = \grosspayoff(p_{\dpiter}(\action|\state),\utilitysymbolbig_{\dpiter}) - \grosspayoff(p_{\dpitertwo}(\action|\state),\utilitysymbolbig_{\dpiter})$. Since GARRI~\eqref{eqn:GARRI} holds, or equivalently, GARP~\eqref{eqn:GARP_nonlinear-alt} holds under the variable map of Theorem~\ref{thrm:eq}, it is trivial to show the matrix A is cyclically consistent. From \cite[Lemma 2]{FO09} and \cite[Sections 2 and 3]{FOS04}, there exist scalars $\bar{\runcostinst}_\dpiter$ and $\lambda_\dpiter>0$ that satisfy the following set of Afriat-type~\eqref{eqn:Lagrange_nonlinearafriat} feasibility inequalities:
\begin{equation}\label{eqn:proof-corl-garri-feasibility}
    \bar{\runcostinst}_\dpitertwo- \bar{\runcostinst}_\dpiter - \lambda_\dpiter~(~\grosspayoff(p_{\dpiter}(\action|\state),\utilitysymbolbig_{\dpiter})-\grosspayoff(p_{\dpitertwo}(\action|\state),\utilitysymbolbig_{\dpiter})~)\leq 0~\quad \forall~\dpiter,\dpitertwo. 
\end{equation}
In complete analogy to \eqref{eqn:proof_th_2} in Appendix~\ref{app:proof_argmin}, we modify \eqref{eqn:proof-corl-garri-feasibility} into a form that resembles \eqref{eqn:Lagrange_nonlinearafriat_alt} whose feasibility is equivalent to GARP~\eqref{eqn:GARP_nonlinear-alt} by performing the following change of variables:\\
Without loss of generality, we restrict $\bar{\runcostinst}_\dpiter$ to be finite for all $\dpiter=1,2,\ldots,\numdp$. Let $M_\runcostinst<\infty$ denote an arbitrary positive scalar that uniformly bounds $\{\bar{\runcostinst}_\dpiter\}_{\dpiter=1}^\numdp$ from above. Note that since $\bar{\runcostinst}_\dpiter$ is bounded for all $\dpiter$, such an $M_\runcostinst$ exists and $\bar{\runcostinst}_\dpiter> 0$. We now define the variable $\runcostinst_\dpiter = M - \bar{\runcostinst}_\dpiter$ for all $\dpiter$ and rewrite \eqref{eqn:proof-corl-garri-feasibility} in terms of the new variables $\{\runcostinst_\dpiter\}_{\dpiter=1}^\numdp$:
\begin{equation}\label{eqn:final-manipulation-cost-garri}
\begin{split}
    & \bar{\runcostinst}_\dpitertwo- \bar{\runcostinst}_\dpiter - \lambda_\dpiter(\grosspayoff(p_{\dpiter}(\action|\state),\utilitysymbolbig_{\dpiter})-\grosspayoff(p_{\dpitertwo}(\action|\state),\utilitysymbolbig_{\dpiter}))\leq 0,~\forall~\dpiter,\dpitertwo\\
    \Leftrightarrow & - \bar{\runcostinst}_\dpitertwo- (-\bar{\runcostinst}_\dpiter) - \lambda_\dpiter(\grosspayoff(p_{\dpitertwo}(\action|\state),\utilitysymbolbig_{\dpiter})-\grosspayoff(p_{\dpiter}(\action|\state),\utilitysymbolbig_{\dpiter}))\geq 0,~\forall~\dpiter,\dpitertwo\\
    \Leftrightarrow & (M_\runcostinst-\bar{\runcostinst}_\dpitertwo)- (M_\runcostinst-\bar{\runcostinst}_\dpiter) - \lambda_\dpiter(\grosspayoff(p_{\dpitertwo}(\action|\state),\utilitysymbolbig_{\dpiter})-\grosspayoff(p_{\dpiter}(\action|\state),\utilitysymbolbig_{\dpiter}))\geq 0,~\forall~\dpiter,\dpitertwo\\
    \Leftrightarrow & \runcostinst_\dpitertwo- \runcostinst_\dpiter - \lambda_\dpiter(\grosspayoff(p_{\dpitertwo}(\action|\state),\utilitysymbolbig_{\dpiter})-\grosspayoff(p_{\dpiter}(\action|\state),\utilitysymbolbig_{\dpiter}))\geq 0,~\forall~\dpiter,\dpitertwo \\
    \underset{\text{\footnotesize (if NIAS holds)}}{\Leftrightarrow}~~ & \boxed{\runcostinst_\dpitertwo- \runcostinst_\dpiter - \lambda_\dpiter(\grosspayoff(p_{\dpitertwo}(\action|\state),\utilitysymbolbig_{\dpiter})-\sum_{\state,\action}\prior(\state)p_\dpiter(\action|\state)\utilitysymbolbig_\dpiter(\state,\action))\geq 0,~\forall~\dpiter,\dpitertwo \equiv \eqref{eqn:niac-garri-feasibility}}.
    \end{split}
\end{equation}
\noindent \textbf{Statement (3) $\Rightarrow$ (4).} Consider the reconstructed information acquisition cost $\RIcost_{recon}(p(\action|\state))=\max_{\dpiter} \{\runcostinst_\dpiter + \lambda_{\dpiter}(\grosspayoff(p(\action|\state),\utilitysymbolbig_{\dpiter})-\grosspayoff(p_{\dpiter}(\action|\state),\utilitysymbolbig_{\dpiter}))\}$. Clearly, $\RIcost_{recon}$ is monotone and continuous since it is a point-wise maximum of monotone continuous functions. We now show that $\RIcost_{recon}(p(\action|\state)) = \max_{\dpiter} \{\runcostinst_\dpiter + \lambda_{\dpiter}(\grosspayoff(p(\action|\state),\utilitysymbolbig_{\dpiter})-\grosspayoff(p_{\dpiter}(\action|\state),\utilitysymbolbig_{\dpiter}))\}$ rationalizes the dataset $\dataset_{RRI}$~\eqref{eqn:dataset-Bayesian}:\\
Fix index $\dpiter$. Then, $\RIcost_{recon}(p_\dpiter(\action|\state))\geq \runcostinst_\dpiter$ by definition. Also, from \eqref{eqn:final-manipulation-cost-garri}, we have that:
\begin{equation*}
    \runcostinst_\dpiter \geq \runcostinst_\dpitertwo + \lambda_\dpitertwo (\grosspayoff(p_\dpiter(\action|\state),\utilitysymbolbig_{\dpitertwo}) - \grosspayoff(p_\dpitertwo(\action|\state),\utilitysymbolbig_{\dpitertwo}))~\forall \dpitertwo\neq\dpiter
\end{equation*}
Therefore, it is clear that $\RIcost_{recon}(p_\dpiter(\action|\state)) = \runcostinst_\dpiter$~\eqref{eqn:final-manipulation-cost-garri}. Now, let $p(\action|\state)$ denote any feasible response in decision problem $\dpiter$, that is, $\grosspayoff(p(\action|\state),\utilitysymbolbig_\dpiter)\geq\grosspayoff(p_\dpiter(\action|\state),\utilitysymbolbig_\dpiter)$. Then:\\
\begin{equation}\label{eqn:rationalizability-garri}
\begin{split}
\RIcost_{recon}(p(\action|\state))& = \max_{\dpiter} \{\runcostinst_\dpiter + \lambda_{\dpiter}(\grosspayoff(p(\action|\state),\utilitysymbolbig_\dpiter)-\grosspayoff(p_\dpiter(\action|\state),\utilitysymbolbig_\dpiter))\}\\
& \geq \underbrace{\runcostinst_\dpiter}_{=\RIcost_{recon}(p_\dpiter(\action|\state))} +~ \underbrace{\lambda_{\dpiter}(\grosspayoff(p(\action|\state),\utilitysymbolbig_\dpiter)-\grosspayoff(p_\dpiter(\action|\state),\utilitysymbolbig_\dpiter))}_{\geq 0} \geq \RIcost_{recon}(p_\dpiter(\action|\state))\\
\Rightarrow~& \lambda_\dpiter~\grosspayoff(p_\dpiter(\action|\state),\utilitysymbolbig_\dpiter) - \RIcost_{recon}(p_\dpiter(\action|\state)) \geq \lambda_\dpiter~\grosspayoff(p(\action|\state),\utilitysymbolbig_\dpiter) - \RIcost_{recon}(p(\action|\state)),~\forall~p(\action|\state)\\
\Rightarrow~&\boxed{p_{\dpiter}(\action|\state) =\underset{p(\action|\state)}{\argmax}\quad\lambda_\dpiter~\grosspayoff(p(\action|\state),\utilitysymbolbig_\dpiter) - \RIcost_{recon}(p(\action|\state))}
\end{split}
\end{equation}
Since \eqref{eqn:rationalizability-garri} holds for all $\dpiter=1,2,\ldots,\numdp$, the reconstructed cost $\RIcost_{recon}(p(\action|\state)) = \max_{\dpiter} \{\runcostinst_\dpiter + \lambda_{\dpiter}(\grosspayoff(p(\action|\state),\utilitysymbolbig_\dpiter)-\grosspayoff(p_\dpiter(\action|\state),\utilitysymbolbig_\dpiter))\}$ rationalizes the analyst's dataset $\dataset_{RRI}$~\eqref{eqn:dataset-Bayesian}.\vspace{0.2cm}\\
{\bf Statement (4)$\Rightarrow$(1).}The reconstructed information acquisition cost $\RIcost_{recon}(p(\action|\state))=\max_{\dpiter} \{\runcostinst_\dpiter + \lambda_{\dpiter}(\grosspayoff(p(\action|\state),\utilitysymbolbig_\dpiter)-\grosspayoff(p_\dpiter(\action|\state),\utilitysymbolbig_\dpiter))\}$ is identical to $\RIcostest$ defined in \eqref{eqn:reconstruct_RI}. By construction, the cost $\RIcost_{recon}$ is monotone, continuous (since it is a point-wise maximum of finitely many monotone continuous segments) and rationalizes the Bayesian decision maker's actions~\eqref{eqn:rationalizability-garri}.\hfill\qedsymbol

\subsubsection*{Showing reconstructed information acquisition cost is weakly monotone, mixture feasible and normalized}
Consider the reconstructed information acquisition cost $\RIcostest$~\eqref{eqn:reconstruct_RI}. The cost $\RIcostest$ is ordinal, i.e.\,, any monotone transformation of $\RIcostest$ and $\lambda_\dpiter$ rationalizes the dataset $\dataset_{RRI}$~\eqref{eqn:dataset-Bayesian} equally well. Hence, without loss of generality, we normalize $\RIcostest$~\eqref{eqn:reconstruct_RI} as:
\begin{align}
    \RIcostest_{norm}(p(\action|\state)) & =  \max_{\dpiter}~\{ \runcostinst_{\dpiter} +\lambda_{\dpiter}\left( \grosspayoff(p(\action|\state),\utilitysymbolbig_\dpiter) - \grosspayoff(p_\dpiter(\action|\state),\utilitysymbolbig_\dpiter)\right)\} - \RIcostest^{\ast},~\text{where}\nonumber\\
    \RIcostest^{\ast} & = \max_{\dpiter}~\{ \runcostinst_{\dpiter} +\lambda_{\dpiter}\left( \grosspayoff(p_0(\action|\state),\utilitysymbolbig_\dpiter) - \grosspayoff(p_\dpiter(\action|\state),\utilitysymbolbig_\dpiter)\right)\}. \label{eqn:cost_est}
\end{align}
In~\eqref{eqn:cost_est}, $\grosspayoff$ is the surrogate expected utility defined in \eqref{eqn:exputil-surrogate} and $p_0(\action|\state)$ is the non-informative (uniform conditional probability) action selection policy, i.e.\,, $p_0(\action|\state)=1/|\actionset|$ for all $\action,\state$. Combining \eqref{eqn:niac-garri-feasibility} and \eqref{eqn:cost_est} above gives $\RIcostest_{norm}(p_{\dpiter}(\action|\state))=\runcostinst_{\dpiter}-\RIcostest^{\ast}$. To show $\RIcostest_{norm}$~\eqref{eqn:cost_est} rationalizes dataset $\dataset_{RRI}$~\eqref{eqn:dataset-Bayesian} in Corollary~\ref{corl:niac-to-garp}, fix index $\dpiter$ and consider any action selection policy $p(\action|\state)$ such that $\RIcostest_{norm}(p(\action|\state))\leq \RIcostest_{norm}(p_\dpiter(\action|\state))$. By definition~\eqref{eqn:cost_est}, $0\geq \RIcostest_{norm}(p(\action|\state))-\RIcostest_{norm}(p_\dpiter(\action|\state))\geq \lambda_\dpiter(\grosspayoff(p(\action|\state),\utilitysymbolbig_\dpiter)-\grosspayoff(p_\dpiter(\action|\state),\utilitysymbolbig_\dpiter))$, which implies $\grosspayoff(p(\action|\state),\utilitysymbolbig_\dpiter)\leq\grosspayoff(p_\dpiter(\action|\state),\utilitysymbolbig_\dpiter)$. This inequality holds for all $\dpiter$. Hence, $\RIcostest_{norm}$~\eqref{eqn:cost_est} rationalizes the $\dataset_{RRI}$~\eqref{eqn:dataset-Bayesian}.

We now use Lemma~\ref{lem:partialorder_relax} to show that the reconstructed cost $\RIcostest_{norm}$ \eqref{eqn:cost_est} is: (i) weakly monotone in information, mixture feasible and normalized as theorized in \cite[Theorem 2]{CD15}.
\begin{compactenum}
\item[K1.] Weak monotonicity in information. {\em The cost $\RIcostest_{norm}$ is weakly monotonic in information if for any two action selection policies $p(\action|\state), \hat{p}(\action|\state)$, we have  $\RIcostest_{norm}(\hat{p}(\action|\state)) \leq \RIcostest_{norm}({p}(\action|\state))$, when $p(\action|\state) \BD \hat{p}(\action|\state)$, where $\BD$ stands for `Blackwell dominates'.}\\
Condition K1 can be viewed as a monotonicity condition with respect to the Blackwell order.\\
\noindent {\bf Proof.} Since $p(\action|\state) \BD \hat{p}(\action|\state)$, Lemma~\ref{lem:partialorder_relax} ensures
the following inequalities hold:
\begin{align*}
    \RIcostest_{norm}(\hat{p}(\action|\state)) & =  \max_{\dpiter}~\{ \runcostinst_{\dpiter} +\lambda_{\dpiter}\left( \grosspayoff(\hat{p}(\action|\state),\utilitysymbolbig_\dpiter) - \grosspayoff(p_\dpiter(\action|\state),\utilitysymbolbig_\dpiter)\right)\} - \RIcostest^{\ast}\\
    & \leq \max_{\dpiter}~\{ \runcostinst_{\dpiter} +\lambda_{\dpiter}\left( \grosspayoff(p(\action|\state),\utilitysymbolbig_\dpiter) - \grosspayoff(p_\dpiter(\action|\state),\utilitysymbolbig_\dpiter)\right)\} - \RIcostest^{\ast} = \RIcostest_{norm}({p}(\action|\state))\\
    \Rightarrow & \boxed{\RIcostest_{norm}(\hat{p}(\action|\state)) \leq \RIcostest_{norm}(p(\action|\state))}
\end{align*}
\item Mixture feasibility. {\em The cost $\RIcostest_{norm}$ is mixture feasible if for action selection policies $p(\action|\state),p'(\action|\state),p''(\action|\state)$ related as $p(\action|\state) = \theta p'(\action|\state) + (1-\theta) p''(\action|\state),~\theta>0$, cost $\RIcostest_{norm}$ satisfies $\RIcostest_{norm}(p(\action|\state)) \leq \theta~\RIcostest_{norm}(p'(\action|\state)) + (1-\theta)~\RIcostest_{norm}(p''(\action|\state))$.}\\
\noindent {\bf Proof.}
\begin{align*}
    \RIcostest_{norm}(p(\action|\state))+ \RIcostest^{\ast}= &\max_{\dpiter}~\{ \runcostinst_{\dpiter} +\lambda_{\dpiter}\left( \grosspayoff(p(\action|\state),\utilitysymbolbig_\dpiter) - \grosspayoff(p_\dpiter(\action|\state),\utilitysymbolbig_\dpiter)\right)\} \\
    = &\max_{\dpiter}~\{ \runcostinst_{\dpiter} +\lambda_{\dpiter}\left( \grosspayoff(\theta p'(\action|\state) + (1-\theta) p''(\action|\state),\utilitysymbolbig_\dpiter) - \grosspayoff(p_\dpiter(\action|\state),\utilitysymbolbig_\dpiter)\right)\}\\
    \leq & \max_{\dpiter}~\{ \runcostinst_{\dpiter} +\lambda_{\dpiter}\left( \theta~\grosspayoff(p'(\action|\state),\utilitysymbolbig_\dpiter) + (1-\theta)\grosspayoff(p''(\action|\state),\utilitysymbolbig_\dpiter) - \grosspayoff(p_\dpiter(\action|\state),\utilitysymbolbig_\dpiter)\right)\}\\
    &~(\text{since the surrogate expected utility }\grosspayoff(\cdot,\utilitysymbolbig_\dpiter)\text{ is convex in }p(\action|\state))\\
   \leq&~ \theta~\max_{\dpiter}~\{ \runcostinst_{\dpiter} +\lambda_{\dpiter}\left( \grosspayoff(p'(\action|\state),\utilitysymbolbig_\dpiter) - \grosspayoff(p_\dpiter(\action|\state),\utilitysymbolbig_\dpiter)\right)\}\\
   +&(1-\theta)~\max_{\dpiter}~\{ \runcostinst_{\dpiter} +\lambda_{\dpiter}\left( \grosspayoff(p''(\action|\state),\utilitysymbolbig_\dpiter) - \grosspayoff(p_\dpiter(\action|\state),\utilitysymbolbig_\dpiter)\right)\}\\
   &~(\text{since the max operation is convex})\\
   \Rightarrow & \RIcostest_{norm}(p(\action|\state)) \leq \theta (\RIcostest_{norm}(p'(\action|\state)) + \RIcostest^\ast) + (1-\theta) (\RIcostest_{norm}(p''(\action|\state)) + \RIcostest^\ast) - \RIcostest^\ast\\
  \Rightarrow &\boxed{\RIcostest_{norm}(p(\action|\state))  \leq \theta~\RIcostest_{norm}(p'(\action|\state)) + (1-\theta)~\RIcostest_{norm}(p''(\action|\state))} 
\end{align*}
\item Normalization. The cost $\RIcostest_{norm}$ is normalized if $\RIcostest_{norm}(p_0(\action|\state))=0$, where $p_0(\action|\state) = 1/|\actionset|$ (uninformative action selection policy).\\
\noindent{\bf Proof.} This holds true from the definition of $\RIcostest_{norm}$ in~\eqref{eqn:cost_est}.
\end{compactenum}

\end{document}